\documentclass[onecolumn,3p,preprint,a4paper,10pt]{elsarticle}
\usepackage{amsmath,natbib,amssymb}

\defcitealias{pm05}{PM05}

\begin{document}
%opening
\title{Constrained Hyperbolic Divergence Cleaning for Smoothed Particle Magnetohydrodynamics}

\author{Terrence S. Tricco}
\ead{terrence.tricco@monash.edu}

\author{Daniel J. Price}
\ead{daniel.price@monash.edu}

\address{Monash Centre for Astrophysics \& School of Mathematical Sciences, Monash University, Melbourne Vic, 3800, Australia}

\begin{abstract}
We present a constrained formulation of Dedner et al's hyperbolic/parabolic divergence cleaning scheme for enforcing the $\nabla\cdot{\bf B} = 0$ constraint in Smoothed Particle Magnetohydrodynamics (SPMHD) simulations. The constraint we impose is that energy removed must either be conserved or dissipated, such that the scheme is guaranteed to decrease the overall magnetic energy. This is shown to require use of conjugate numerical operators for evaluating $\nabla\cdot{\bf B}$ and $\nabla \psi$ in the SPMHD cleaning equations. The resulting scheme is shown to be stable at density jumps and free boundaries, in contrast to an earlier implementation by Price \& Monaghan (2005). Optimal values of the damping parameter are found to be $\sigma$ = 0.2--0.3 in 2D and $\sigma$ = 0.8--1.2 in 3D. With these parameters, our constrained Hamiltonian formulation is found to provide an effective means of enforcing the divergence constraint in SPMHD, typically maintaining average values of $h \vert\nabla\cdot{\bf B}\vert / \vert{\bf B}\vert$ to 0.1--1\%, up to an order of magnitude better than artificial resistivity without the associated dissipation in the physical field. Furthermore, when applied to realistic, 3D simulations we find an improvement of up to two orders of magnitude in momentum conservation with a corresponding improvement in numerical stability at essentially zero additional computational expense.
\end{abstract}

\maketitle

\section{Introduction}

A key problem in numerical magnetohydrodynamics (MHD) is maintenance of the divergence constraint, $\nabla \cdot {\bf B} = 0$ from Maxwell's equations.  If this is not maintained, a spurious force parallel to the magnetic field appears which can lead to numerical instability.  A variety of methods have been developed to combat numerical divergence error, including Brackbill and Barnes \cite{1980JCoPh..35..426B} projection method, Evans and Hawley's \cite{1988ApJ...332..659E} constrained transport, and Powell's \cite{1994arsm.rept.....P,pea99} eight wave approach or variants thereof.  In general, these methods either aim to ``clean'' any divergence of the magnetic field that has been generated, or to alter the MHD formulation so that the divergence constraint is satisfied by construction.  T{\'o}th \cite{toth00} provides an excellent comparison of these schemes.

However, even methods such as constrained transport which guarantee divergence free magnetic fields only do so in a particular discretisation. This means that numerical artefacts may still be present in different discretisations --- such as those used in the force terms.  The goal of all methods aimed at maintaining the divergence-free constraint is not to keep $\nabla\cdot{\bf B}$ exactly zero, but rather to prevent the growth of these numerical artefacts.

Several attempts have been made to enforce the divergence constraint in the Smoothed Particle Hydrodynamics (SPH) implementation of magnetohydrodynamics (SPMHD).  One method is to formulate the magnetic field in terms of the Euler potentials \cite{1970AmJPh..38..494S} $\alpha$ and $\beta$, setting $\bf{B} = \nabla \alpha \times \nabla \beta$, which guarantees zero divergence of the magnetic field by construction.  Due to the Lagrangian nature of SPH, the scalar variables are advected exactly, which means the magnetic field can be reconstructed simply from the particle positions relative to the initial conditions.  While this method has found reasonable success (e.g. \cite{2009MNRAS.397..733K, 2007MNRAS.377...77P}), the range of problems it is suitable for is limited.  For example, winding motions cannot be modelled past one rotation as the field is essentially ``reset'' with each turn.  Price \cite{price10} investigated use of the vector potential formulation of the magnetic field as a way to overcome these limitations while still retaining the guarantee of zero physical divergence in the field.  However, formulating the equations of motion in terms of the vector potential were found to be unstable, and there were significant difficulties were found with the time evolution of the vector potential, from which Price concluded that this was not a viable approach. 

Divergence cleaning schemes have also been used in SPMHD.  The simplest approach is to rely on artificial resistivity to restrict growth of the divergence of the magnetic field.  Artificial resistivity was introduced by Price and Monaghan \cite[][henceforth PM05]{pm05}) to capture shocks and discontinuities in the magnetic field (see \cite{price08} for a general discussion on discontinuities in SPH), and corresponds to adding physical diffusion terms of the form \cite{pm04a,price12}
\begin{equation}
 \left( \frac{\partial {\bf B}}{\partial t} \right)_\text{resist} \equiv \eta \nabla^{2} {\bf B} = \eta \nabla \left(\nabla \cdot \bf{B}\right)  - \eta \nabla  \times ( \nabla \times {\bf B}),
 \label{eq:resist}
\end{equation}
where $\eta$ is the resistivity parameter (in SPMHD $\eta \sim \alpha_{\rm B} v_{\rm sig} h$ where $h$ is the smoothing length, $v_{\rm sig}$ is the maximum signal velocity and $\alpha_{\rm B}$ is a dimensionless parameter of order unity). This means that a form of divergence cleaning may be obtained as a byproduct at no additional computational expense.  B{\"u}rzle et al \cite{2011MNRAS.412..171B} found this to be the case in star formation simulations, where maintaining the divergence constraint has proved challenging \cite{pf10b}. The caveat to relying on artificial resistivity to control divergence error is that it diffuses both physical and unphysical components of the magnetic field (Eq.~\ref{eq:resist}).  Resistivity should not be increased in strength just to address divergence errors, as doing so will also weaken the physical field.

 Dedner et al.'s \cite{2002JCoPh.175..645D} hyperbolic divergence cleaning scheme has found popular use in both Eulerian (ie, \cite{2010JCoPh.229.2117M}, \cite{2009ApJ...696...96W}) and Lagrangian codes (\cite{2011MNRAS.414..129G}, \cite{2011MNRAS.tmp.1536P}).  To facilitate cleaning of divergence errors, an additional field $\psi$ is coupled to the magnetic field. The Dedner et al. scheme was originally adapted to SPMHD by \citetalias{pm05}, but was not adopted for two main reasons: i) the reduction in divergence error was relatively small (a factor of $\sim 2$--$3$) and ii) on certain test cases it was found that it could lead to an increase in the divergence error. As such, its use was not recommended \cite[c.f.][]{price12}.
 
 Our aim in this paper is to provide a formulation of hyperbolic divergence cleaning for SPMHD that is guaranteed to be stable and ensures a negative definite contribution to the magnetic energy. This means that the divergence cleaning is guaranteed to decrease the errors associated with non-zero divergence of the magnetic field, in turn leading to a method that is suitable for general use in SPMHD simulations.
 
We begin with a review of the MHD equations and the divergence constraint for continuum and SPMHD systems (\S\ref{sec:mhd}).  In \S\ref{sec:hyperbolic}, we discuss hyperbolic cleaning as part of the ideal MHD equations, and in \S\ref{sec:continuum-energy-conservation}, define an energy term associated with the $\psi$ field. Using this energy term, we derive a new form for the $\psi$-evolution equation which conserves total energy (\S\ref{sec:idealmhdenergy}).  In \S\ref{sec:discretised-hyperbolic}, the discretisation of hyperbolic cleaning into SPMHD is discussed and we show how the constraint of energy conservation can be used to construct a formulation that is numerically stable.  In particular, this leads to a requirement for the discretisation of $\nabla\cdot{\bf B}$ and $\nabla\psi$ used in the induction and $\psi$-evolution equations to form a conjugate pair (\S\ref{sec:spmhd-clean-diff} and \S\ref{sec:spmhd-clean-symm}).  Importantly, we prove that the dissipative (parabolic) term in the evolution of $\psi$ gives a negative definite contribution to magnetic energy (\S\ref{sec:negdef}).  Our new, constrained formulation of hyperbolic cleaning in SPMHD is then applied to a suite of test problems designed to evaluate all aspects of the algorithm and to derive parameter ranges suitable for general use (\S\ref{sec:tests}). The final test (\S\ref{sec:jet}) is drawn from our current work on applying the method to star formation problems and shows that our technique performs well in practice, dramatically improving the accuracy and robustness of realistic SPMHD simulations in three dimensions. The results are discussed and summarised in \S\ref{sec:summary}.

\section{Smoothed particle magnetohydrodynamics}
\label{sec:mhd}
\subsection{Equations of ideal magnetohydrodynamics}

The ideal MHD equations are formulated by the coupling of the Euler equations of fluid flow with Maxwell's equations and the Lorentz force, under the assumption of the fluid being a perfect conductor with zero net charge.  Electric field contributions to the dynamics are therefore considered negligible, and the resulting equations of motion are given by the continuity, momentum, and induction equations,
\begin{eqnarray}
\frac{{\rm d}\rho}{{\rm d}t} & = & -\rho \nabla \cdot {\bf v} , \label{eq:cty} \\
 \frac{{\rm d} \bf{v}}{{\rm d}t} & = & - \frac{\nabla P}{\rho} + \frac{1}{\rho} \nabla \cdot \bf{M}, \\
 \frac{{\rm d}\bf{B}}{{\rm d}t} & = & \left(\bf{B} \cdot \nabla\right) \bf{v} - \bf{B} \left(\nabla \cdot \bf{v}\right), \label{eq:ind}
\end{eqnarray}
where $\rho$ is the density, ${\bf v}$ is the fluid velocity, $P$ is the thermal pressure, ${\bf B}$ is the magnetic field and ${\rm d}/{\rm d}t$ is the comoving (Lagrangian) time derivative. The magnetic force is written as the divergence of the Maxwell stress tensor,
\begin{equation}
\nabla \cdot {\bf M} =  \frac{1}{\mu_0} \nabla\cdot \left( -\tfrac{1}{2} B^2 {\bf I} + {\bf B}{\bf B} \right) =  \frac{1}{\mu_0}  \left[ -\nabla \left( \tfrac{1}{2} B^2 \right) + ({\bf B}\cdot\nabla){\bf B} + {\bf B} (\nabla\cdot{\bf B}) \right].
\label{eq:divM}
\end{equation}
The first term provides a magnetic pressure analogous to thermal pressure, while the second term gives the tension force perpendicular to the magnetic field lines. The third term occurs only for non-zero divergence of the magnetic field and results in an unphysical force parallel to the field lines \cite{1980JCoPh..35..426B, 2000JCoPh.160..649J, 2002JCoPh.179...95D}. 

\subsection{The divergence constraint in MHD}
 The divergence constraint enters the MHD equations only as an initial condition,
\begin{equation}
 \frac{\partial}{\partial t} \left(\nabla \cdot {\bf B}\right) = 0,
\end{equation}
or, with the ``source term'' approach \citep{1994arsm.rept.....P,2000JCoPh.160..649J} corresponding to Eq.~\ref{eq:ind},
\begin{equation}
\frac{\partial (\nabla \cdot {\bf B})}{\partial t} + \nabla\cdot ({\bf v}\nabla \cdot {\bf B}) = 0,
\end{equation}
which is identical in form to the continuity equation (Eq.~\ref{eq:cty}) but with the density replaced by $\nabla \cdot {\bf B}$.

\subsection{Discrete form of ideal MHD equations}
\label{sec:spmhd-equations}
The discrete SPMHD representation of the ideal MHD equations (\ref{eq:cty})--(\ref{eq:ind}) are given by \cite{pm04a,pm05}
\begin{align}
\rho_a = & \sum_b m_b W_{ab} (h_a), \hspace{1cm} h_{a} = h_{\rm fac} \left( \frac{m_{a}}{\rho_{a}}\right)^{1/n_\text{dim}}, \label{eq:sphcty} \\
\frac{{\rm d}{\bf{v}}_a}{{\rm d}t} = 
&- \sum_b m_b \left[\frac{P_a}{\Omega_a \rho_a^2} \nabla_a W_{ab}(h_a) + \frac{P_b}{\Omega_b \rho_b^2} \nabla_b W_{ab}(h_b) \right] \nonumber \\
&+ \sum_b m_b \left[\frac{{\bf M}_a}{\Omega_a \rho_a^2}\cdot \nabla_a W_{ab}(h_a) + \frac{{\bf M}_{b}}{\Omega_b \rho_b^2} \cdot \nabla_b W_{ab}(h_b) \right], \label{eq:spmhd-momentum-eqn} \\
 \frac{{\rm d}{\bf{B}}_a}{{\rm d}t} = & - \frac{1}{\Omega_a \rho_a} \sum_b m_b \left[ {\bf{v}}_{ab} \left( {\bf{B}}_a \cdot \nabla_a W_{ab}(h_a) \right) - {\bf{B}}_a \left( {\bf{v}}_{ab} \cdot \nabla_a W_{ab}(h_a) \right) \right], \label{eq:sphind}
\end{align}
where $W_{ab} (h_{a}) \equiv W(\vert {\bf r}_{a} - {\bf r}_{b} \vert, h_{a})$ is the SPH smoothing kernel, $h$ is the smoothing length, $m$ is the particle mass, the summation is over the particle's neighbours and we use the subscripts $a$ and $b$ to refer to the particle index. In the variable smoothing length formulation of SPH, $\Omega$ is a term that arises due to gradients in the smoothing length and Eq.~\ref{eq:sphcty} is a non-linear set of equations that is solved iteratively for both $h$ and $\rho$ (for details see \cite{pm04b,pm07}).

\subsection{$\nabla\cdot{\bf B}$ in SPMHD}

There are two basic operators for calculating first derivatives in SPH, which we refer to as `difference' and `symmetric' \cite{price12}. For the divergence of the magnetic field, the difference operator is given by
 \begin{equation}
 \left( \nabla \cdot {\bf B} \right)_a = -\frac{1}{\Omega_a \rho_a} \sum_b m_b \left( {\bf B}_a - {\bf B}_b \right) \cdot \nabla_a W_{ab}(h_a) ,
\label{eq:divbdiff}
\end{equation}
while the symmetric operator is
\begin{equation}
 \left( \nabla \cdot {\bf B} \right)_a = \rho_a \sum_b m_b \left[ \frac{{\bf B}_a}{\Omega_a \rho_a^2} \cdot \nabla_a W_{ab}(h_a) + \frac{{\bf B}_b}{\Omega_b \rho_b^2} \cdot \nabla_a W_{ab}(h_b) \right].
\label{eq:divbsym}
\end{equation}

\subsection{Tensile instability correction}
\label{sec:tensile-instability-correction}
\citet{pm04b} showed that the discrete SPMHD momentum equation (Eq.~\ref{eq:spmhd-momentum-eqn}) can be derived self-consistently from a Lagrangian using the discrete form of the continuity and induction equations (i.e., Eqs.~\ref{eq:sphcty} and \ref{eq:sphind}, respectively) as constraints, ensuring that this is a consistent set of equations and that energy and momentum are conserved exactly. However, the conservative form of the momentum equation (Eq.~\ref{eq:spmhd-momentum-eqn}) inherently contains the unphysical force parallel to the field lines (c.f. Eq.~\ref{eq:divM}). The effect is that a numerical instability --- the ``tensile instability'' --- occurs when $\tfrac{1}{2} {\bf{B}}^2 > P$, causing particles to unphysically clump together.  This can be effectively countered using B{\o}rve et al's \cite{2001ApJ...561...82B} approach to subtract the ${\bf{B}} (\nabla \cdot {\bf{B}})$ source term from the magnetic force,
\begin{equation}
 \left(\frac{{\rm d}{\bf{v}}_a}{{\rm d}t}\right)_{\nabla \cdot \bf{B}} = - {\hat{\beta}} {\bf B}_a \sum_b m_b \left[ \frac{{\bf B}_a}{\Omega_a \rho_a^2} \cdot \nabla_a W_{ab}(h_a) + \frac{{\bf B}_b}{\Omega_b \rho_b^2} \cdot \nabla_a W_{ab}(h_b) \right] .
\label{eq:tensile-instability-correction}
\end{equation} 
which is added to Eq.~\ref{eq:spmhd-momentum-eqn}.  Since the instability manifests only when $\tfrac{1}{2} {\bf{B}}^2 > P$, \citet{bot04} introduce an adjustable parameter $\hat{\beta}$, showing that it is sufficient to use $\hat{\beta} = \tfrac{1}{2}$ to correct the instability in the magnetic pressure-dominated regime. Indeed, recently \cite{2011arXiv1112.0340B} have recommended using $\hat{\beta} = \tfrac{1}{2}$ for general SPMHD calculations. However, we find in this paper (\S\ref{sec:halfdivb}) that using $\hat{\beta} < 1$ can produce numerical artefacts (c.f.~Fig.~\ref{fig:halfdivb}). We therefore strongly recommend using $\hat{\beta} = 1$ and adopt this throughout the paper unless otherwise specified. Note that with $\hat{\beta} = 1$ the induction and momentum equations are formally equivalent to Powell's eight wave approach \cite{1994arsm.rept.....P}.

\subsection{Dissipative terms in SPMHD}
\label{sec:resist}
Dissipative terms are added to SPMHD in the form of artificial viscosity, resistivity, and thermal conduction in order to treat discontinuities and shocks.  The form of artificial viscosity and thermal conduction used in this work is that of Monaghan's \cite{1997JCoPh.136..298M}, derived by analogy with Riemann solvers. The form of artificial resistivity used is that of \citet{pm05}, given by
\begin{equation}
 \left(\frac{{\rm d}{\bf B}_a}{{\rm d}t}\right)_\text{diss} = \rho_a \sum_b m_b \frac{\alpha_B v_\text{sig}}{\overline{\rho}_{ab}^2} ({\bf B}_a - {\bf B}_b) \hat{{\bf r}} \cdot \nabla_a W_{ab}.
\end{equation}
Each particle has $\alpha_B$ set individually, and is allowed to vary according in the range $\alpha_{B} \in [0,1]$ according to 
\begin{equation}
 \frac{{\rm d}\alpha_{B,a}}{{\rm d}t} = - \frac{\alpha_{B,a}}{\tau} + \max \left(\frac{ |\nabla \times {\bf B}_a|}{\sqrt{\mu_0 \rho_a}} , \frac{ |\nabla \cdot{\bf B}_a|}{\sqrt{\mu_0 \rho_a}}\right) ,
\end{equation}
where $\tau = h_a / \sigma_B v_\text{sig}$ with $\sigma_B = 0.1$.  Thus resistivity is only strong at large gradients in the magnetic field.

\section{Hyperbolic divergence cleaning}
\label{sec:hyperbolic}

\subsection{Hyperbolic divergence cleaning for the MHD equations}

Hyperbolic divergence cleaning involves the introduction of a new scalar field, $\psi$, that is coupled to the magnetic field by a term appearing in the induction equation,
\begin{equation}
 \left( \frac{{\rm d}\bf{B}}{{\rm d}t} \right)_\psi = - \nabla \psi,
\label{eq:induction_psi}
\end{equation}
and the field $\psi$ evolves according to
\begin{equation}
 \frac{{\rm d} \psi}{{\rm d} t} = -c_h^2 \nabla \cdot {\bf{B}} - \frac{\psi}{\tau}.
\label{eq:psi_evolution}
\end{equation}
In the comoving frame of the fluid, Eq.~\ref{eq:induction_psi} and \ref{eq:psi_evolution} combine to produce a damped wave equation
\begin{equation}
 \frac{\partial^{2} (\nabla \cdot {\bf B})}{\partial t^2} - c_{h}^{2} \nabla^2 (\nabla \cdot {\bf B}) + \frac{1}{\tau} \frac{\partial (\nabla \cdot {\bf B})}{\partial t} = 0.
\end{equation}
 The equation above shows that this approach spreads divergence of the magnetic field like a wave away from a source, diluting the initial divergence over a larger area, enabling the parabolic (diffusion) term, $-\psi/\tau$, to act more effectively in reducing it to zero. The wave speed, $c_{h}$, is chosen to be the fastest speed permissible by the time step, typically equal to the speed of the fast MHD wave. A key consideration is setting the damping strength correctly to achieve critical damping of the wave, which maximises the benefit of wave propagation without damping being too weak.  Dedner et al. suggested using $1/\tau = c_h c_r$ where $c_{r} = 0.18$, though this is problematic as $c_{r}$ is not a dimensionless quantity.  Instead, PM05 define 
\begin{equation}
\frac{1}{\tau} \equiv \frac{\sigma c_{h}}{h},
\label{eq:sigmadef}
\end{equation}
where $h$ is the smoothing length and $\sigma$ is a dimensionless quantity specifying the damping strength. \citetalias{pm05} found that optimal cleaning was obtained for $\sigma \in [0.4,0.8]$ in their tests.  A similar form was also adopted by Mignone and Tzeferacos \cite{2010JCoPh.229.2117M} in their Eulerian code, who suggested values $\sigma \in [0,1]$.

\subsection{Energy associated with the $\psi$ field}
\label{sec:continuum-energy-conservation}

For later purposes it will be useful to define an energy term associated with the $\psi$ field, $e_{\psi}$ (here defined as the energy per unit mass). Specifically, the energy should be defined such that, in the absence of damping terms, any change in magnetic energy should be balanced by a corresponding change in $e_{\psi}$. This is not merely a book-keeping exercise, as it will enable us to construct a formulation of hyperbolic divergence cleaning in SPMHD that is guaranteed to be stable.

If we consider the closed system of equations formed by equations \ref{eq:induction_psi} and \ref{eq:psi_evolution}, the total energy of the system can be specified according to
\begin{equation}
E = \int \left[ \frac{B^2}{2 \mu_0 \rho} + e_\psi \right] \rho {\rm d}V.
\label{eq:tote}
\end{equation}
Conservation of energy in this subsystem implies
\begin{equation}
\frac{{\rm d}E}{{\rm d}t} = \int \left[ \frac{\bf B}{\mu_0 \rho} \cdot \left( \frac{{\rm d} \bf B}{{\rm d}{t}} \right)_\psi + \frac{{\rm d} e_\psi}{{\rm d}t} \right]  \rho {\rm d}V = 0,
\end{equation}
where we have used the fact that ${\rm d}(\rho {\rm d}V)/{\rm d}t = 0$.
We assume that the time derivative of $e_{\psi}$ can be related to the time derivative of $\psi$, giving
\begin{equation}
\int \left[ \frac{\bf B}{\mu_0\rho} \cdot \left( \frac{{\rm d} \bf B}{{\rm d}{t}} \right)_\psi + \chi \frac{{\rm d} \psi}{{\rm d} t}  \right] \rho{\rm d}V = 0,
\end{equation}
where $\chi$ is an unspecified variable to be determined. Using Eqs.~\ref{eq:induction_psi} and \ref{eq:psi_evolution} in the absence of damping gives
\begin{equation}
\label{eq:psi_energy_equal}
 \int \left[ -\frac{\bf{B}}{\mu_0\rho} \cdot \nabla \psi - \chi c_h^2 \nabla \cdot {\bf{B}} \right] \rho {\rm d}V = 0.
\end{equation}
Integrating the first term by parts, we obtain
\begin{equation}
 \int \left[ \frac{\psi}{\mu_0\rho}  - \chi c_h^2 \right] (\nabla \cdot {\bf{B}}) \rho {\rm d}V -  \frac{1}{\mu_0} \int_s \psi {\bf{B}} \cdot {\rm d}\hat{\bf{s}} = 0.
\label{eq:psi_energy_derivation}
\end{equation}
We take the surface integral in equation \ref{eq:psi_energy_derivation} to be zero.  If the bounding surface is taken to be at infinity, then this assumption is reasonable as it should be expected that the amplitude of a divergence wave would be diluted to zero at such a limit.  For closed systems, it is not clear how the surface term should be treated. However, similar surface terms appear in the standard SPH formulation and are treated by the addition of diffusion terms to capture discontinuities \cite{price08}. For this reason we investigated adding an artificial $\psi$-diffusion term to account for $\psi$-discontinuities, but found no particular advantage to using this in practice (see \ref{sec:dissipation_term}).

From (\ref{eq:psi_energy_derivation}) we conclude that energy conservation requires $\chi \equiv \psi / \mu_0 \rho c_h^2$, and therefore that the specific energy of the $\psi$ field should be defined according to
\begin{equation}
e_\psi \equiv \frac{\psi^2}{2 \mu_0 \rho c_h^2}.
\end{equation}

\subsubsection{Energy conservation as part of the ideal MHD equations}
\label{sec:idealmhdenergy}

Considering total energy conservation with hyperbolic divergence cleaning included as part of the set of ideal MHD equations, additional terms relating to ${\rm d}\rho/{\rm d}t$ appear in the preceding analysis (along with kinetic and other energy terms). Any terms not involving $\psi$ do not need to be considered as they conserve energy together \cite[see][]{pm04b}, so energy conservation reduces to the condition
\begin{equation}
\label{eq:psi_energy_density_variation}
\int \left[ \frac{\bf B}{\mu_0 \rho} \cdot \left( \frac{{\rm d} \bf B}{{\rm d}{t}} \right)_\psi + \frac{\psi}{\mu_0 \rho c_h^2} \frac{{\rm d}\psi}{{\rm d}t} - \frac{\psi^2}{2 \mu_0 \rho_a^2 c_h^2} \frac{{\rm d}\rho}{{\rm d}t} \right]  \rho {\rm d}V = 0.
\end{equation}
The first two terms balance each other, however, the third term remains. There are several possible approaches to ensuring total energy conservation with respect to this term. One approach which we explored was to derive the MHD$+$cleaning equations from a Lagrangian that includes the $e_{\psi}$ term. The result is that an additional isotropic pressure term, $-\tfrac12 \psi^{2}/(\mu_{0} c_{h}^{2})$, appears in the momentum equation. Since it is undesirable to change the physical forces in the system, we instead adopt a simpler approach, which is to slightly modify the evolution equation for $\psi$.

From the continuity equation (Eq.~\ref{eq:cty}), we can deduce that the third term in Eq.~\ref{eq:psi_energy_density_variation} will be balanced by replacing Eq.~\ref{eq:psi_evolution} with
\begin{equation}
 \frac{{\rm d} \psi}{{\rm d} t} = -c_h^2 \nabla \cdot {\bf{B}} - \frac{\psi}{\tau} - \tfrac{1}{2} \psi \nabla \cdot {\bf v}.
\label{eq:psi_evolution_halfdivv}
\end{equation}

\section{Hyperbolic divergence cleaning in SPMHD}
\label{sec:discretised-hyperbolic}

\subsection{Hyperbolic divergence cleaning in SPMHD}
 Hyperbolic divergence cleaning in SPMHD can be constructed for either the difference (Eq.~\ref{eq:divbdiff}) or symmetric (Eq.~\ref{eq:divbsym}) measure of $\nabla \cdot {\bf B}$ by using the appropriate operator in Eq.~\ref{eq:psi_evolution_halfdivv}. While both measure the divergence of the magnetic field, they do not provide the same measurement.  For example, if a random distribution of particles is given a uniform magnetic field, the difference form will measure precisely zero --- since the magnetic field is equal for all particles --- but the symmetric form will not because it will reflect the disordered particle arrangement. Thus, it may be expected that the difference operator in general gives a more accurate measure of $\nabla\cdot{\bf B}$ and should be the operator used for cleaning.
On the other hand, it is the symmetric form which is used in the momentum equation (Eq.~\ref{eq:spmhd-momentum-eqn}) and correspondingly in the tensile instability correction (Eq.~\ref{eq:tensile-instability-correction}), and cleaning in this operator may be more effective at improving the conservation of momentum. Thus in the context of divergence cleaning, it is not clear {\it a priori} which of the two should be preferred. This is one of the questions we will seek to answer in our tests.

 It is also not clear how the operator for $\nabla \psi$ should be chosen. In \citetalias{pm05} a difference operator was used for both $\nabla\cdot{\bf B}$ and $\nabla \psi$. However, the choice of operator for $\nabla\psi$ turns out to be an important issue in ensuring a stable method.

\subsection{Energy conservation of discretised hyperbolic divergence cleaning}
\label{sec:sph_energy_conserv}
 The key constraint we wish to impose on our divergence cleaning scheme is that the total magnetic energy should never increase due to cleaning. That is, any magnetic energy transferred into the $\psi$-field should either be conserved or dissipated. Specifically, in the absence of damping terms, the propagation of divergence waves should conserve energy, not only in the continuum limit but also in the discrete system. We can thus use the $e_{\psi}$ derived in \S\ref{sec:continuum-energy-conservation} to derive stable formulations of hyperbolic divergence cleaning for SPMHD --- for either difference or symmetric $\nabla\cdot{\bf B}$ operators.
  
 As in \S\ref{sec:continuum-energy-conservation}, we first consider only the subsystem formed by Eqs.~\ref{eq:induction_psi} and \ref{eq:psi_evolution}. This means that for the moment we do not consider additional terms related to ${\rm d}\rho/{\rm d}t$ (these are discussed in \S\ref{sec:spmhdenergy}). The total energy of the subsystem (Eq.~\ref{eq:tote}) can be discretised by writing the integral as a sum and replacing the mass element $\rho {\rm d}V$ with the particle mass $m$, giving
\begin{equation}
 E = \sum_a m_a \left[ \frac{B_a^2}{\mu_0 \rho_{a}} + \frac{\psi_a^2}{\mu_0 \rho_{a} c_h^2} \right] .
\end{equation}
Assuming that the total energy of the subsystem is conserved, we have
\begin{equation}
\frac{{\rm d}E}{{\rm d}t} = \sum_a m_a \left[ \frac{{\bf B}_a}{\mu_0 \rho_{a}} \cdot \left( \frac{{\rm d}{\bf B}_a}{{\rm d}t} \right)_\psi + \frac{\psi_a}{\mu_0 \rho_{a} c_h^2} \frac{{\rm d}\psi_a}{{\rm d}t} \right ] = 0.
\label{eq:dedtspmhd}
\end{equation}

\subsubsection{Hyperbolic cleaning with difference operator for $\nabla\cdot{\bf B}$}
\label{sec:spmhd-clean-diff}
If we choose to clean using the difference operator for $\nabla \cdot {\bf B}$, then the SPMHD version of Eq.~\ref{eq:psi_evolution} in the absence of the damping term is given by
\begin{equation}
\frac{d\psi_{a}}{{\rm d}t} = c_{h}^{2} \frac{1}{\Omega_a \rho_a} \sum_b m_b \left( {\bf B}_a - {\bf B}_b \right) \cdot \nabla_a W_{ab}(h_a).
\label{eq:psievo-divb-diff}
\end{equation}
Using this in Eq.~(\ref{eq:dedtspmhd}), we have
\begin{equation}
\sum_a \frac{m_a}{\mu_0 \rho_a} {\bf{B}}_a \cdot \left( \frac{{\rm d}{\bf B}_a}{{\rm d}t} \right)_\psi  = - \sum_a \frac{m_a}{\mu_0 \rho_a} \psi_a \frac{1}{\Omega_a \rho_a} \sum_b m_b \left( {\bf B}_a - {\bf B}_b \right) \cdot \nabla_a W_{ab}(h_a) .
\end{equation}
Expanding the right hand side into two separate terms gives
\begin{align}
 \sum_a \frac{m_a}{\mu_0 \rho_a} {\bf B}_a \cdot \left( \frac{{\rm d}{\bf B}_a}{{\rm d}t} \right)_\psi  = &- \sum_a \sum_b \frac{m_a m_b}{\mu_0 \Omega_a \rho_a^2} \psi_a {\bf B}_a \cdot \nabla_a W_{ab}(h_a) \nonumber \\
&+ \sum_a \sum_b \frac{m_a m_b}{\mu_0 \Omega_a \rho_a^2} \psi_a {\bf B}_b \cdot \nabla_a W_{ab}(h_a) ,
\end{align}
where by swapping the arbitrary summation indices $a$ and $b$ in the second term on the right hand side and using the anti-symmetry of the kernel gradient ($\nabla_{a} W_{ab} = -\nabla_{b} W_{ba}$), we can simplify to find
\begin{equation}
 \sum_a \frac{m_a}{\mu_0 \rho_a} {\bf B}_a \cdot \left( \frac{{\rm d}{\bf B}_a}{{\rm d}t} \right)_\psi = - \sum_a \frac{m_a}{\mu_0 \rho_a} {\bf B}_a \cdot \left\{ \rho_a \sum_b m_b \left[ \frac{\psi_a}{\Omega_a \rho_a^2} \nabla_a W_{ab}(h_a) + \frac{\psi_b}{\Omega_b \rho_b^2} \nabla_a W_{ab}(h_b) \right] \right\} .
\end{equation}
This gives the SPMHD version of Eq.~\ref{eq:induction_psi} in the form
\begin{equation}
\left( \frac{{\rm d}{\bf B}_a}{{\rm d}t} \right)_\psi = -\rho_a \sum_b m_b \left[ \frac{\psi_a}{\Omega_a \rho_a^2} \nabla_a W_{ab}(h_a) + \frac{\psi_b}{\Omega_b \rho_b^2} \nabla_a W_{ab}(h_b) \right].
\label{eq:gradpsisym}
\end{equation}

Thus, by choosing the difference operator for $\nabla \cdot {\bf B}$, the symmetric operator for $\nabla \psi$ is imposed. That is, the total energy of the hyperbolic divergence cleaning scheme is only conserved if the operators are chosen to form a conjugate pair \cite[c.f.][]{cr99,price10,price12}. This is an important improvement over the \citetalias{pm05} implementation which used a difference operator for both. We demonstrate in \S\ref{sec:tests} that indeed the use of conjugate operators significantly improves the robustness and stability of our cleaning algorithm in practice.

\subsubsection{Hyperbolic cleaning with symmetric operator for $\nabla\cdot{\bf B}$}
\label{sec:spmhd-clean-symm}
 An energy-conserving formulation can also be constructed for divergence cleaning with the symmetric operator. That is, with Eq.~\ref{eq:psi_evolution} discretised according to
\begin{equation}
\frac{d\psi_{a}}{{\rm d}t} = -c_{h}^{2} \rho_a \sum_b m_b \left[ \frac{{\bf B}_a}{\Omega_a \rho_a^2} \cdot \nabla_a W_{ab}(h_a) + \frac{{\bf B}_b}{\Omega_b \rho_b^2} \cdot \nabla_a W_{ab}(h_b) \right],\
\label{eq:psievo-divb-symm}
\end{equation}
the discrete version of Eq.~\ref{eq:induction_psi} which must be used to conserve energy is constrained to be
\begin{equation}
\left( \frac{{\rm d}{\bf B}_a}{{\rm d}t} \right)_\psi = \frac{1}{\Omega_a \rho_a} \sum_b m_b \left( \psi_a - \psi_b \right) \nabla_a W_{ab}(h_a),
\label{eq:gradpsidiff}
\end{equation}
which again forms a conjugate pair.

\subsubsection{Hyperbolic cleaning as part of the SPMHD equations}
\label{sec:spmhdenergy}

In \S\ref{sec:idealmhdenergy}, the evolution equation for $\psi$ was modified to include a $-\tfrac{1}{2} \psi (\nabla\cdot{\bf v})$ term (Eq.~\ref{eq:psi_evolution_halfdivv}).  This was done to conserve energy in the presence of ${\rm d}\rho/{\rm d}t$ terms.  The discretised form of $\nabla\cdot{\bf v}$ in Eq.~\ref{eq:psi_evolution_halfdivv} should therefore be the same as that used in the SPH continuity equation (see \cite{2005RPPh...68.1703M}), which leads to
\begin{equation}
 - \tfrac{1}{2} \psi_a (\nabla \cdot {\bf v}_a) = \frac{\psi_a}{2\Omega_a\rho_a} \sum_b m_b ({\bf v}_a - {\bf v}_b) \cdot \nabla_a W_{ab} (h_{a}).
\end{equation}

\subsection{Energy loss due to damping}
\label{sec:negdef}
For completeness, it is important to prove that the damping term in Eq.~\ref{eq:psi_evolution_halfdivv} will result in a negative definite energy change. Inserting the damping term into the total change of $\psi$ energy, we see that
\begin{equation}
\left(\frac{{\rm d}E}{{\rm d}t}\right)_{\rm damp} = \sum_a m_a \frac{\psi_a}{\mu_0 \rho_a c_h^2} \left( \frac{{\rm d}\psi_a}{{\rm d}t} \right)_{\text{damp}} = - \sum_a m_a \frac{\psi_a^2}{\mu_0 \rho_a c_h^2 \tau} ,
\end{equation}
which is indeed negative definite. These energy changes could be balanced with equivalent increases in thermal energy to keep the total energy constant. The issue with doing this is that the heat generated is not necessarily deposited in the same location as it was removed from the magnetic field, due to the transport of divergence errors inherent in the hyperbolic cleaning scheme. Thus, we do not add such heat gains as part of our method, although the term above can be used to keep track of the energy loss due to divergence cleaning.

\section{Tests}
\label{sec:tests}
 We have designed our numerical tests to examine the following key aspects of our constrained hyperbolic divergence cleaning algorithm:
 \begin{enumerate}
 \item[i)] The importance of the energy-conserving, ``constrained'' formulation compared to a non-conservative, or ``unconstrained'', approach,
 \item[ii)] Whether or not cleaning using the symmetric $\nabla\cdot{\bf B}$ operator (\S\ref{sec:spmhd-clean-symm}) provides any advantage over use of the difference operator (\S\ref{sec:spmhd-clean-diff}), e.g. by improving momentum conservation,
 \item[iii)] Optimal parameter choices for $\sigma$,
\item[iv)] The practical effect of including the $-\frac12\psi \nabla\cdot {\bf v}$ term (Eq.~\ref{eq:psi_evolution_halfdivv}). 
 \end{enumerate}
 In particular, we have investigated these aspects both in isolation using simple idealised setups, as well as their combined effects in more realistic 2 and 3D simulations. Our goal is to verify the robustness of the algorithm for practical application in astrophysics, though it offers a general solution to maintaining the divergence constraint in SPMHD. 
 
 As well as examining the divergence of the magnetic field using the operators given by Eq.~\ref{eq:divbdiff} and \ref{eq:divbsym}, we measure the divergence error in the standard manner for SPMHD with the dimensionless quantity,
 \begin{equation}
\frac{h \vert \nabla\cdot{\bf B} \vert}{\vert {\bf B} \vert}.
 \end{equation}
To prevent artificially high values where $\vert {\bf B} \vert \to 0$, a small parameter $\epsilon$ is added to $\vert {\bf B} \vert$ in the denominator, where $\epsilon \sim 1\%$ of the maximum B-field value.  We find this is only necessary for the Orszag-Tang vortex problem.

 All of the tests have been performed using a Leapfrog integrator with magnetic field integrated alongside the velocity and timesteps set according to the standard condition $\Delta t < \min_{a} (C_{\rm cour} h_{a}/v_{{\rm sig},a})$, where $C_{\rm cour} = 0.2$ and $v_{{\rm sig},a}$ is the MHD fast wave speed on each particle. We therefore use $c_{h} = \max_{a} (v_{{\rm sig},a})$ in the hyperbolic cleaning, except for the final test (\S\ref{sec:jet}) where $c_{h}$ is individual to each particle.  The damping parameter is chosen to be $\sigma=0.4$ ($\sigma=0.8$ for the final test), except for cases when it is varied to find optimal values. Unless otherwise indicated, we use the standard SPH cubic spline kernel for all tests with $h_{\rm fac} = 1.2$ in Eq.~\ref{eq:sphcty} corresponding to $\sim 18$ neighbours in 2D and $\sim 58$ neighbours in 3D. The magnetic field is specified in units such that $\mu_{0} = 1$ in the code \cite[c.f.][]{pm04a}.  Artificial resistivity is only used where noted, in which case it is applied as described in \S\ref{sec:resist}.

 \subsection{Divergence advection}
\label{sec:divBadvection}

  The simplest test we consider consists of divergence in the magnetic field artificially induced in the initial conditions and advected by a uniform flow. The test is performed in a two dimensional periodic domain with three dimensional magnetic and velocity fields (2.5D). The first version of this test is identical to the `divergence advection problem' from \cite{2002JCoPh.175..645D}, as generalised by \citetalias{pm05}. We use this to illustrate the basic features of the hyperbolic/parabolic cleaning approach and to examine the optimal choice of $\sigma$ when the divergence error has a scale comparable to the numerical resolution.

\begin{figure}
 \centering
\includegraphics[width=\textwidth]{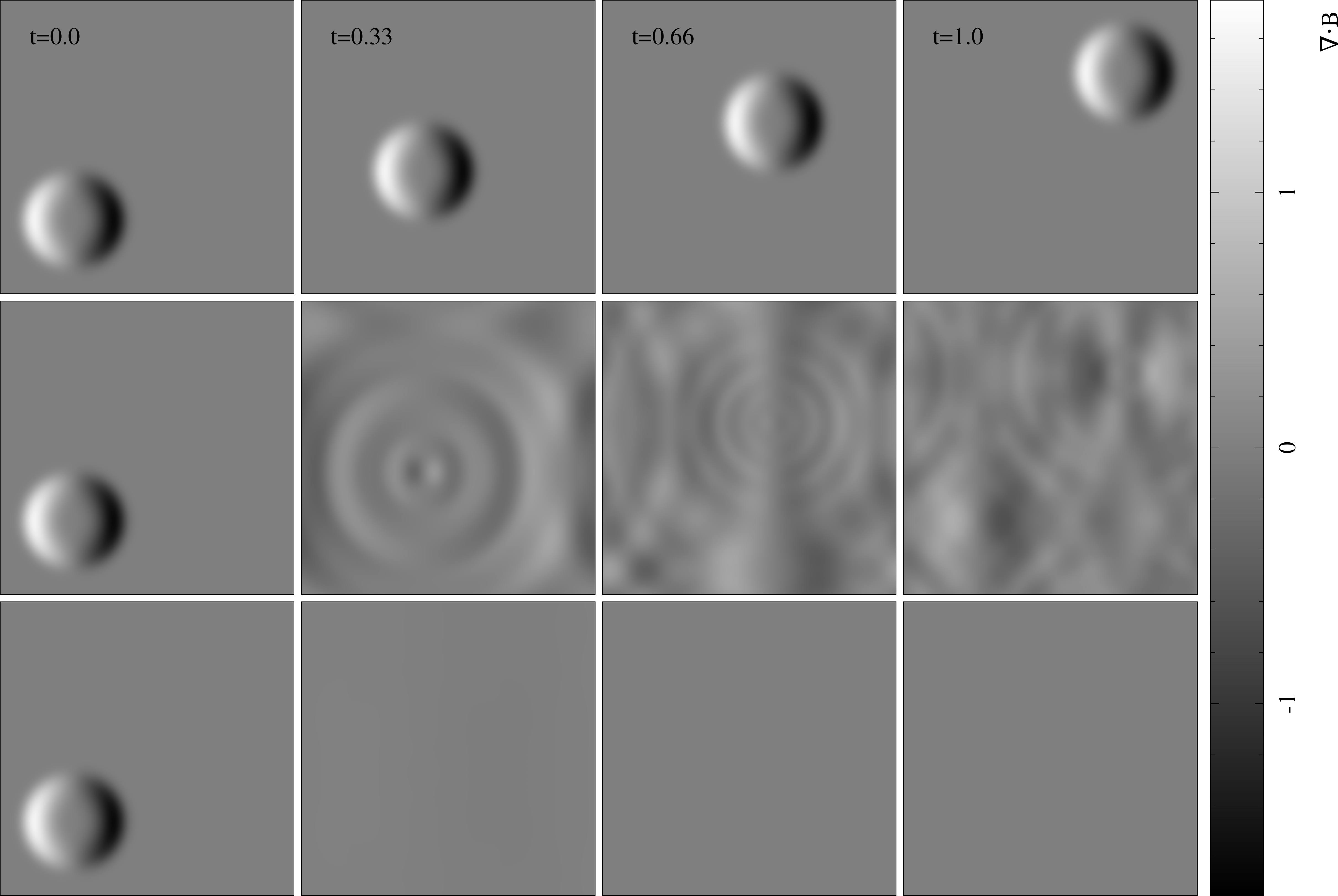}
\caption{A fluid with uniform velocity has a blob of divergence introduced to the initial conditions. In the top row, no cleaning is applied and the divergence blob is advected exactly with the flow.  Undamped cleaning (purely hyperbolic) is applied to the centre row and the divergence in the magnetic field is spread through the system as a system of interacting waves.  In the bottom row, damped cleaning (mixed hyperbolic/parabolic) is utilised and the divergence in the magnetic field is rapidly removed.}
\label{fig:advection}
 \end{figure}

\subsubsection{Setup}
The domain is a square area of fluid in the region $x,y \in [-0.5, 1.5]$.  The system has uniform density $\rho = 1$, with pressure $P = 6$ and $\gamma = 5/3$.  The velocity field is ${\bf v} = [1,1]$ and $B_z = 1 / \sqrt{4 \pi}$.  A perturbation is created in the x-component of the magnetic field of the form
\begin{equation}
\label{eq:adv-divergence-perturbation}
B_x = \frac{1}{\sqrt{4 \pi}} \left[ \left(r / r_0\right)^8 - 2 \left( r/r_0 \right)^4 + 1 \right]; \hspace{1cm} r < r_{0},
\end{equation}
where $r \equiv \sqrt{x^{2} + y^{2}}$ and $r_0$ specifies the radial extent. We set up the problem using $50 \times 50$ particles on a square lattice, giving $h = 1.2 \Delta x = 0.048$.

\subsubsection{Results}
Fig.~\ref{fig:advection} shows renderings of $\nabla\cdot{\bf B}$ at various times from three calculations: no cleaning, undamped cleaning (purely hyperbolic), and damped cleaning (mixed hyperbolic/parabolic) with $r_{0} = 1/\sqrt{8}$, following \cite{2002JCoPh.175..645D}. These three calculations illustrate the basic ideas behind the divergence cleaning scheme: In the absence of any cleaning (top), the magnetic field and its divergence perturbation is advected without change on the particles. With the addition of hyperbolic cleaning (middle), the divergence errors are spread in a wave-like manner throughout the domain. Finally, the addition of the parabolic damping term (bottom row) acts to rapidly diffuse the divergence error to zero.

\begin{figure}
\centering
 \includegraphics[width=0.45\textwidth]{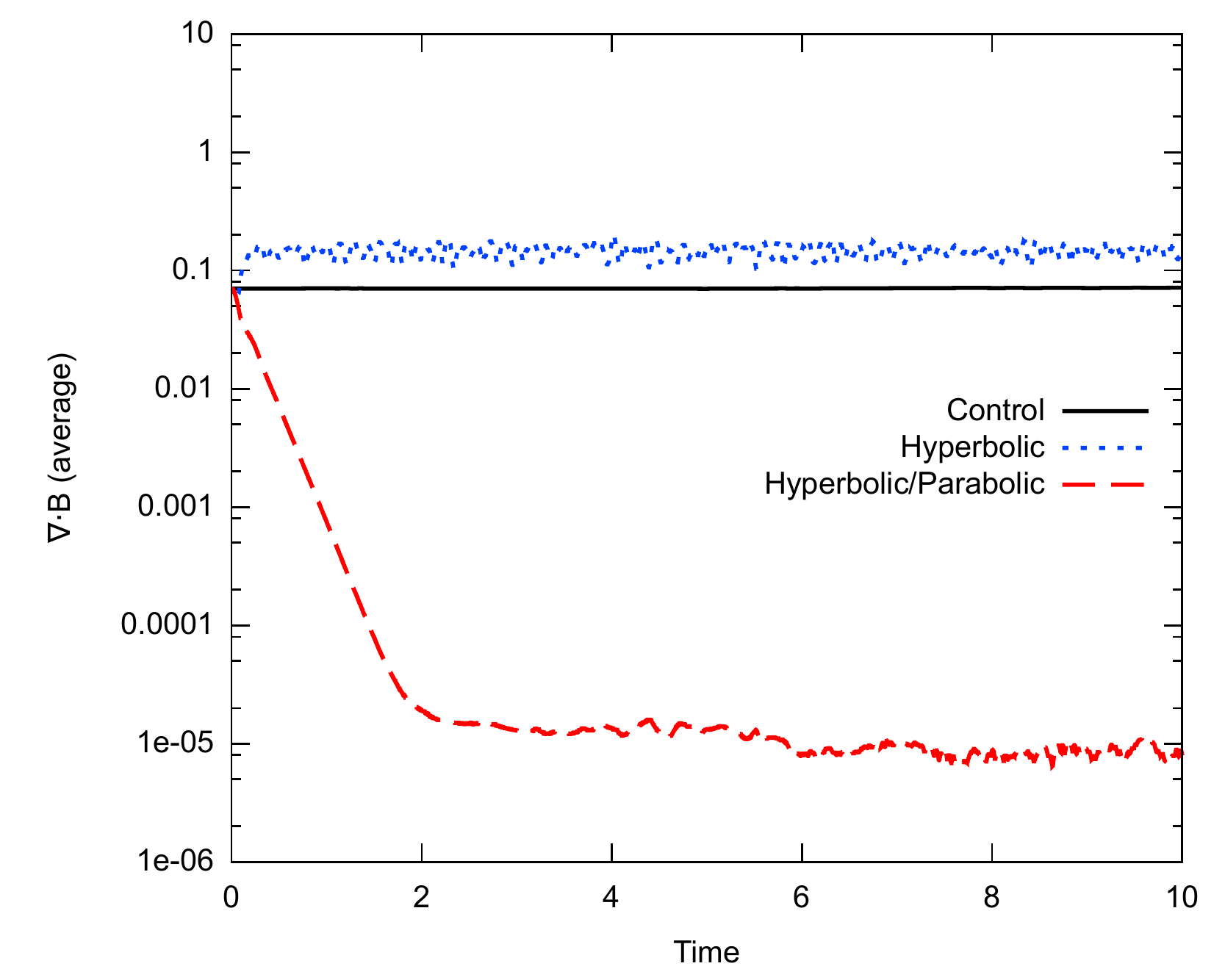}
 \includegraphics[width=0.45\textwidth]{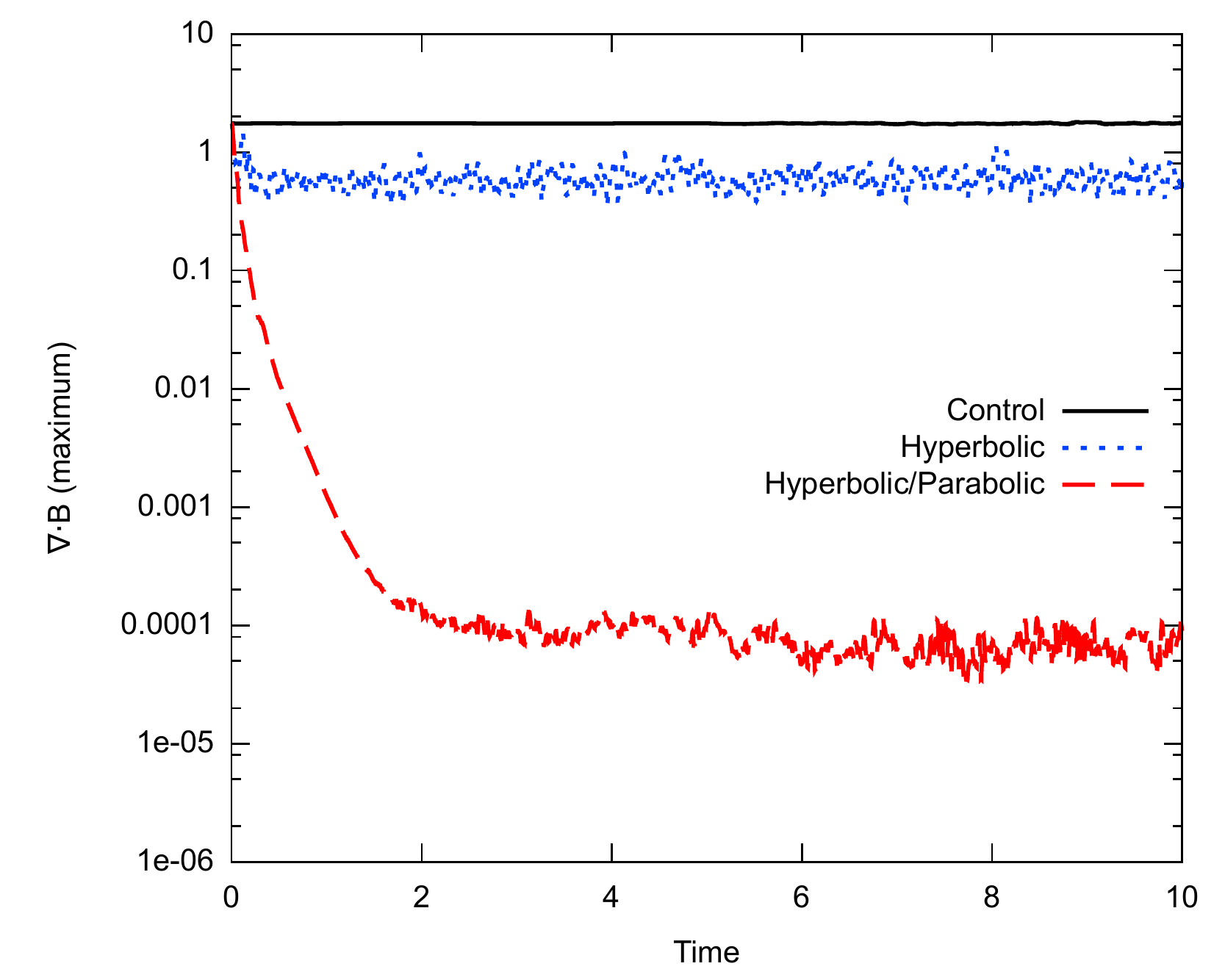}
\caption{Average and maximum $\nabla \cdot {\bf B}$ in code units, measured using the difference operator (Eq.~\ref{eq:divbdiff}), as a function of time for the divergence advection test with $r_{0} = 1/\sqrt{8}$.  Without cleaning, the divergence for this simple problem remains constant.  Using undamped cleaning (purely hyperbolic), the maximum divergence is reduced with an increase in average throughout the system.  With damped cleaning (mixed hyperbolic/parabolic), both average and maximum are rapidly reduced.}
\label{fig:advection-divb}
\end{figure}

This is demonstrated more quantitatively in Fig.~\ref{fig:advection-divb}, which shows the average and maximum values of $\vert\nabla\cdot{\bf B}\vert$ as a function of time for the three calculations. While purely hyperbolic cleaning can be seen to quickly reduce the maximum divergence error, the average error increases. The parabolic damping means that both the average and maximum values are reduced by an order of magnitude in roughly one wave crossing time ($t\sim 0.3$), and by roughly 5 orders of magnitude after several crossing times ($t \gtrsim 2$). After this time the divergence error continues to decrease, but at a much slower rate (this is more obvious in Fig.~\ref{fig:tuning} for the case $r_{0} = h$).  We attribute the turnover in the decay rate to the rapid removal of the short wavelength errors by the cleaning scheme, leaving only slowly decaying long-wavelength modes.  We have confirmed this interpretation by verifying that the transition to a slow decay is independent of timestepping, resolution and is similar using the quintic \cite{price12} instead of the cubic spline kernel.
  
\subsubsection{Optimal choice of damping parameter in 2D}
\begin{figure}
\centering
 \includegraphics[width=0.45\textwidth]{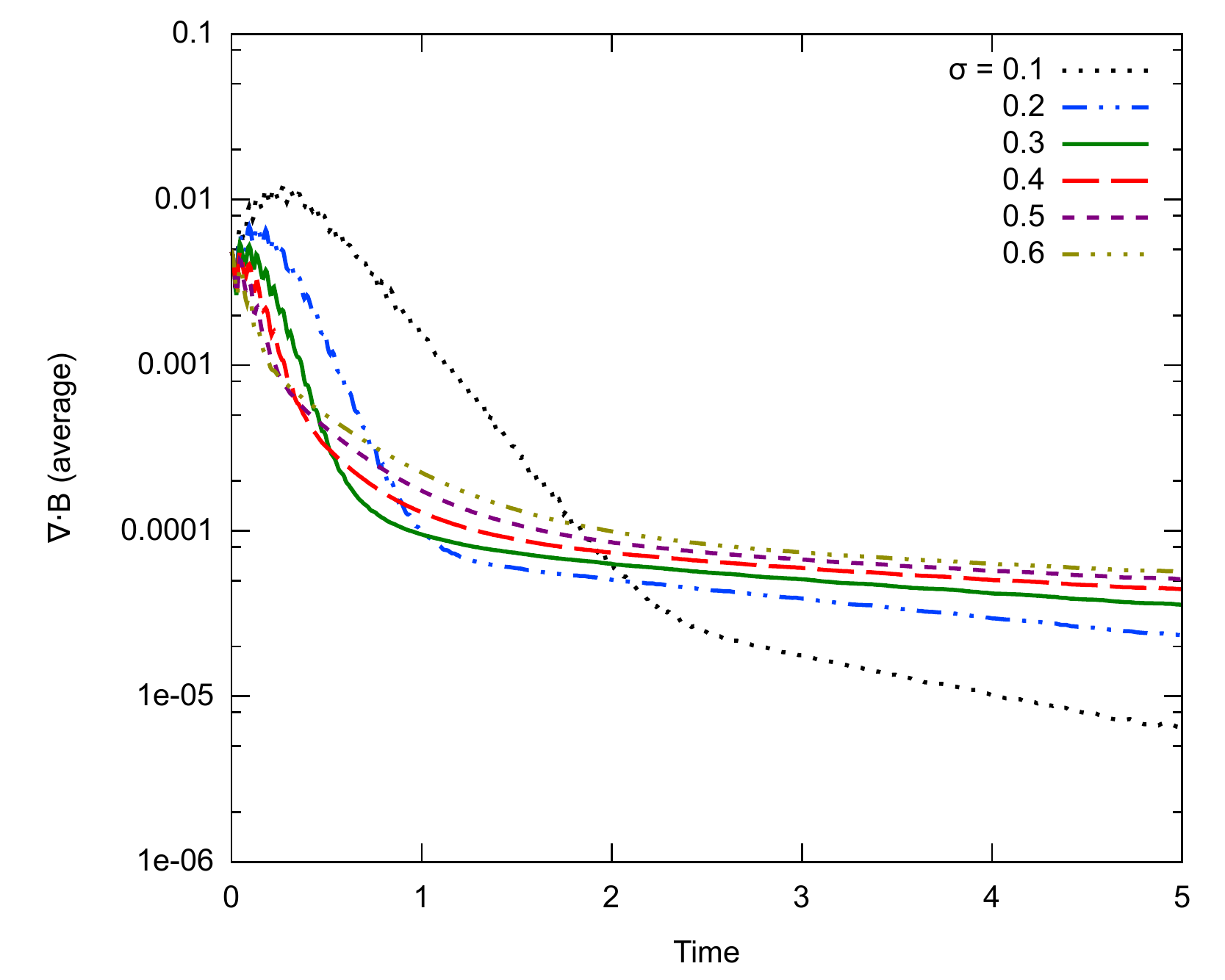}
 \includegraphics[width=0.45\textwidth]{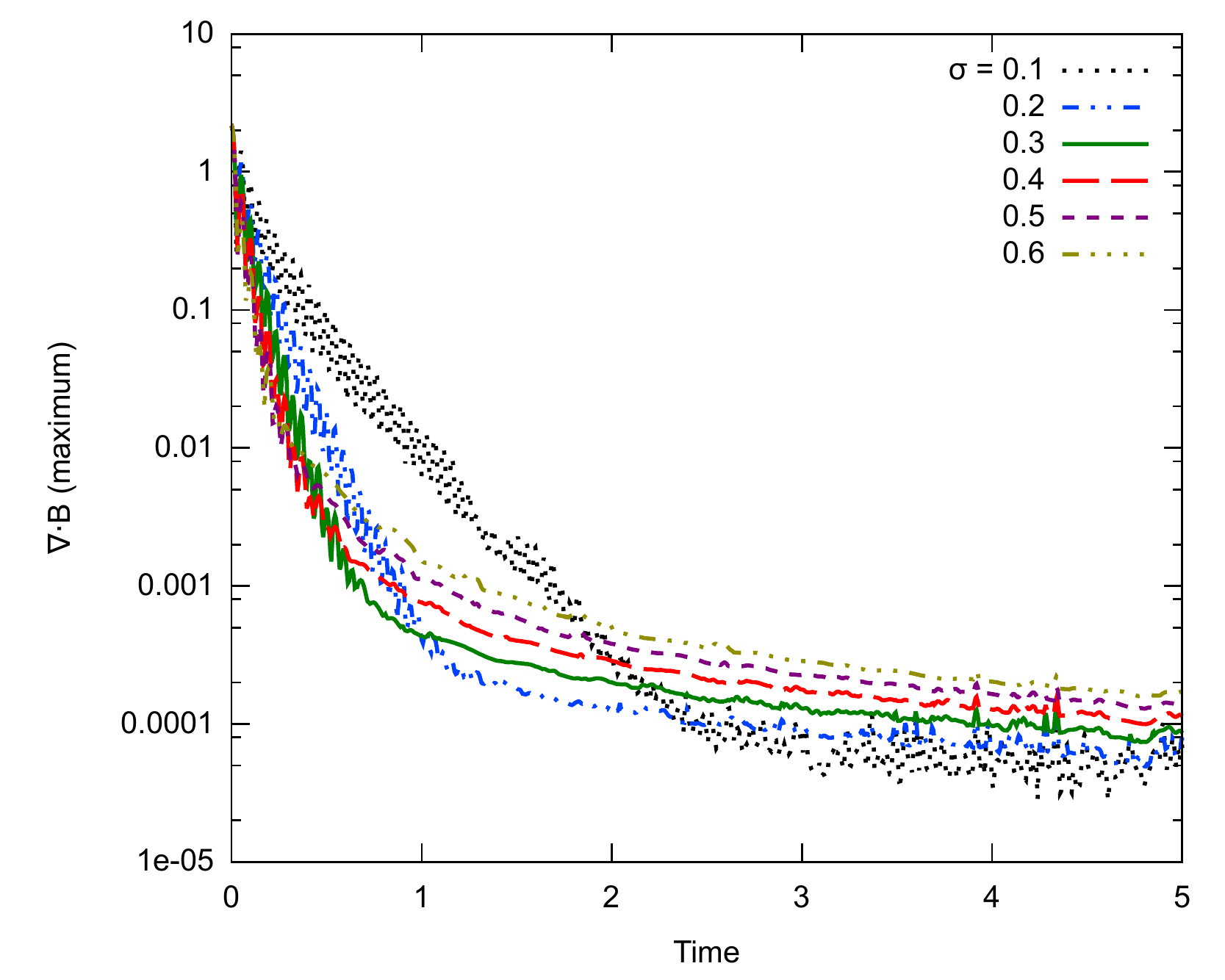}
\caption{The effect of varying the damping parameter $\sigma$ on the average and maximum $\nabla \cdot {\bf B}$ for the divergence advection test with $r_{0} = h$.  The best results for 2D are obtained for values between 0.2--0.3.}
\label{fig:tuning}
\end{figure}

 As noted by \citetalias{pm05}, the optimal choice of damping parameter, $\sigma$, for this problem with $r_{0} = 1/\sqrt{8}$ is somewhat misleading, since in reality one expects divergence errors arising in simulations to have length scales of order the smoothing length. Thus, Fig.~\ref{fig:tuning} shows the average and maximum $\nabla\cdot{\bf B}$ in a series of calculations employing $r_{0} = h$ and values of $\sigma$ between 0.1 and 0.6. The results are similar to those shown in Fig.~\ref{fig:advection-divb}, with best results obtained in this 2D case using $\sigma \sim$ 0.2--0.3.

\subsection{Static cleaning test: density jump}
\label{sec:test-density-jump}

\begin{figure}
 \centering
\includegraphics[width=\textwidth]{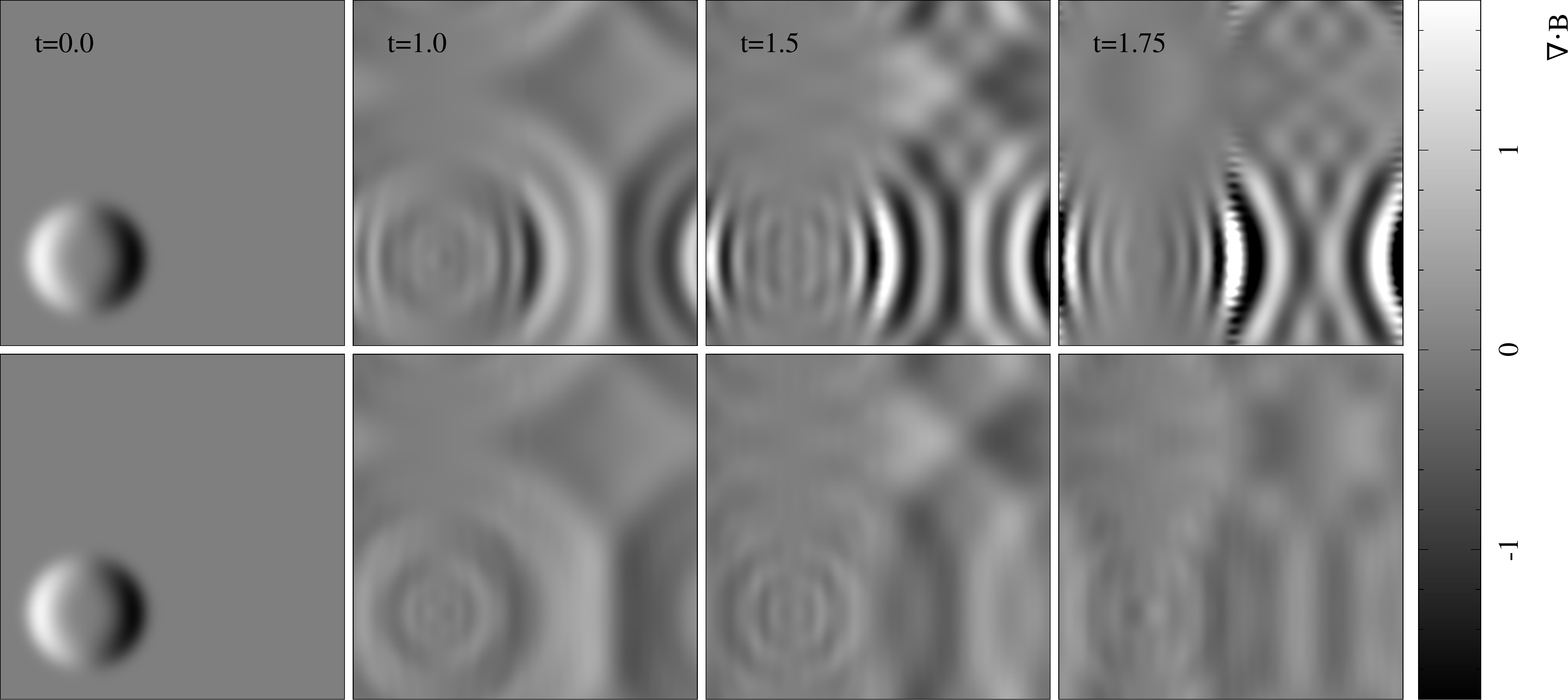}
\caption{Results of the static cleaning test across a 2:1 density jump. Undamped non-conservative cleaning (top) increases the divergence of the magnetic field at the density jump, in turn leading to numerical instability (Fig.~\ref{fig:leftright-div-plots}). Using our constrained divergence cleaning method (bottom), the waves cross the density boundary without issue and the scheme remains stable.}
\label{fig:leftright}
 \end{figure}
 
 Our second test is a variant on the divergence advection problem, with identical setup ($r_{0} = 1/\sqrt{8}$) except that the right half of the domain has its density increased by a factor of two. The idea is to examine the reflection and refraction of the divergence waves as they transition between media of differing densities, as may frequently occur in applications of SPMHD. To simplify the test, we solve only the subset of equations given by Eqs.~\ref{eq:induction_psi}--\ref{eq:psi_evolution} --- that is, the system can only evolve due to divergence cleaning.

\subsubsection{Setup}
 The setup is performed in 2D with $25 \times 50$ particles on a square lattice in the left half of the domain ($x < 0.5$, $\rho = 1$), and $35 \times 70$ particles placed in the right half of the domain ($x > 0.5$, $\rho = 2$), with all particles of equal mass, giving a 2:1 density jump at $x=0.5$. The actual density on the particles is found in the usual manner by iterating the smoothing length and density self-consistently as described in \S\ref{sec:spmhd-equations}. The velocity field is set to zero, all other system parameters are set as previously for the divergence advection test (\S\ref{sec:divBadvection}), and periodic boundary conditions are employed.

\subsubsection{Results}
\begin{figure}
\centering
 \includegraphics[width=0.45\textwidth]{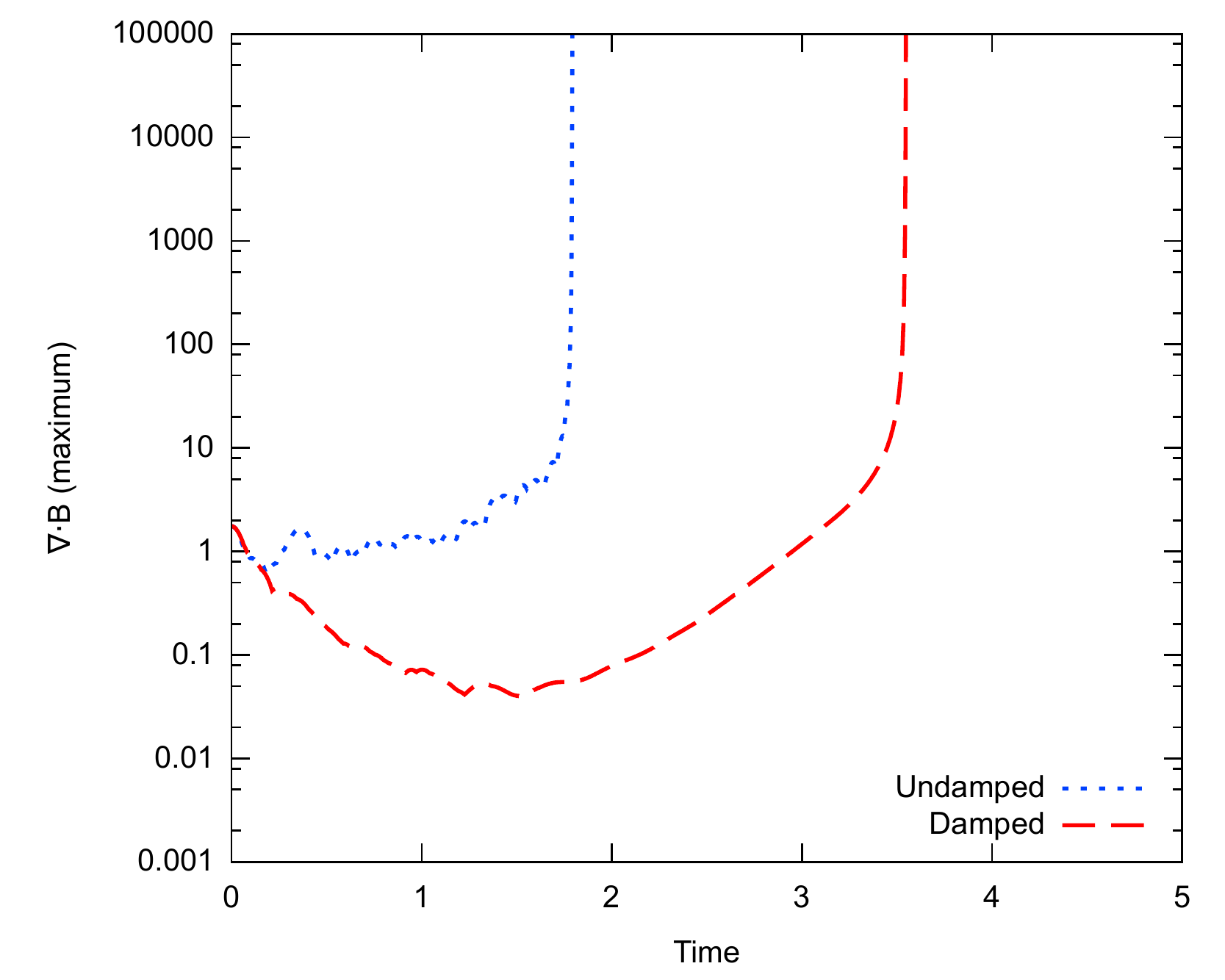}
 \includegraphics[width=0.45\textwidth]{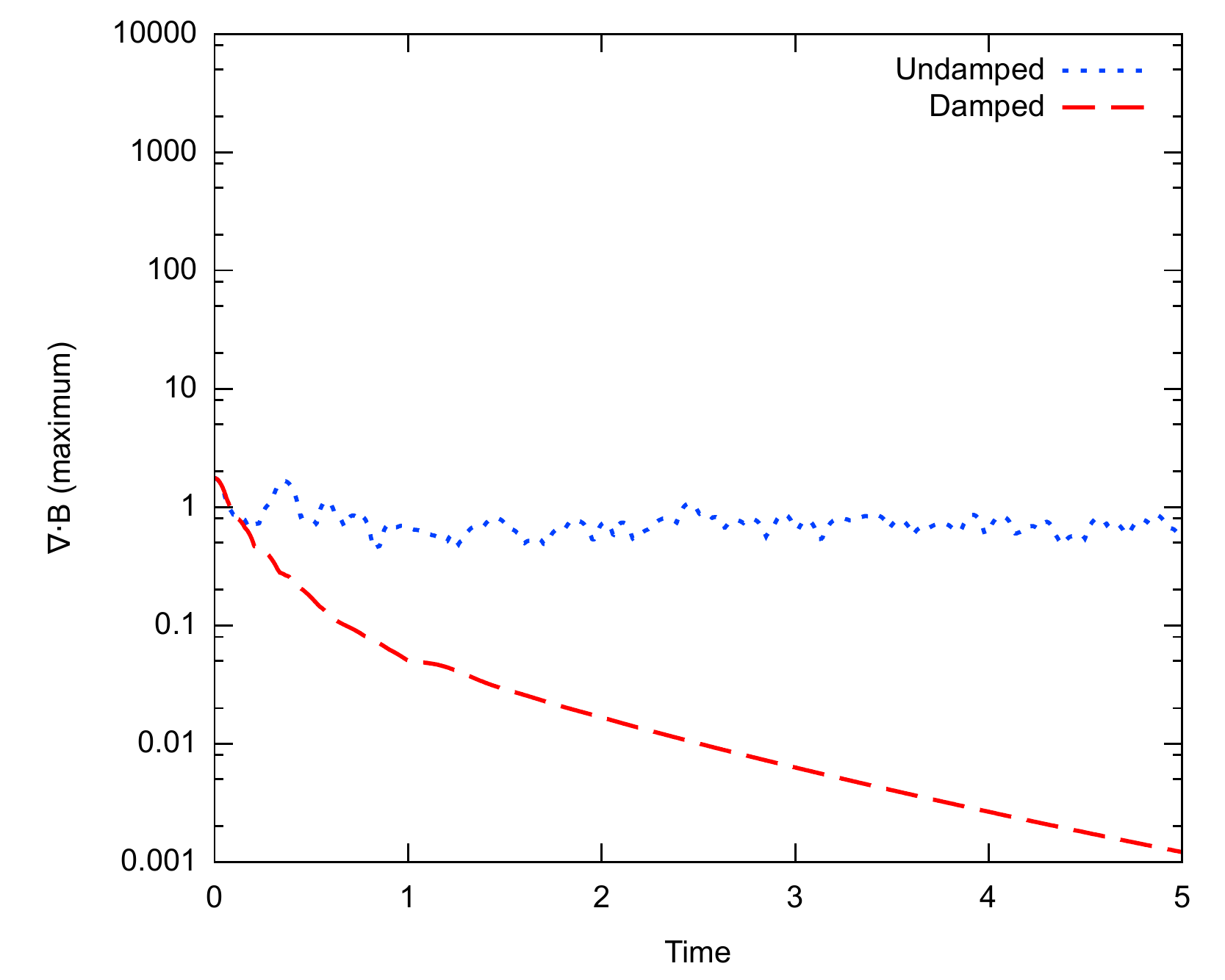}
\caption{Maximum values of $\nabla \cdot {\bf B}$ (difference) for the density jump test for the non-conservative formulation (left) and the new constrained divergence cleaning (right).  The interaction between the divergence waves and the density jump for the non-conservative formulation is unstable, for both damped and undamped cleaning.  Using constrained divergence cleaning is stable across the density jump, with damped cleaning reducing $\nabla \cdot {\bf B}$ as in previous tests.}
\label{fig:leftright-div-plots}
\end{figure}

 Fig.~\ref{fig:leftright} shows the propagation of purely hyperbolic ($\sigma = 0$) divergence waves in this test using i) the non-energy conserving formulation with difference operators for both $\nabla \cdot {\bf B}$ (Eq.~\ref{eq:divbdiff}) and $\nabla \psi$ (Eq.~\ref{eq:gradpsidiff}), and ii) our new constrained hyperbolic divergence cleaning scheme with a difference operator for $\nabla \cdot {\bf B}$ and the conjugate, symmetric operator for $\nabla \psi$ (Eq.~\ref{eq:gradpsisym}). The corresponding time evolution of the maximum $\vert\nabla\cdot{\bf B}\vert$ is shown in Fig.~\ref{fig:leftright-div-plots}. Using the unconstrained formulation, the interaction of the divergence wave with the density jump causes amplification of the divergence errors (top row of Fig.~\ref{fig:leftright}), in turn leading to exponential growth in the total energy and numerical instability (left panel of Fig.~\ref{fig:leftright-div-plots}). By contrast, our new conservative formulation remains stable and continues to reduce the divergence error throughout the domain (bottom row of Fig.~\ref{fig:leftright} and right panel of Fig.~\ref{fig:leftright-div-plots}).

\subsection{Static cleaning test: free boundaries}
\label{sec:test-free-boundaries}

\begin{figure}
 \centering
\includegraphics[width=\textwidth]{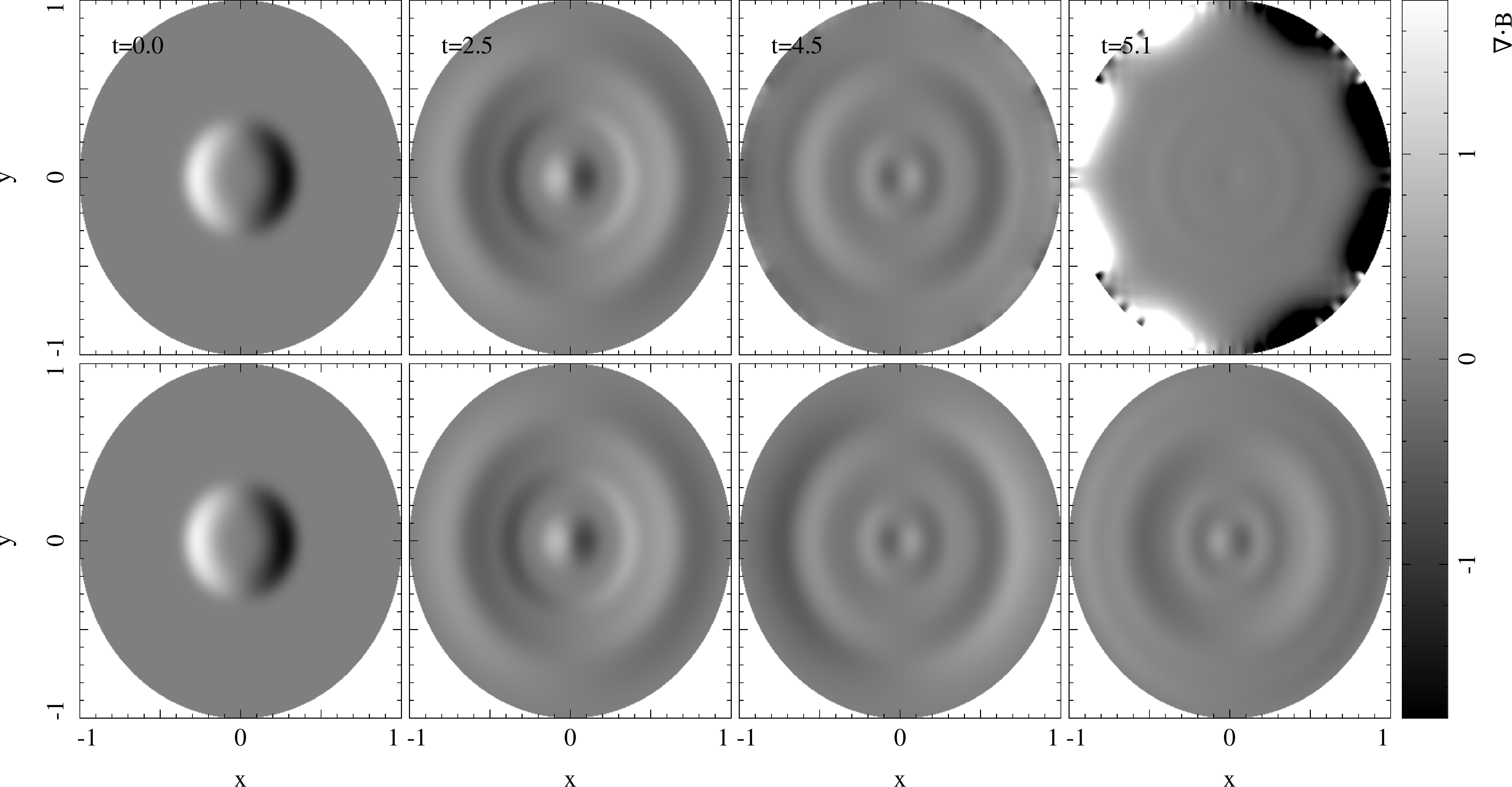}
\caption{$\nabla \cdot {\bf B}$ of the static cleaning test using free boundaries.  In the case of non-conservative cleaning (top row), the interaction of the divergence waves with the boundary cause unchecked divergence growth.  Using constained cleaning (bottom row), the boundary interaction is not problematic.}
\label{fig:boundary_circle_test}
 \end{figure}

 A further variant of the divergence advection test we consider replaces the periodic boundaries by a free boundary, since many applications of SPMHD involve free boundaries (e.g. the merger of two neutron stars \cite{pr06}, or studies of galaxy interactions \cite{kotarbaetal10,kotarbaetal11}).

\subsubsection{Setup}

The setup is identical to the divergence advection problem (\S\ref{sec:divBadvection}) with $r_{0} = 1/\sqrt{8}$, except that the domain is a circular area of fluid with $\rho = 1$ for $r \le 1$ and $\rho = 0$ (no particles) for $r > 1$, set up using a total of 1976 particles placed on a cubic lattice.  The divergence perturbation is introduced at the centre of the circle, and the velocity field is set to zero. Rather than impose an external confining potential, we solve only Eqs.~\ref{eq:induction_psi}--\ref{eq:psi_evolution} without the full MHD equations, as in \S\ref{sec:test-density-jump}.

\subsubsection{Results}
 Fig.~\ref{fig:boundary_circle_test} shows the results of purely hyperbolic cleaning ($\sigma=0$) for this case. As in Fig.~\ref{fig:leftright}, the top row shows the unconstrained and non-conservative difference/difference formulation, while the bottom row shows results using the conservative difference/symmetric combination. Similar results are also found in this case, with divergence errors piling up at the free boundary in the non-conservative formulation leading to numerical instability, but our constrained formulation remaining stable.

\subsection{2D Blast wave in a magnetised medium}
\label{sec:blast}
We now turn to tests that are more representative of the dynamics encountered in typical astrophysical simulations, beginning with a blast wave expanding in a magnetised medium. In this case the initial magnetic field is divergence-free, meaning that the only divergence errors are those created by numerical errors during the course of a simulation --- rather than the artificial errors we have induced in the previous tests. Based on the results from the previous tests, in this and subsequent tests we apply cleaning \emph{only} using constrained, energy-conserving formulations --- that is, with conjugate operators for $\nabla\cdot{\bf B}$ and $\nabla\psi$. We use this problem to the examine the effectiveness of the divergence cleaning in the presence of strong shocks, as well as to investigate whether cleaning should be performed using the difference or symmetric $\nabla \cdot {\bf B}$ operator.  As with the divergence advection test, a key goal is to find optimal values for the damping parameter $\sigma$.

\subsubsection{Setup}
\label{sec:blast-setup}
The implementation of the blast wave follows that of \citet{2000ApJ...530..508L}.  The domain is a unit square with periodic boundaries, set up with $512 \times 590$ particles on a hexagonal lattice with $\rho = 1$.  The fluid is at rest with magnetic field $B_x = 10$.  The pressure of the fluid is set to $P = 1$, with $\gamma = 1.4$, except a region of the centre of radius $0.125$ has its pressure increased by a factor of 100 by increasing its thermal energy. An adiabatic equation of state is used.

\subsubsection{Results}

\begin{figure}
 \centering
\includegraphics[width=\textwidth]{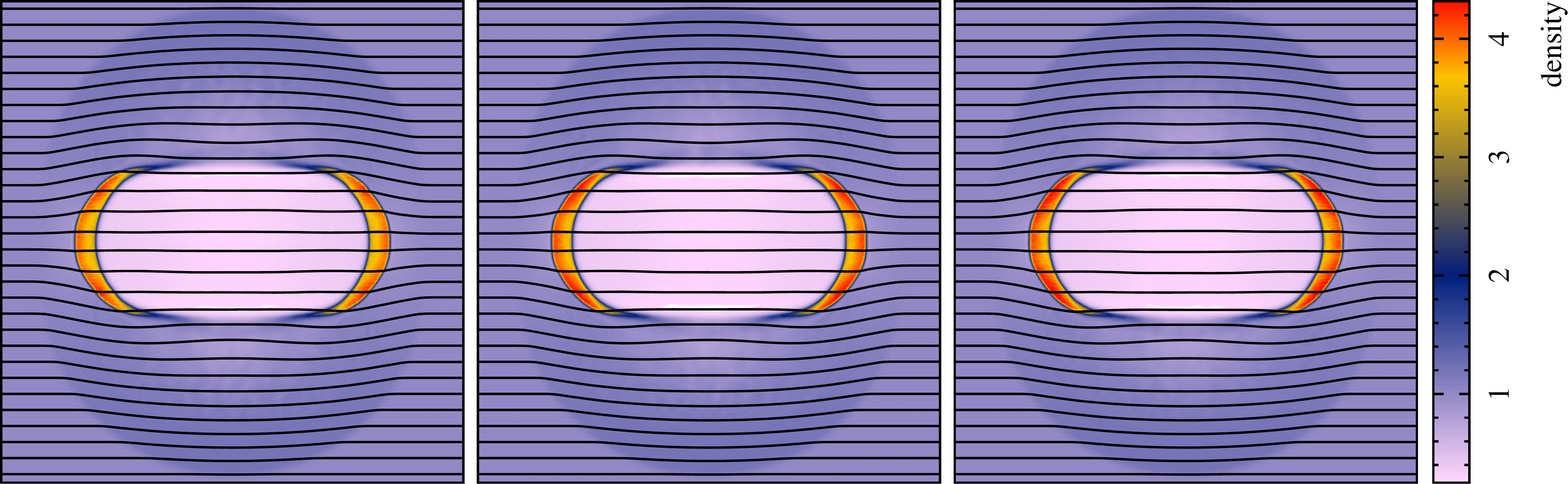}
\caption{Renderings of the density together with overlaid magnetic field lines in the MHD blast wave problem at $t=0.03$, showing the control case with no resistivity and no cleaning (left), with resistivity (centre), and with divergence cleaning (right). Only minor differences in the density evolution are evident.}
\label{fig:blast-compilation-density}
\end{figure}

\begin{figure}
\centering
\includegraphics[width=0.45\textwidth]{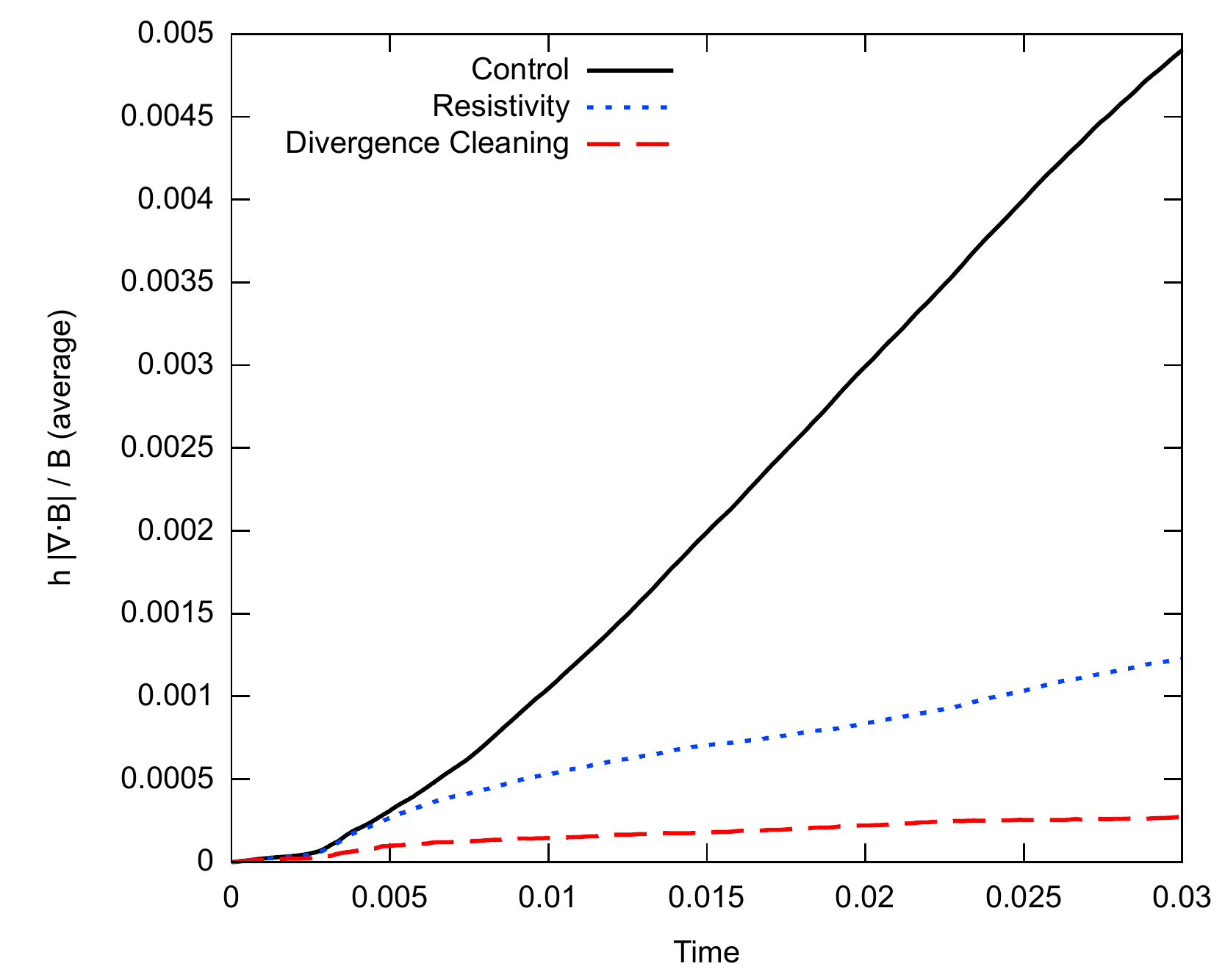}
\includegraphics[width=0.45\textwidth]{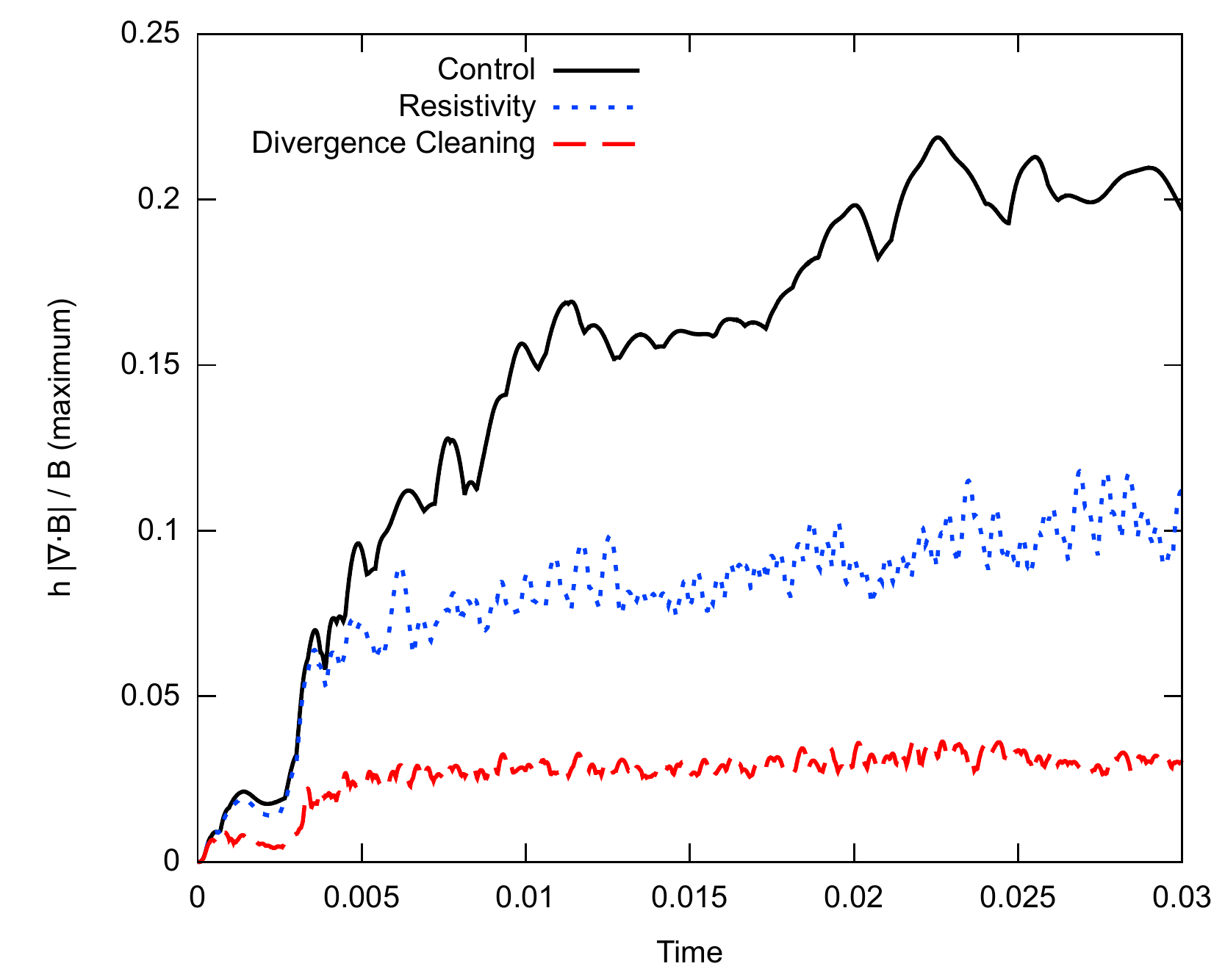}
\caption{Average and maximum of $h \vert \nabla \cdot {\bf B}\vert / \vert{\bf B}\vert$ as a function of time for the blast wave test. At $t=0.03$, resistivity has reduced the average error by a factor of 4 compared to the control case, while divergence cleaning has reduced the average divergence error by a factor of 20. The maximum error has been reduced by a factor of 2 and 8, respectively.}
\label{fig:blast-divb}
\end{figure}

Fig.~\ref{fig:blast-compilation-density} shows the density and magnetic field lines at $t=0.03$ for i) the control case without cleaning and no artificial resistivity (left), ii) including artificial resistivity (centre) and iii) no resistivity, but cleaned using the difference operator (right).  At this time, the MHD fast shock has expanded to fill the domain, yet has not crossed the periodic boundaries to begin interacting with itself, and the three cases show only minimal differences in density structure. The average and maximum divergence error as a function of time are shown in Fig.~\ref{fig:blast-divb}.  Although the density renderings at $t=0.03$ are quite similar, we can see that adding divergence cleaning has reduced the average and maximum divergence error by a factor of 20 and 8, respectively at $t=0.03$, compared to the control case, with factors of 5 and 4 improvement compared to the case with artificial resistivity alone. Thus, divergence cleaning is even more effective than resistivity at enforcing the divergence constraint.

\subsubsection{Operator choice for $\nabla \cdot {\bf B}$}
\label{sec:blast-symmdivb}

\begin{figure}
 %\centering
\setlength{\tabcolsep}{0.005\textwidth}
\begin{tabular}{cccl}
{\scriptsize Control} & {\scriptsize Difference Cleaned} & {\scriptsize Symmetric Cleaned} & \\
\includegraphics[height=0.305\textwidth]{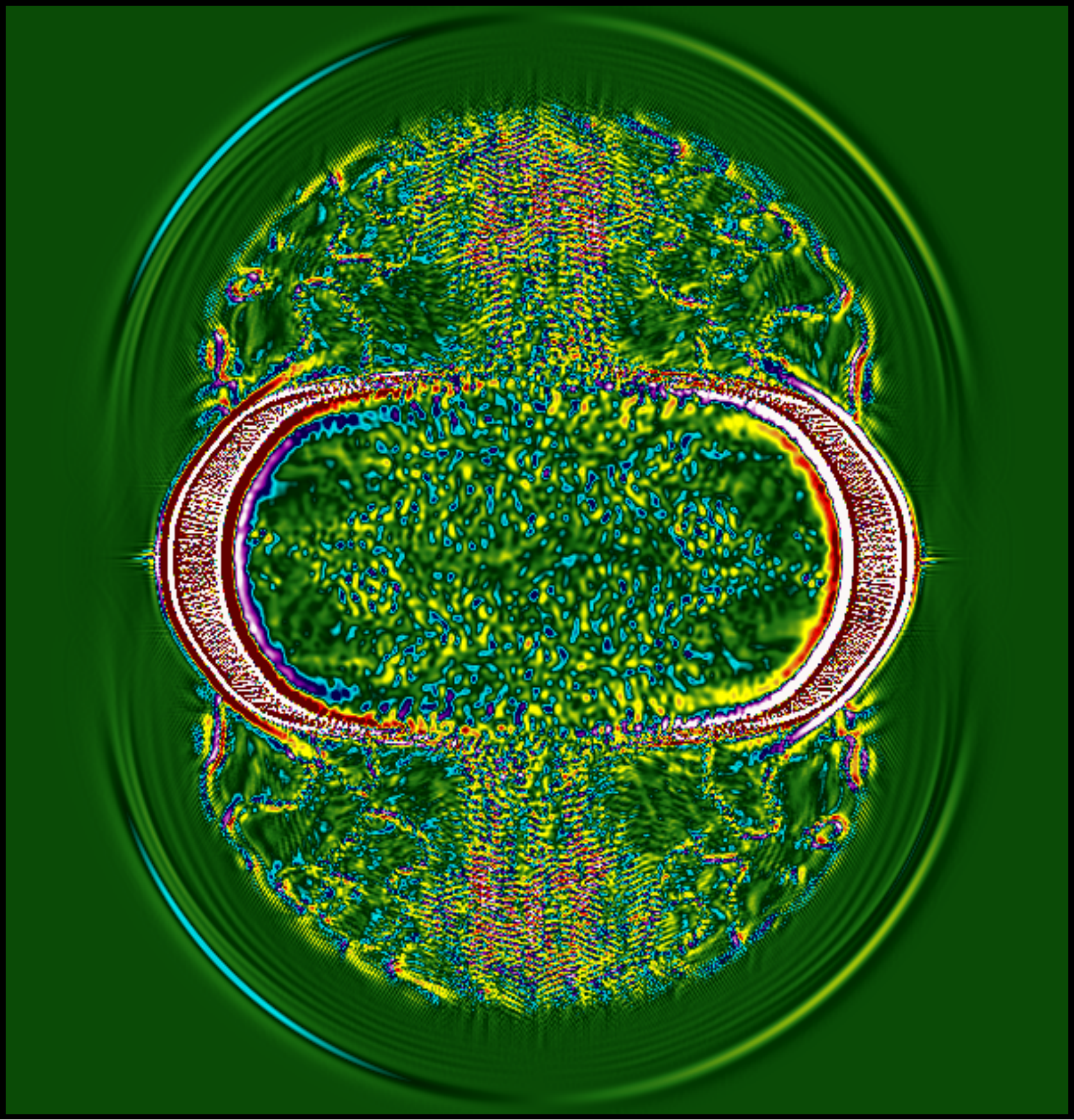} 
 & \includegraphics[height=0.305\textwidth]{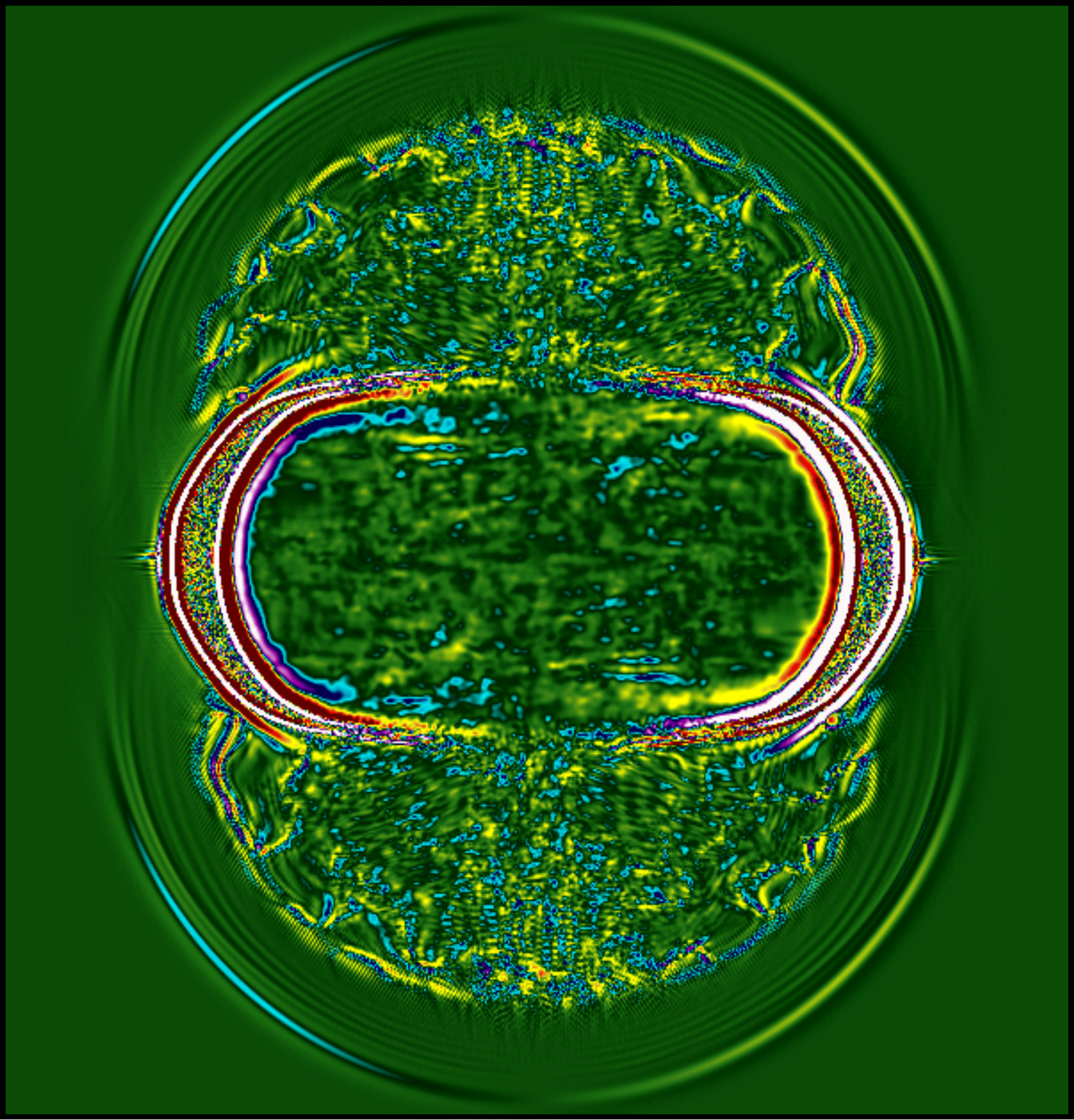}
 & \includegraphics[height=0.305\textwidth]{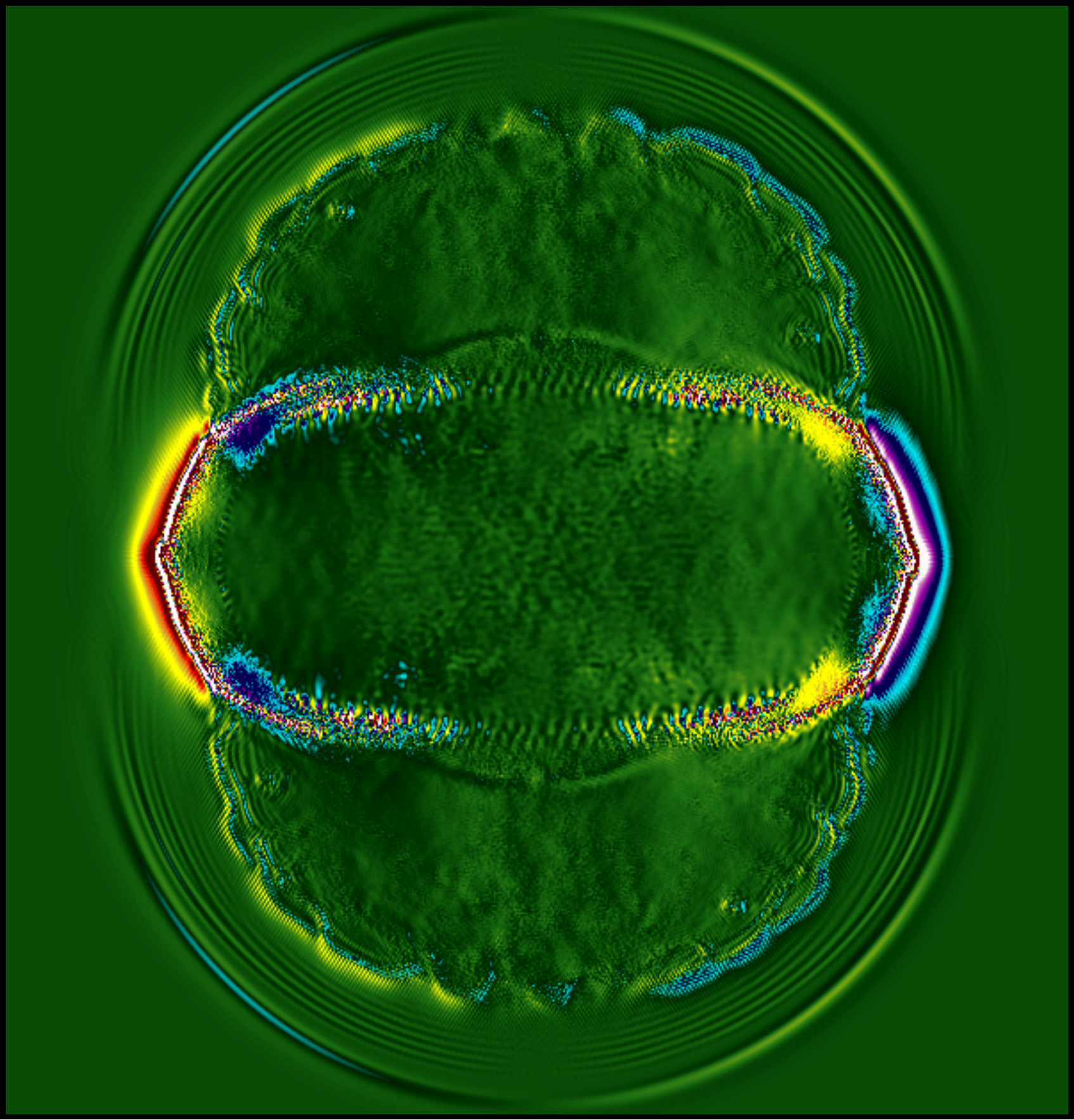}
 & \includegraphics[height=0.305\textwidth]{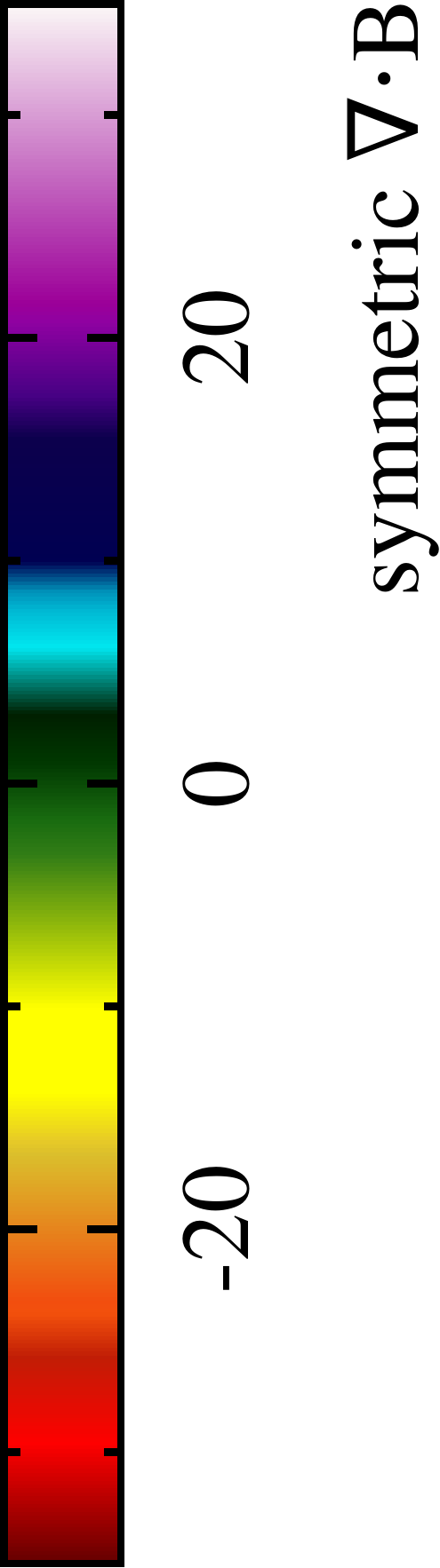}
\end{tabular}
\caption{$\nabla\cdot{\bf B}$ in the blast wave problem at $t=0.03$ measured in code units using the symmetric $\nabla \cdot {\bf B}$ operator, showing the control case (left), $\nabla\cdot{\bf B}$ measured with the opposing operator to that used in the cleaning (centre) and $\nabla\cdot{\bf B}$ measured with the same operator used in the cleaning (right).  Note in particular that the symmetric operator measures a divergence error around the leading edge of the fast MHD wave, even though the field is quite regular.}
\label{fig:blast-compilation-divb-symm}
\end{figure}

\begin{figure}
 %\centering
\setlength{\tabcolsep}{0.005\textwidth}
\begin{tabular}{cccl}
{\scriptsize Control} & {\scriptsize Symmetric Cleaned} & {\scriptsize Difference Cleaned} & \\
\includegraphics[height=0.305\textwidth]{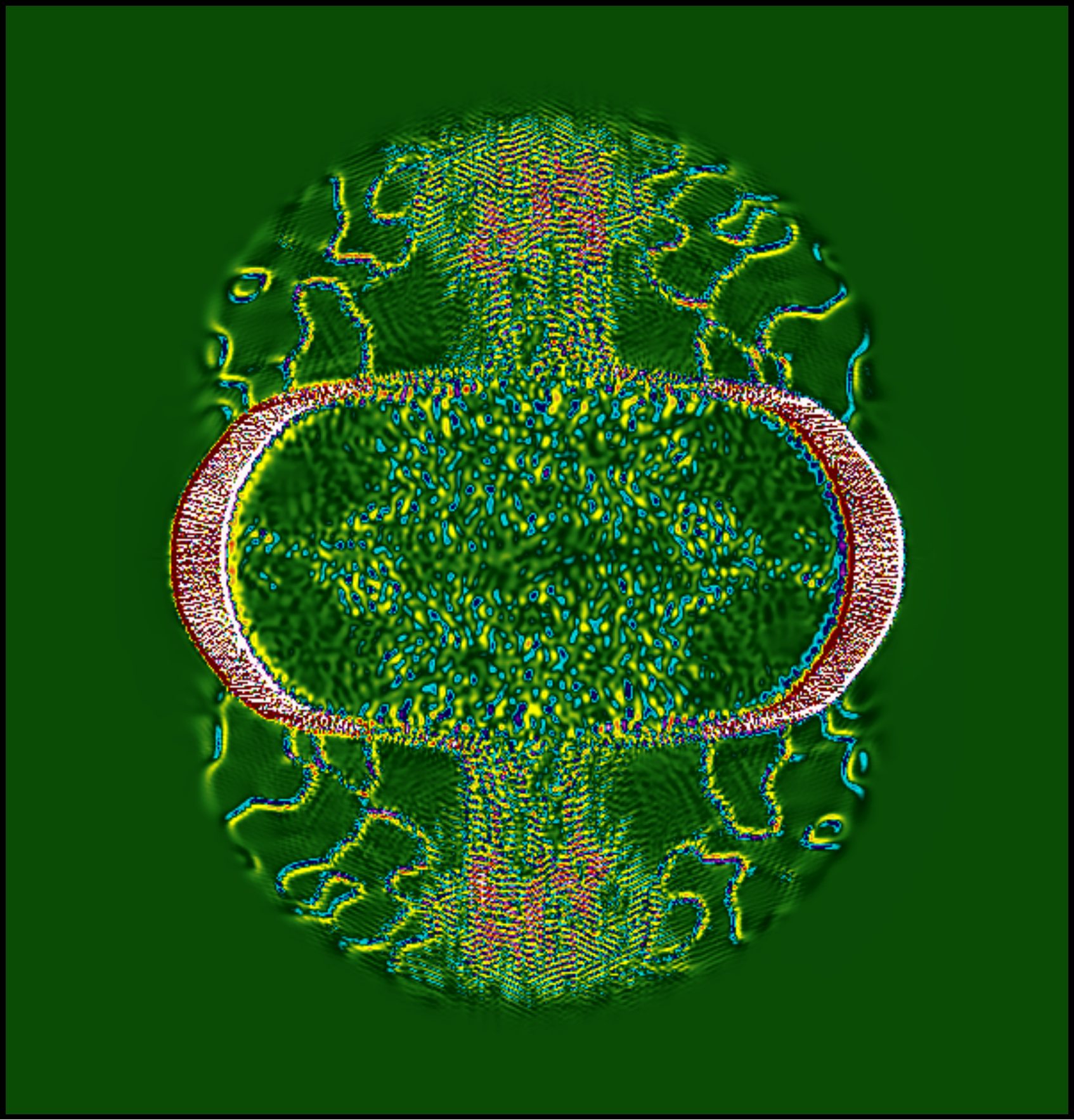} 
 & \includegraphics[height=0.305\textwidth]{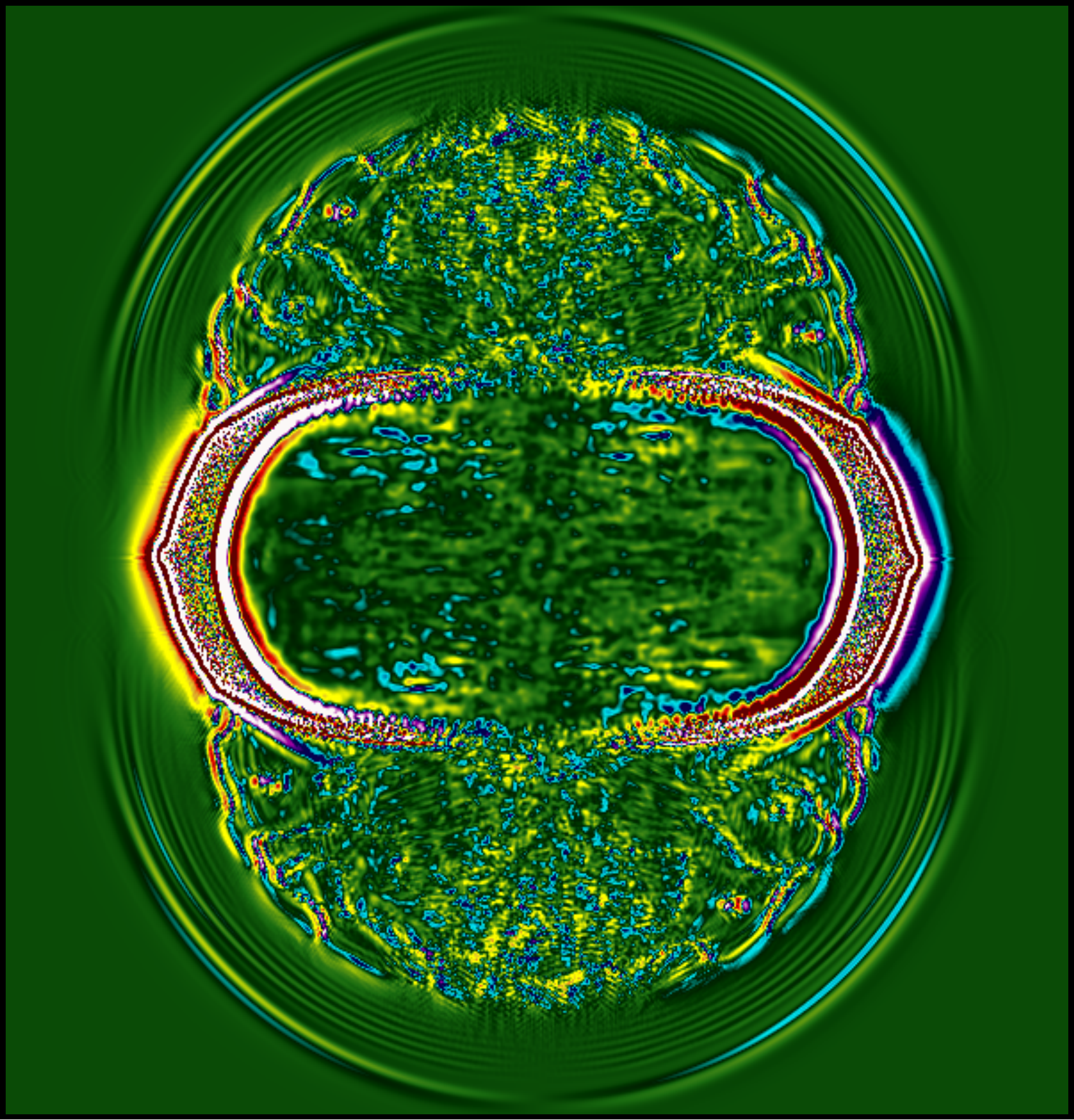}
 & \includegraphics[height=0.305\textwidth]{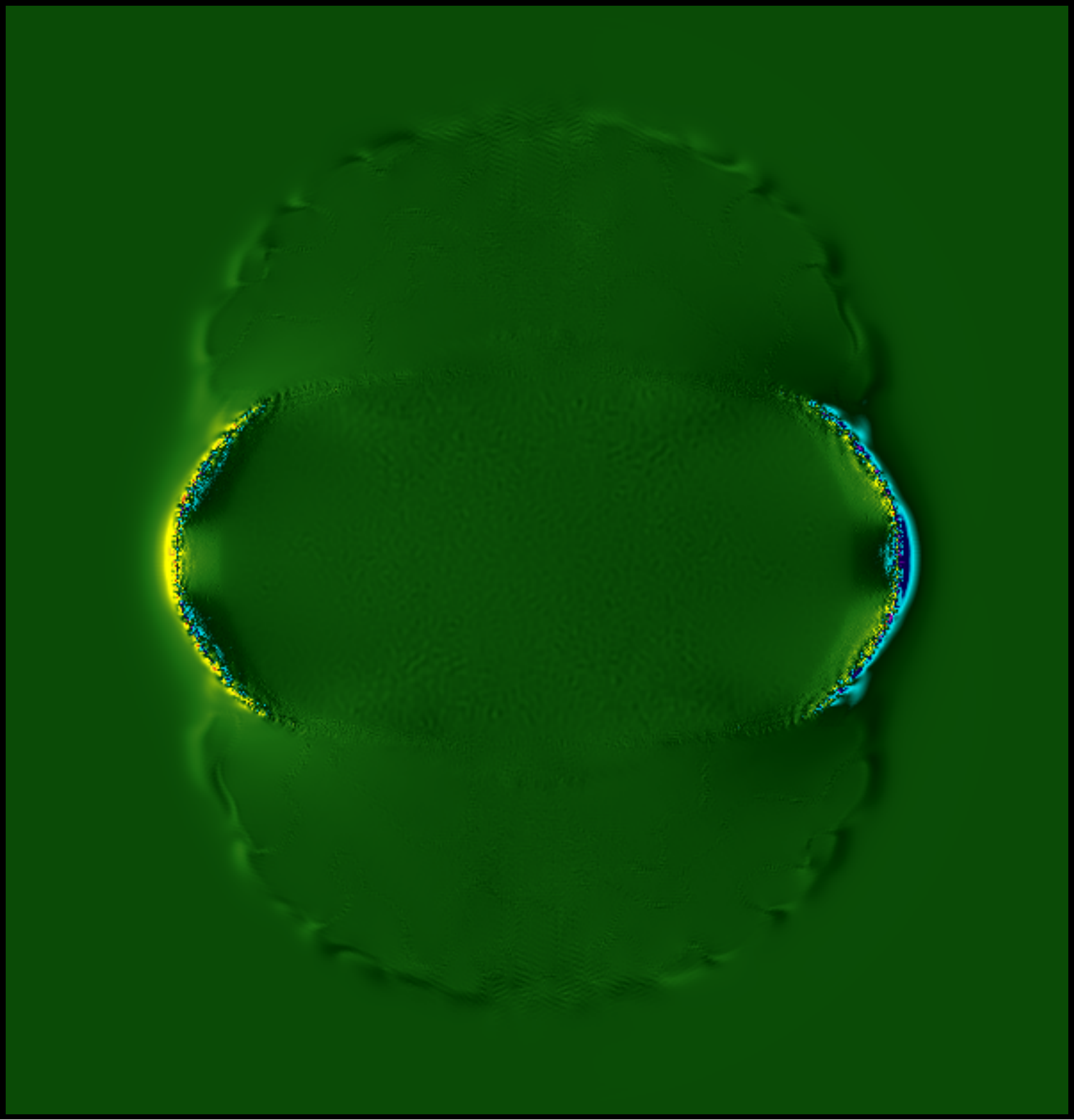}
 & \includegraphics[height=0.305\textwidth]{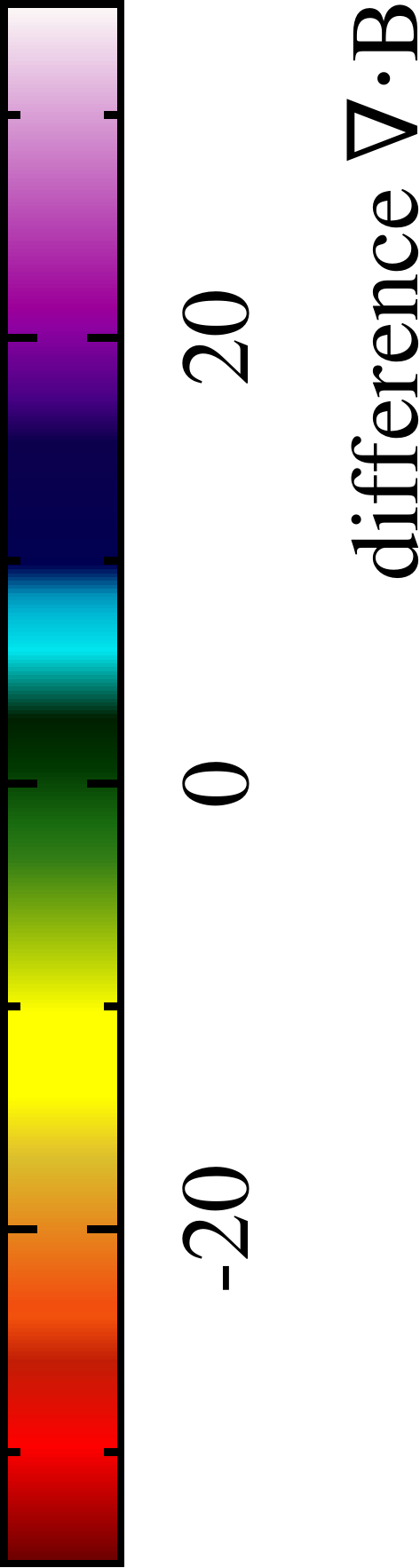}
\end{tabular}
\caption{As in Fig.~\ref{fig:blast-compilation-divb-symm}, but showing $\nabla \cdot {\bf B}$ measured using the difference operator. With this operator, no $\nabla\cdot{\bf B}$ is measured along the leading edge of the magnetic edge for the control and difference-cleaned cases. However, symmetric cleaning produces spurious divergence in this region when measured with the difference operator, because changes have been induced in the magnetic field to compensate for particle disorder.}
\label{fig:blast-compilation-divb-diff}
\end{figure}

To answer the question of whether there is any advantage to cleaning with the symmetric $\nabla \cdot {\bf B}$ operator, the blast wave problem was simulated for three cases: no cleaning; cleaning using the difference operator for $\nabla \cdot {\bf B}$; and cleaning using the symmetric operator. The question is further complicated by fact that the operator used for cleaning may differ from the operator used to measure the error. We therefore show $\nabla \cdot {\bf B}$ for these three cases measured with both the symmetric (Fig.~\ref{fig:blast-compilation-divb-symm}) and difference (Fig.~\ref{fig:blast-compilation-divb-diff}) operators, so that the effect of cleaning using one operator can be seen in both.

The symmetric operator for $\nabla \cdot {\bf B}$ can be seen to pick up a non-zero divergence error on the leading edge of the magnetic wave from the blast (Fig.~\ref{fig:blast-compilation-divb-symm}) despite the fact that the magnetic field shows no error in this region when measured with the difference operator (Fig.~\ref{fig:blast-compilation-divb-diff}). This suggests that the symmetric operator is mainly reflecting the disordered particle arrangement. In turn, it can be seen that in this region, cleaning using the symmetric operator \emph{introduces} divergence error when measured with the difference operator as it attempts to adjust the magnetic field based on the particle arrangement (centre panel of Fig.~\ref{fig:blast-compilation-divb-diff}). Nevertheless, it is true that cleaning with the symmetric operator does produce the greatest reduction in the divergence when measured in the symmetric operator, and may still have potential advantages in terms of momentum conservation (this is examined further in \S\ref{sec:ot}). However, we conclude that cleaning is best performed with the difference operator, since it shows not only the best results when measured with the difference operator (right panel of Fig.~\ref{fig:blast-compilation-divb-diff}), but also an improvement even when measured with the symmetric operator (centre panel of Fig.~\ref{fig:blast-compilation-divb-symm}).

\subsubsection{Optimal damping values}

\begin{figure}
\centering
 \includegraphics[width=0.45\textwidth]{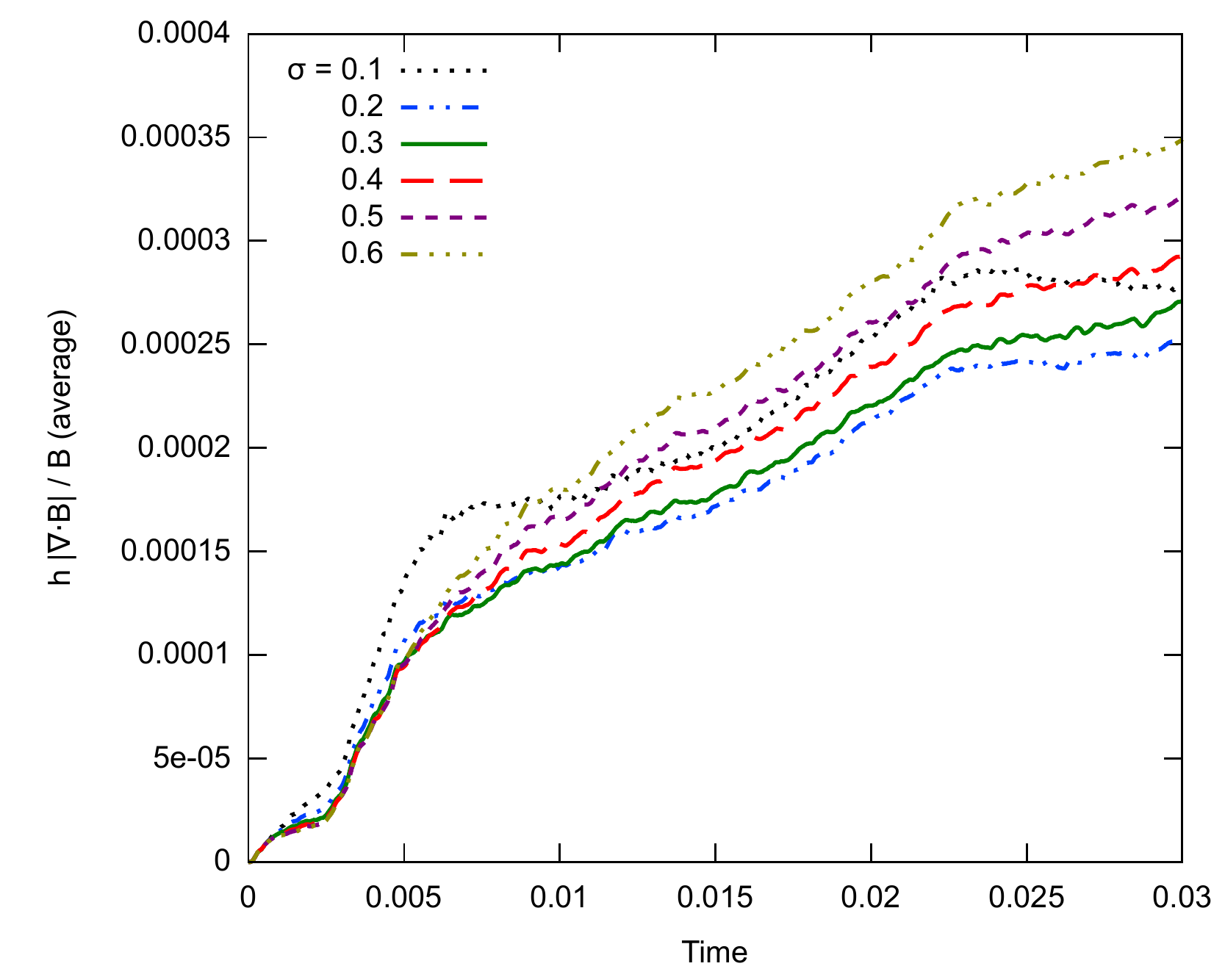}
 \includegraphics[width=0.45\textwidth]{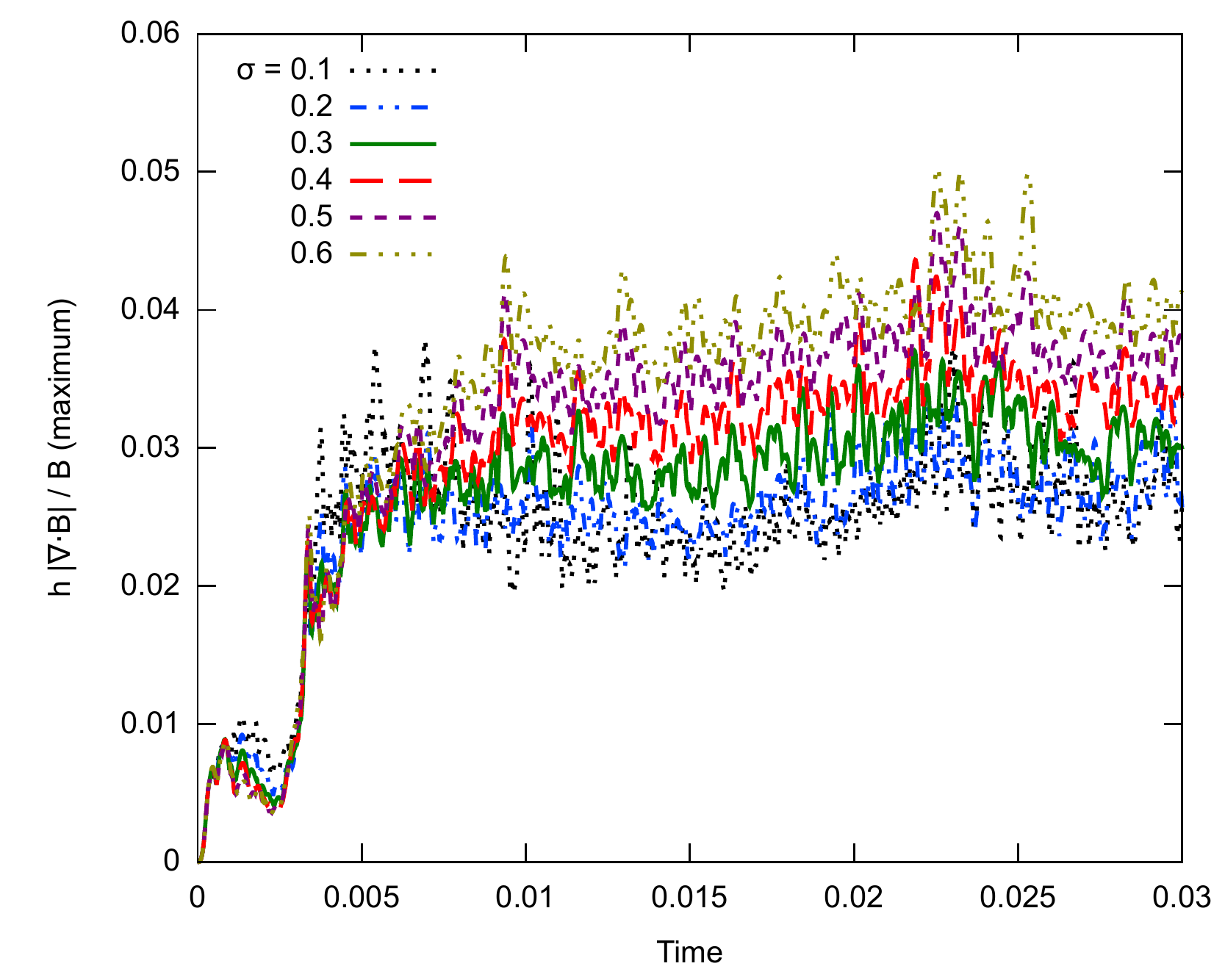}
\caption{Average and maximum $h \vert \nabla \cdot {\bf B}\vert / \vert{\bf B}\vert$ for the blast wave test with varying damping strengths.  The best results are obtained for values of $\sigma$ between 0.2--0.3.}
\label{fig:blast-sigma}
\end{figure}

Fig.~\ref{fig:blast-sigma} shows the average and maximum divergence error as a function of time for differing strengths of the damping parameter $\sigma$ in the range $[0.1, 0.6]$.  The best results are obtained with $0.2 < \sigma < 0.3$, in agreement with the other two dimensional tests.

\subsubsection{Tensile instability correction}
\label{sec:halfdivb}

\begin{figure}
 \centering
\includegraphics[height=0.296\textwidth]{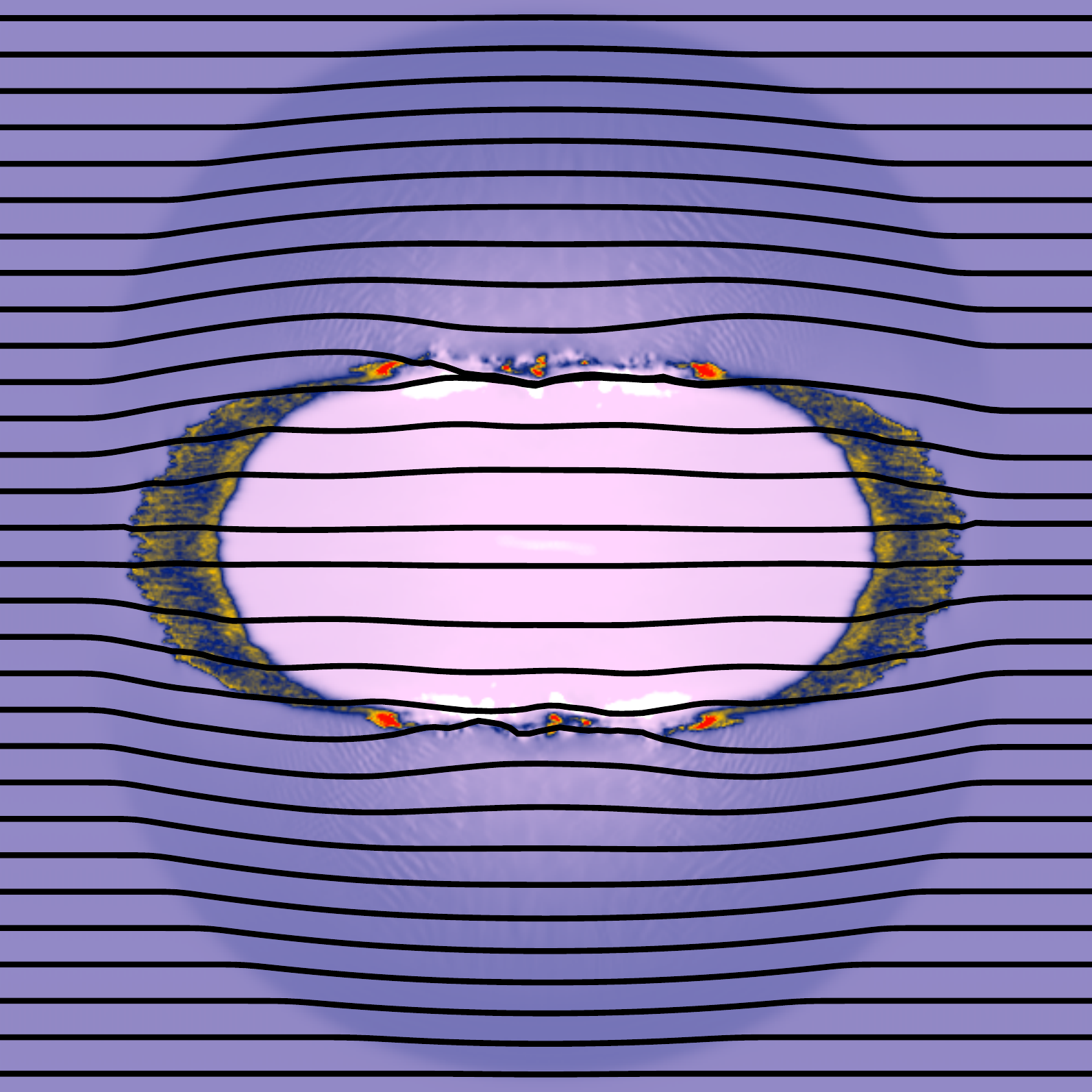}
\includegraphics[height=0.296\textwidth]{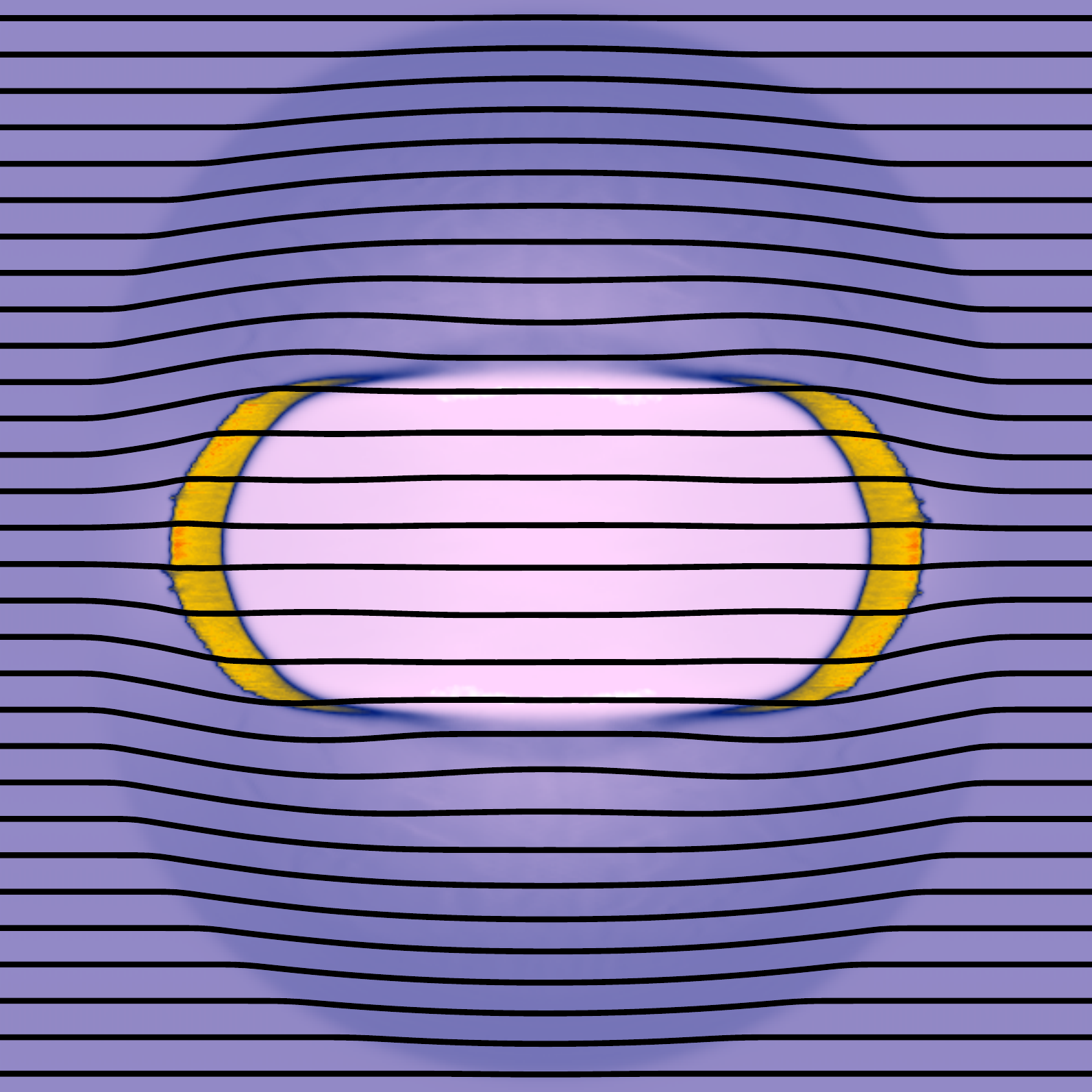}
\includegraphics[height=0.296\textwidth]{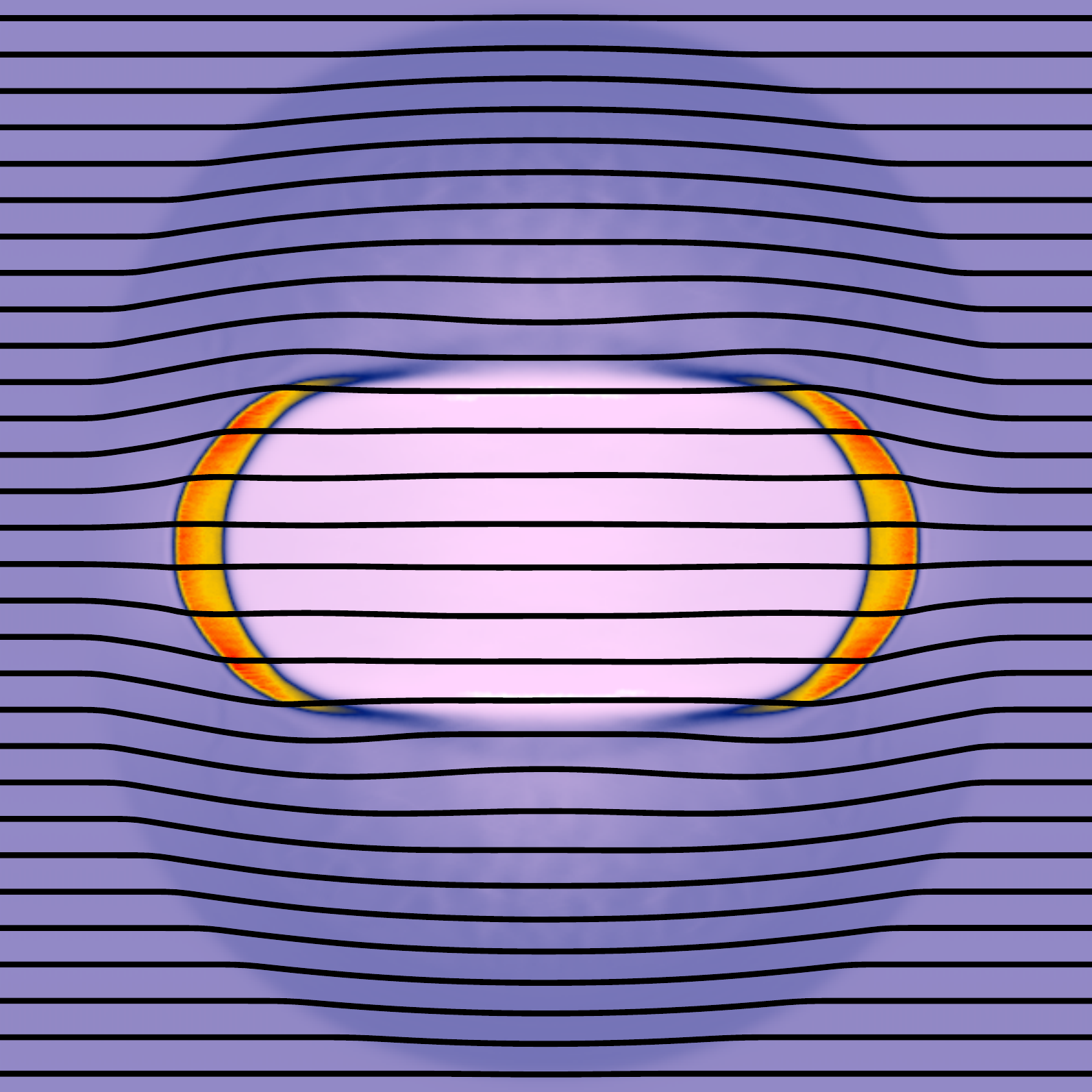}
\includegraphics[height=0.296\textwidth]{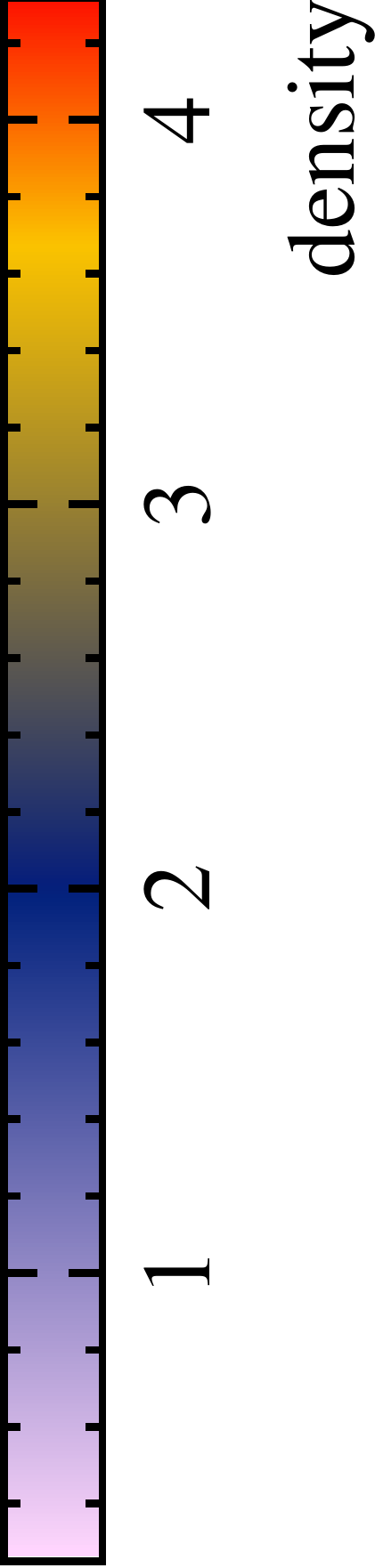}
\caption{Density of the blast wave problem with overlaid magnetic field lines, without any divergence cleaning, but examining the impact of the $-\hat{\beta} {\bf B} (\nabla \cdot {\bf B})$ term used to correct the tensile instability. Results shown use $\hat{\beta} = 0.5$ (left),  $\hat{\beta} = 0.75$ (centre) and $\hat{\beta} = 1.0$ (right).  Using only $\hat{\beta} = 0.5$ is found to result in irregularities along the shock fronts, which are not present using $\hat{\beta} = 1$. Thus, using $\hat{\beta} = 0.5$ is not recommended.}
\label{fig:halfdivb}
\end{figure}
 Finally, we noticed important consequences in this test concerning the $\hat{\beta}{\bf B} (\nabla \cdot {\bf B})$ correction for the tensile instability (\S\ref{sec:tensile-instability-correction}).  Since using $\hat{\beta} = 0.5$ is in principle sufficient to prevent the instability, its use has been suggested by \cite{2011arXiv1112.0340B, price12}.  However, we found this to be problematic in our simulations of the blast wave problem: Fig.~\ref{fig:halfdivb} shows the density with overlaid magnetic field lines at $t=0.03$ using ${\hat{\beta}}$ = 0.5, 0.75 and 1.0 (left to right). With only $\hat{\beta} = 0.5$ (left panel), irregularities can be seen to form in the densest parts of the shockwave. These are not present when performing the full $\hat{\beta} = 1$ subtraction (right panel).

\subsection{Orszag-Tang Vortex}
\label{sec:ot}
The final two dimensional test is the Orszag-Tang vortex \cite{1979JFM....90..129O}, which has been widely used as a test of MHD codes \cite[e.g.][]{2009MNRAS.398.1678D, 2006A&A...457..371F, 2008ApJS..178..137S}.  It consists of a magnetic vortex superimposed onto a velocity vortex generating several classes of interacting shock waves. The complex dynamics provides an excellent test of the constrained hyperbolic divergence cleaning method.  To measure the effectiveness of the method in this case, the results are compared against that of simulations using artificial resistivity (with particle independent strengths as described in \S\ref{sec:blast-setup}) and Euler Potentials as measures of divergence control.  This test is also used to examine whether or not cleaning using the symmetric operator for $\nabla \cdot {\bf B}$ provides any advantage in terms of momentum conservation.  As previously, the damping parameter $\sigma$ is varied to find optimal values.

\subsubsection{Setup}

The problem is set up in a box with dimensions $x,y \in [0,1]$ with periodic boundary conditions.  The initial gas state is set to $\rho = 25 / (36\pi)$, $P = 5/(12\pi)$, $\gamma = 5/3$, with velocity field ${\bf v} = [-\sin(2\pi y), \sin(2\pi x)]$.  The initial magnetic field is ${\bf B} = [-\sin(2\pi y), \sin(4\pi x)]$.  All examples presented use $512 \times 590$ particles initially arranged on a hexagonal lattice.

\subsubsection{Results}

% density: 0.073,  0.350
% magpres: 1.0e-6,  0.280
% divb: -8,  +8

\begin{figure}
 %\centering
\setlength{\tabcolsep}{0.005\textwidth}
\begin{tabular}{ccccl}
{\scriptsize Control} & {\scriptsize Resistivity} & {\scriptsize Euler Potentials} & {\scriptsize Divergence Cleaning} & \\
   \includegraphics[height=0.22\textwidth]{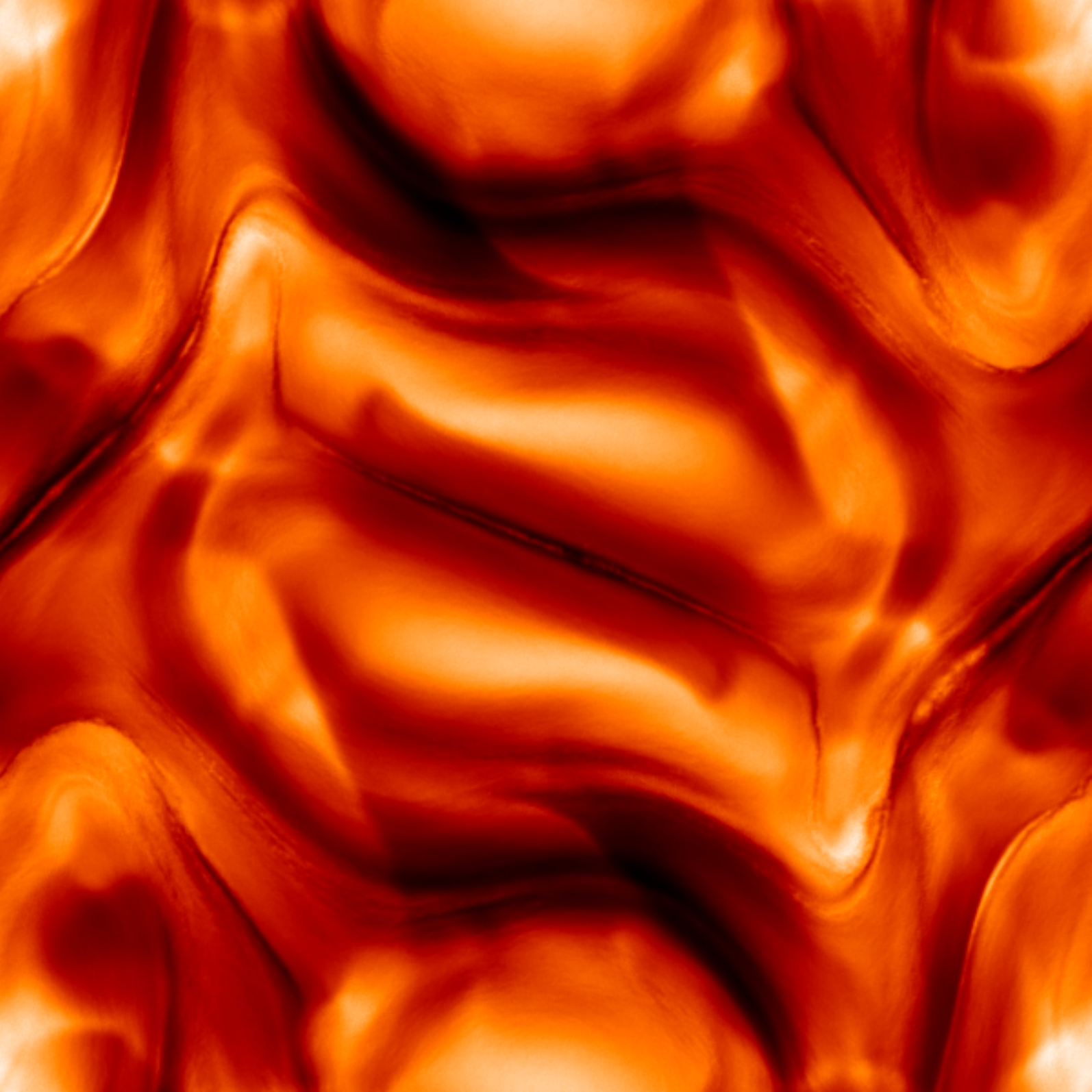} 
 & \includegraphics[height=0.22\textwidth]{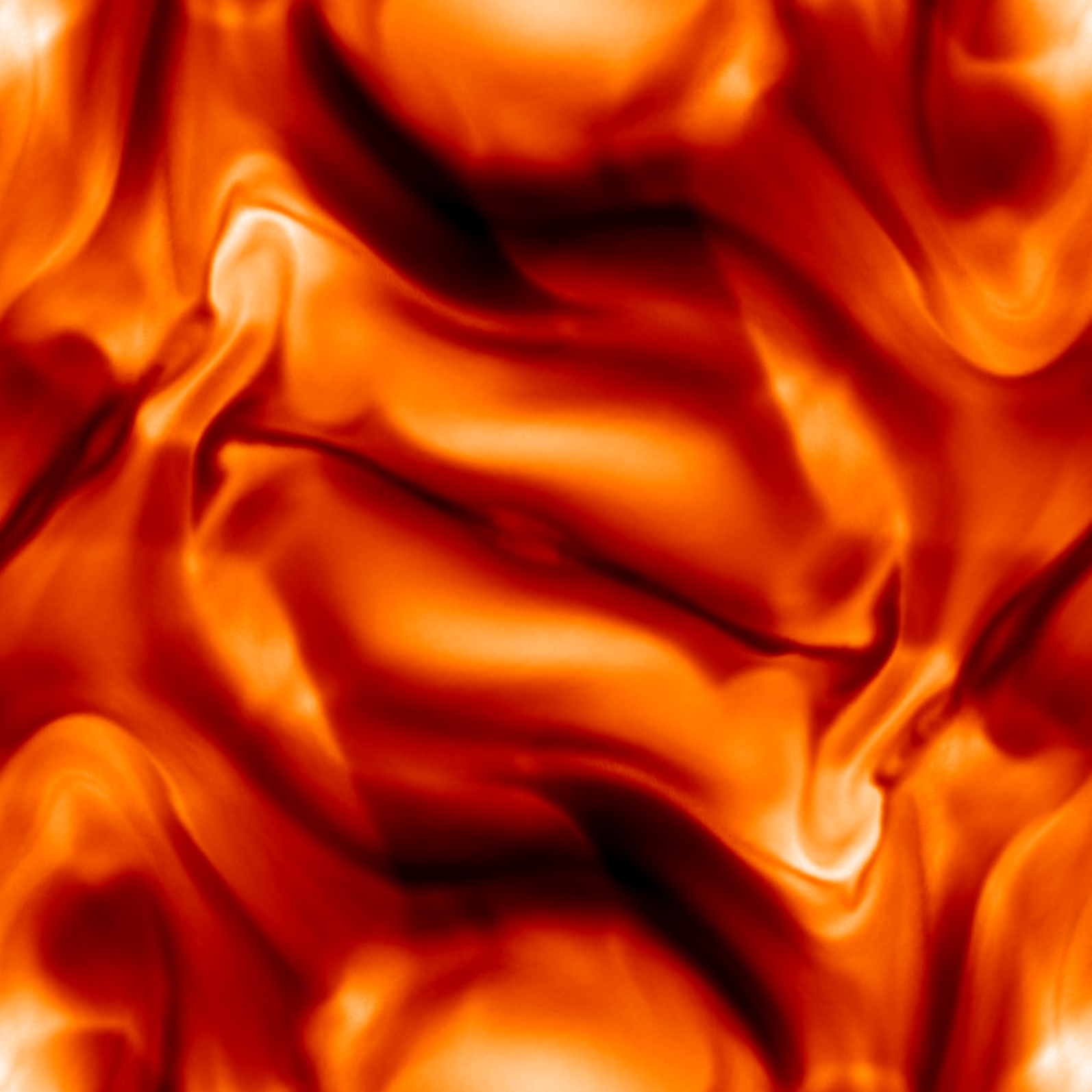}
 & \includegraphics[height=0.22\textwidth]{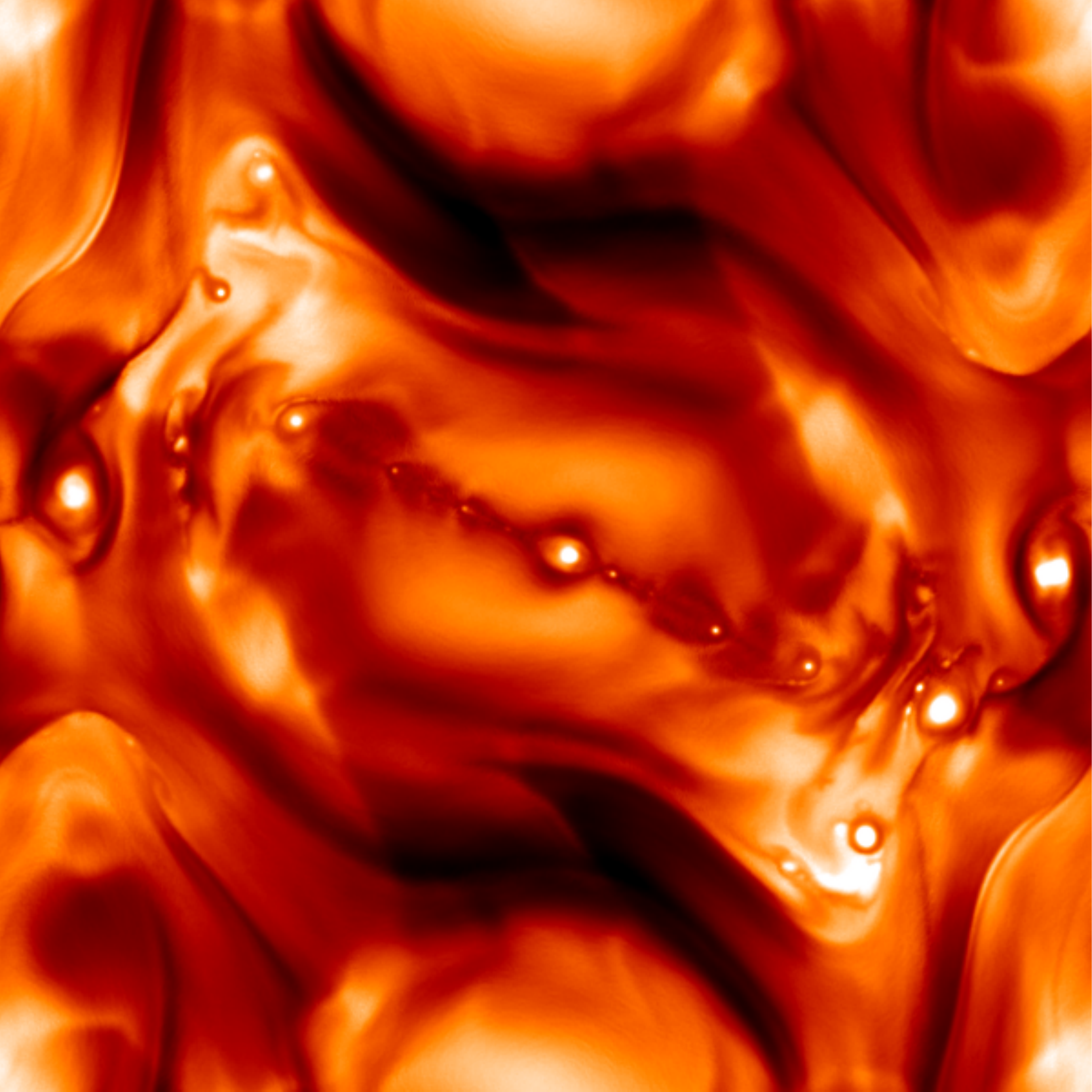}
 & \includegraphics[height=0.22\textwidth]{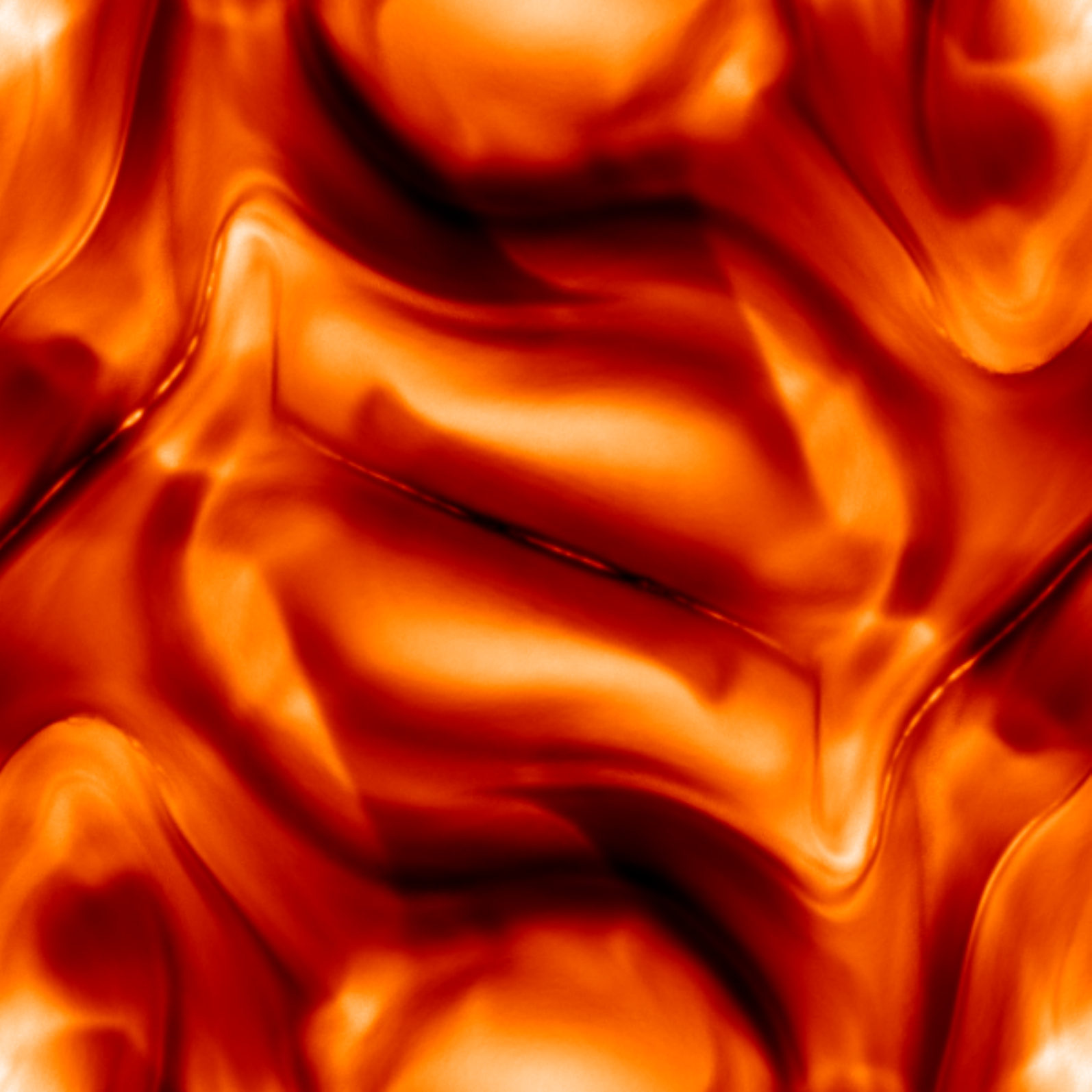}
 & \includegraphics[height=0.22\textwidth]{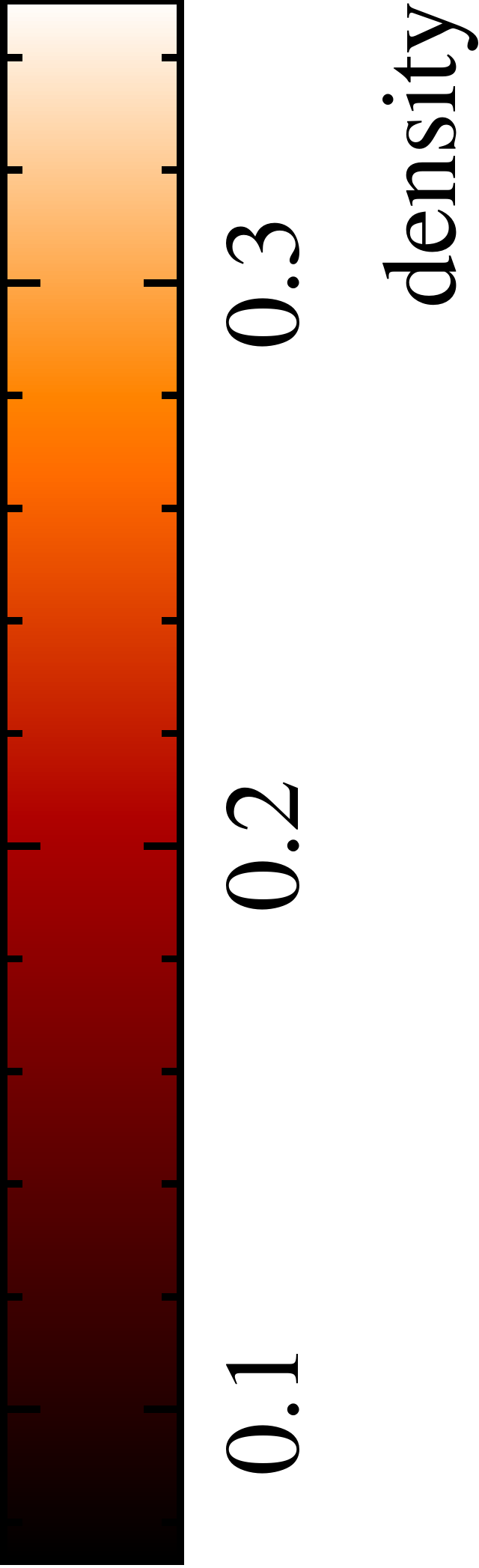}
\\
   \includegraphics[height=0.22\textwidth]{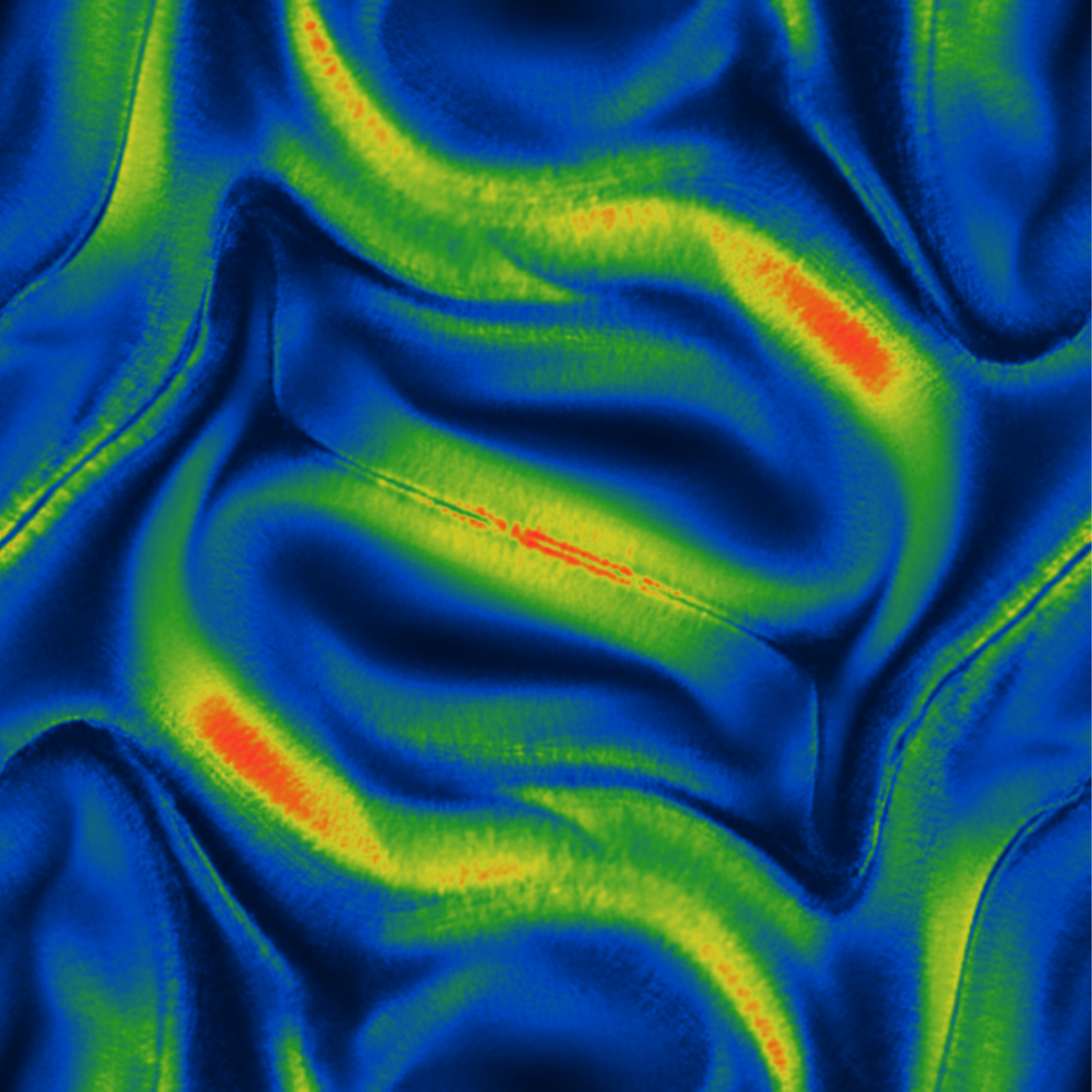}
 & \includegraphics[height=0.22\textwidth]{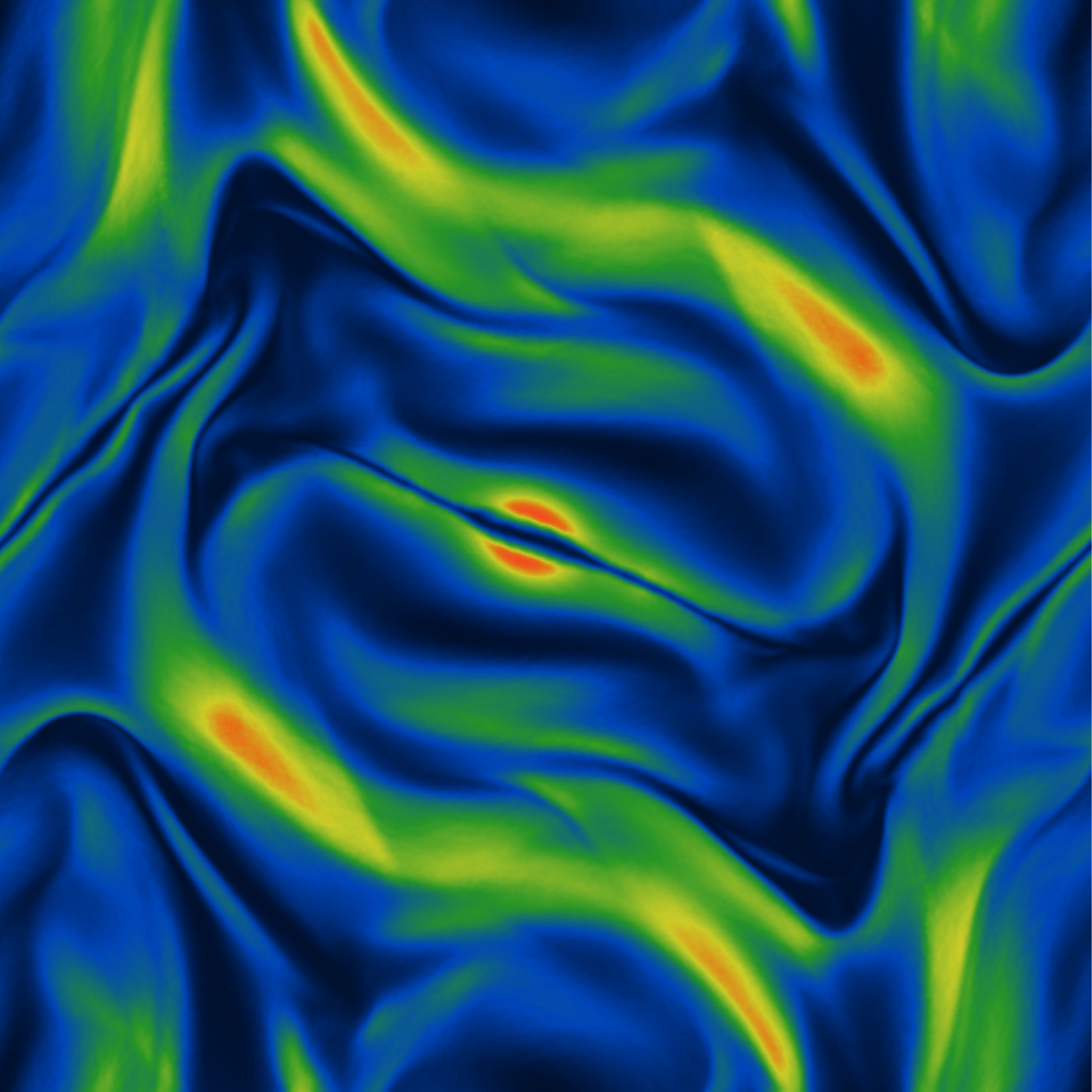}
 & \includegraphics[height=0.22\textwidth]{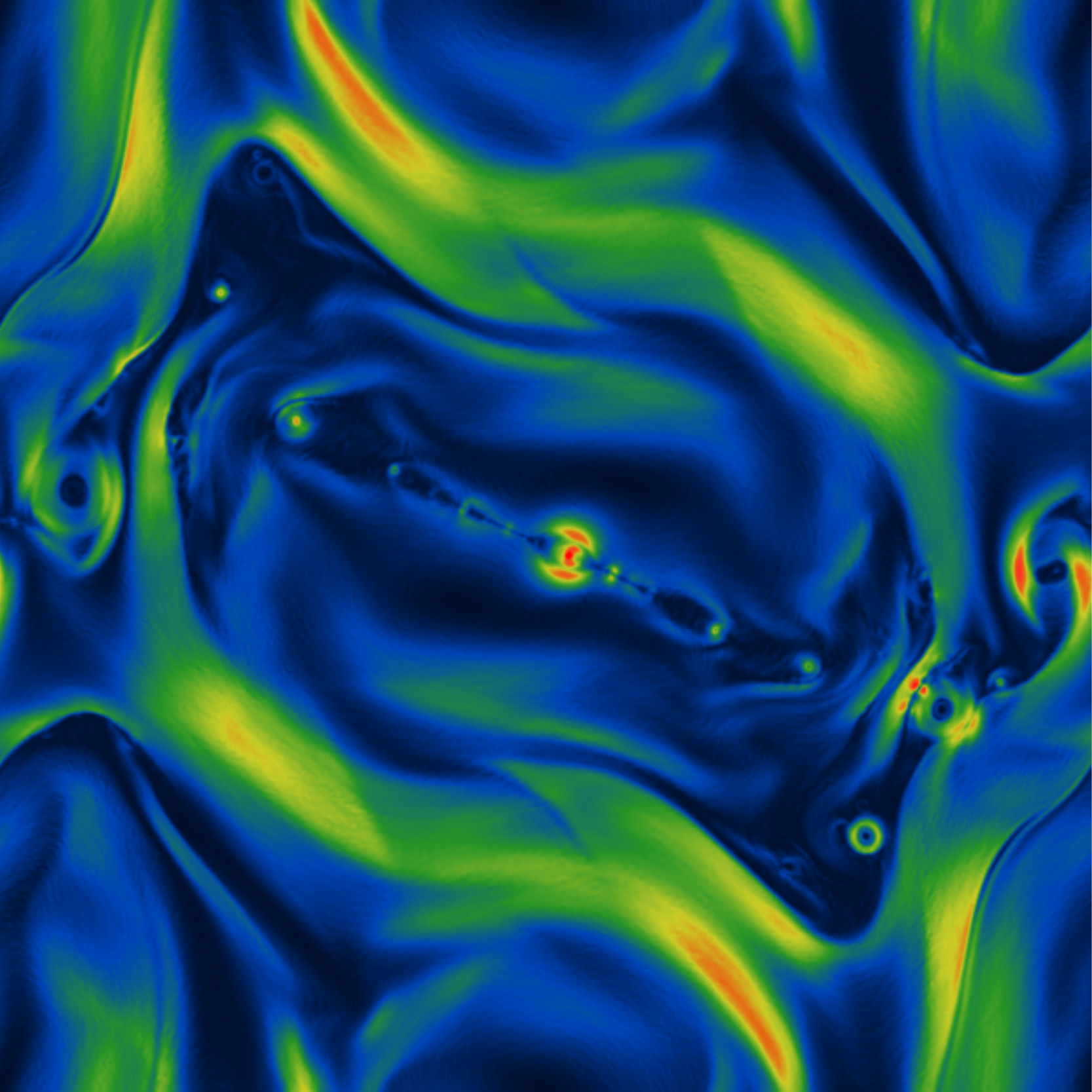}
 & \includegraphics[height=0.22\textwidth]{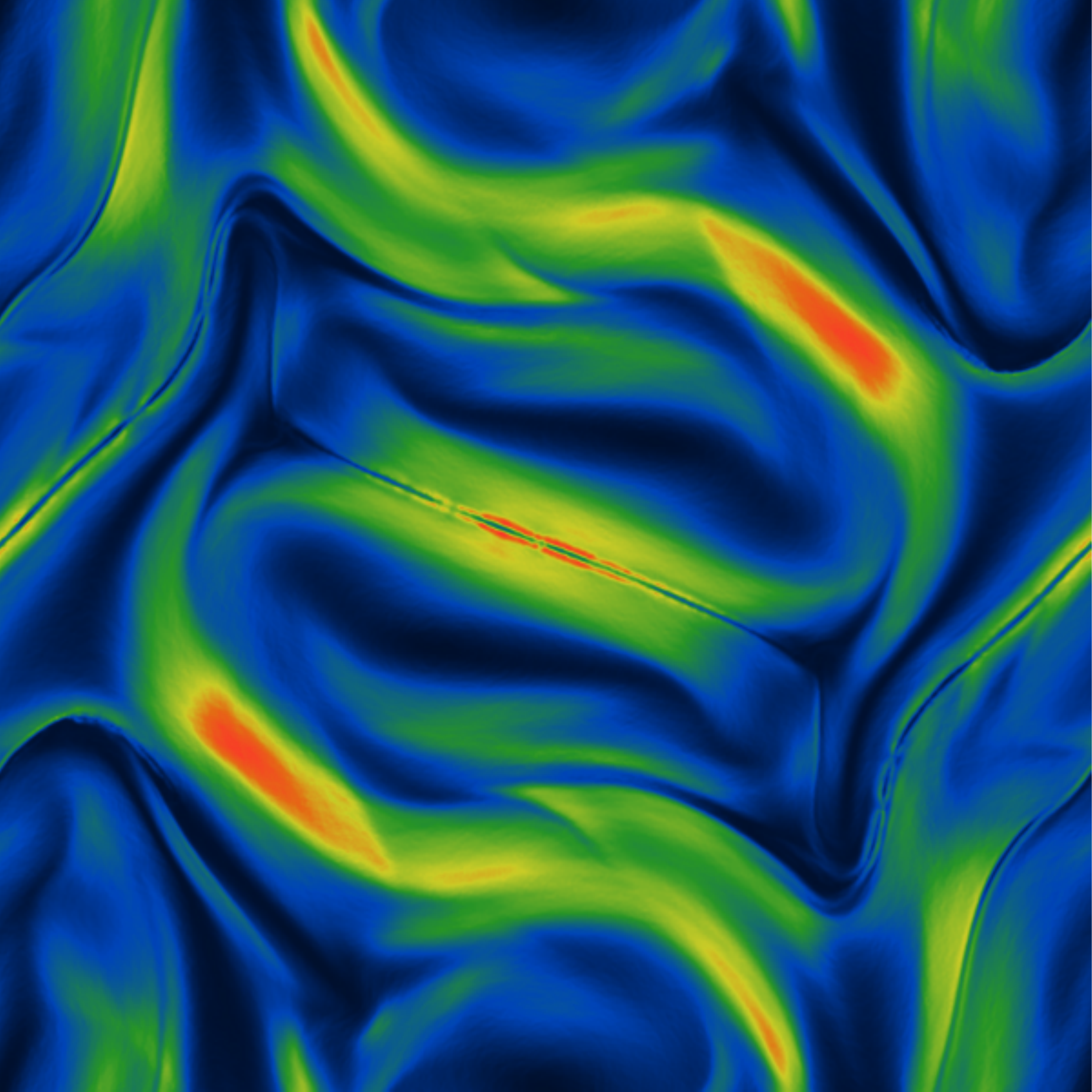}
 & \includegraphics[height=0.22\textwidth]{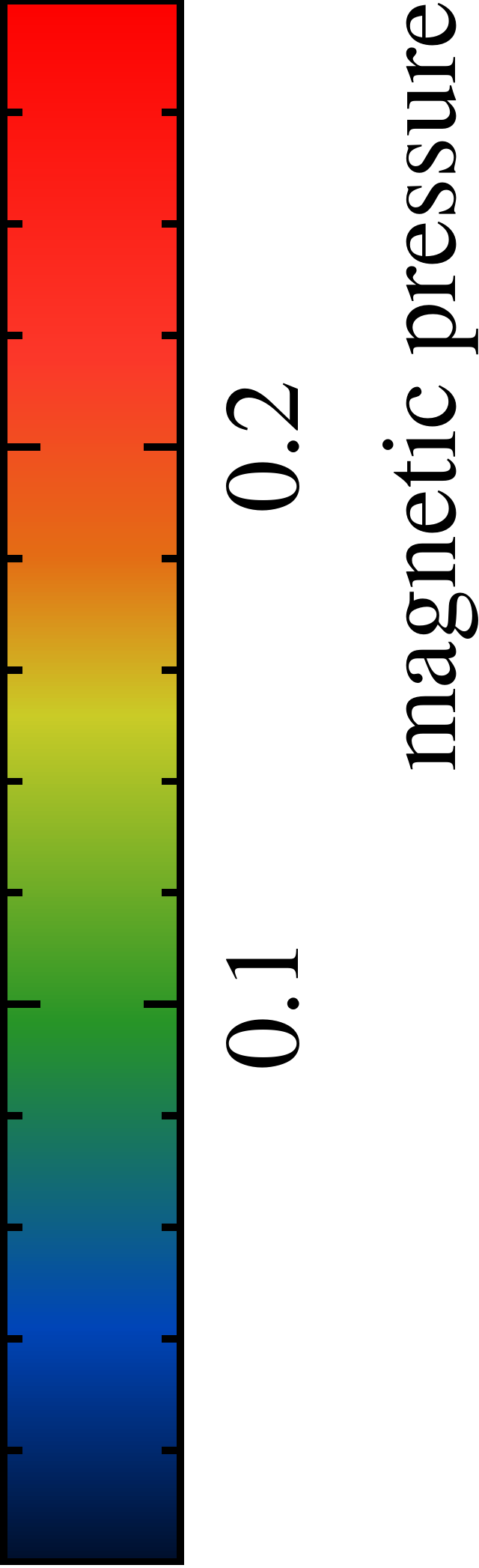}
\\
   \includegraphics[height=0.22\textwidth]{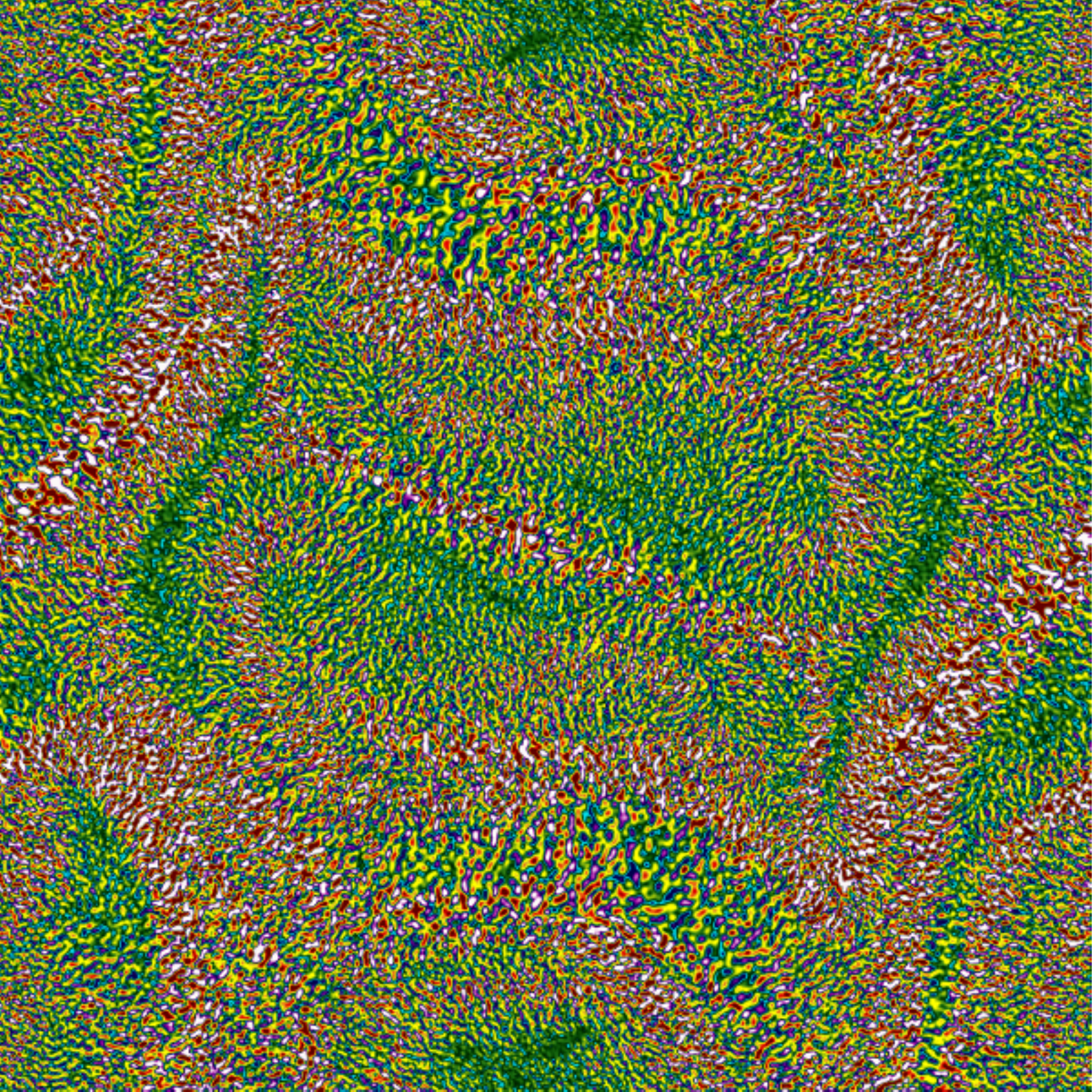}
 & \includegraphics[height=0.22\textwidth]{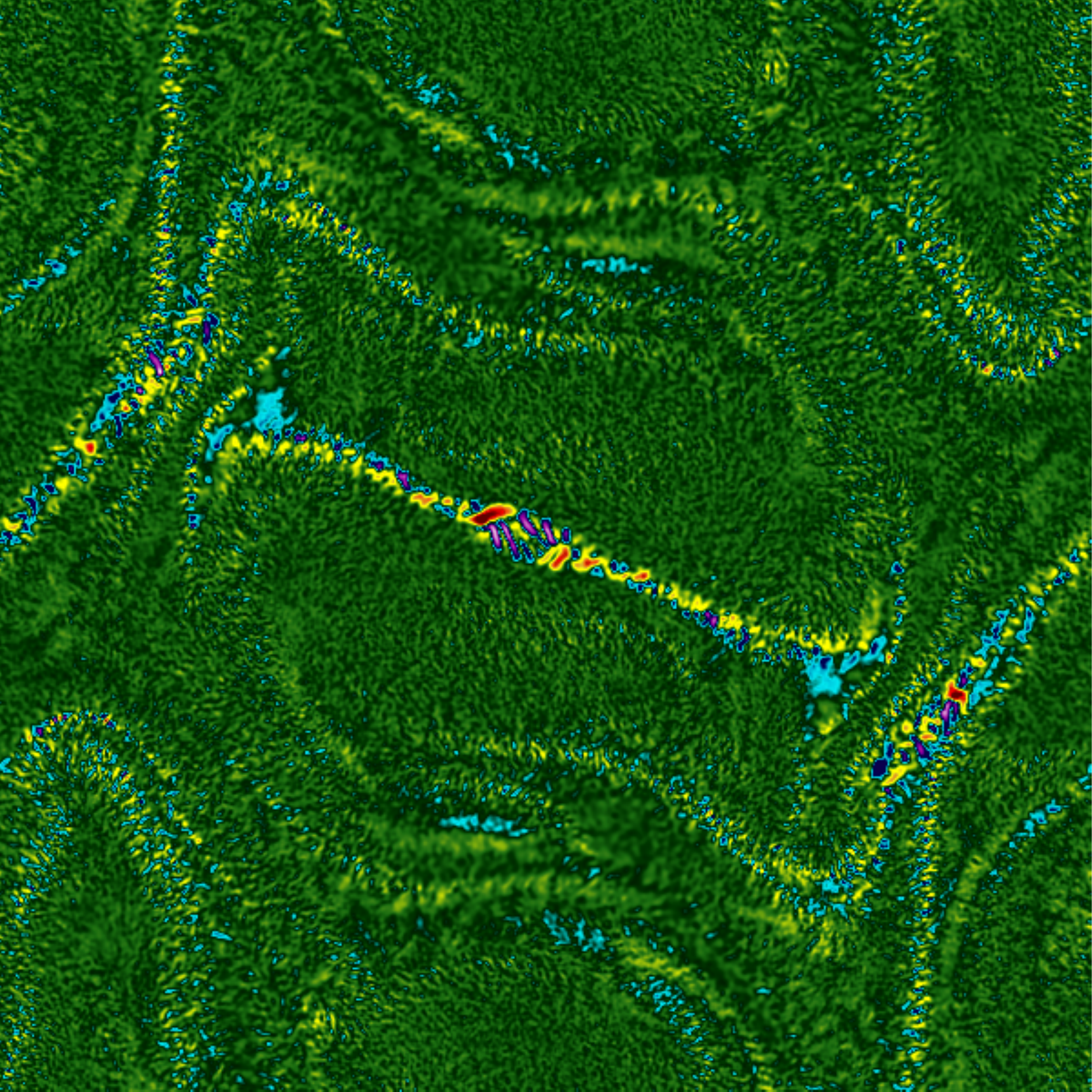}
 & \includegraphics[height=0.22\textwidth]{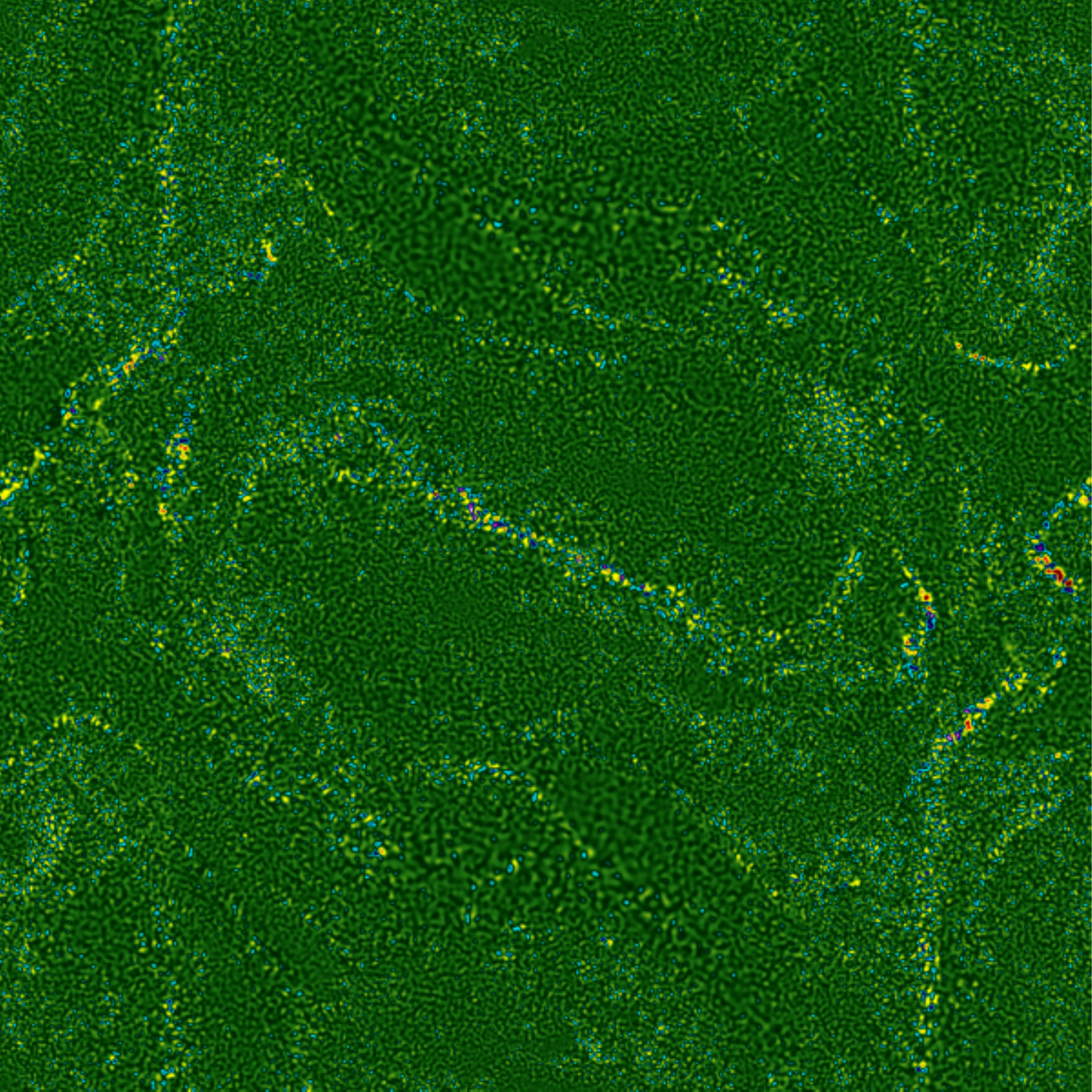}
 & \includegraphics[height=0.22\textwidth]{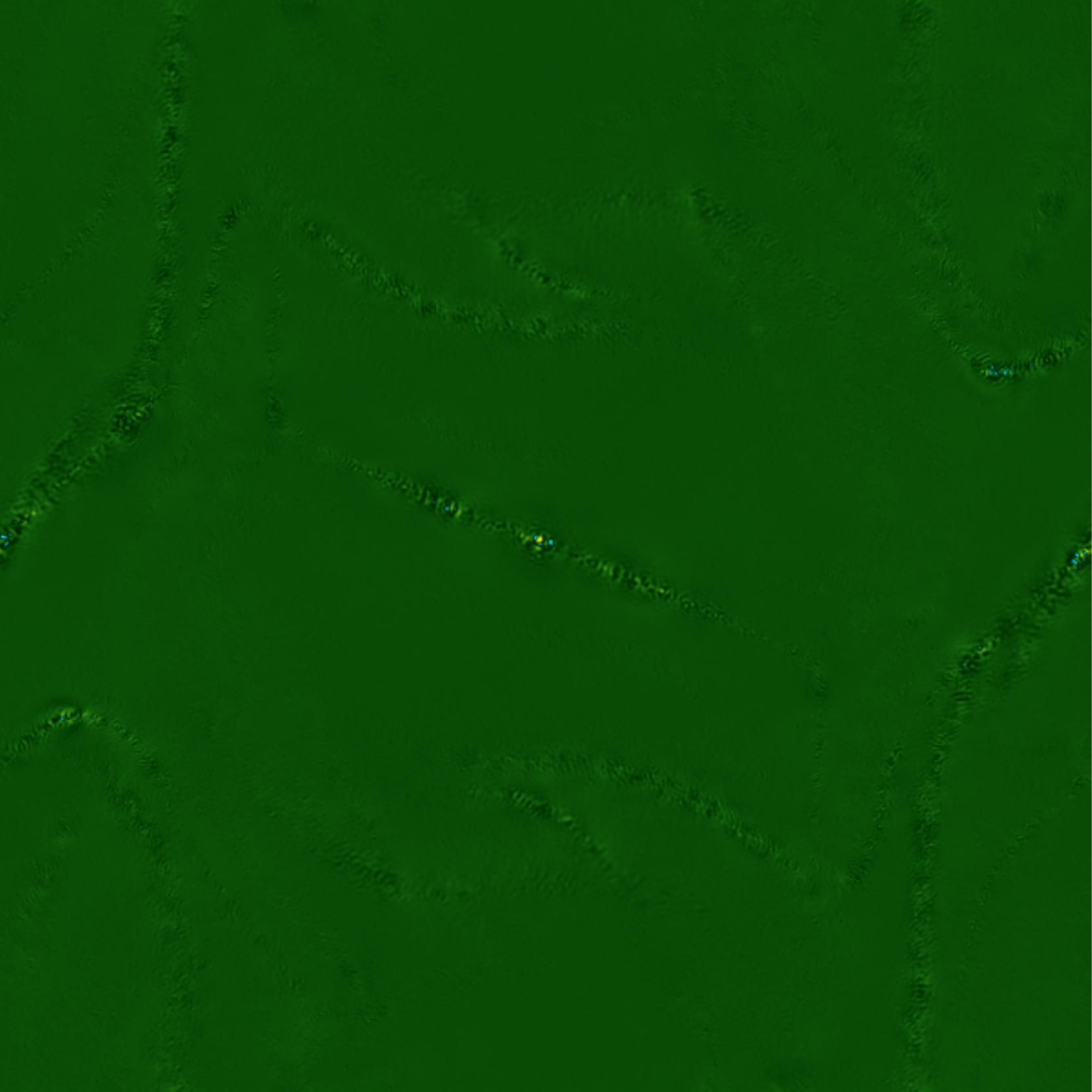}
 & \includegraphics[height=0.22\textwidth]{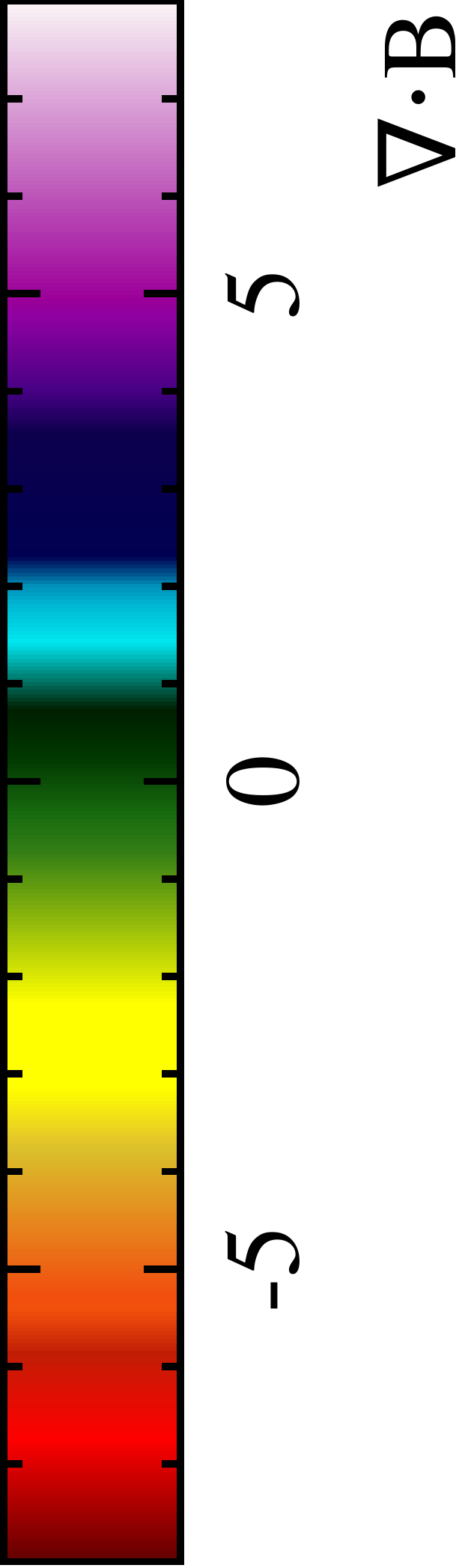}
\end{tabular}
\caption{The density (top row), magnetic pressure (middle row), and the difference measurement of $\nabla \cdot {\bf B}$ (bottom row) in the Orszag-Tang vortex at $t=1.0$ comparing the control case (far left), including artificial resistivity (centre left), evolving the magnetic field using Euler Potentials (centre right), and applying the constrained divergence cleaning method (far right).}
\label{fig:orszag-compilation}
\end{figure}

\begin{figure}
 \centering
\includegraphics[width=0.45\textwidth]{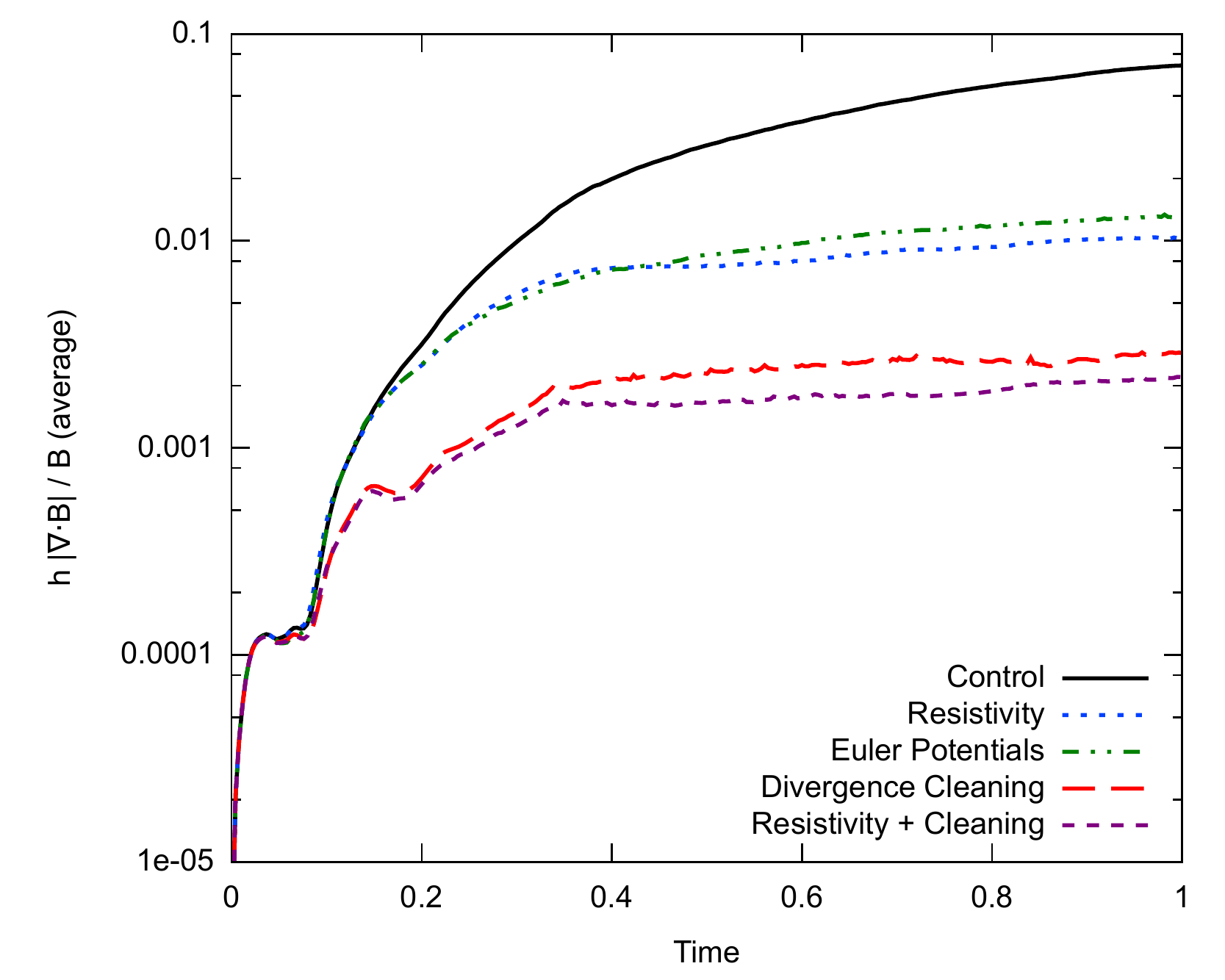}
\includegraphics[width=0.45\textwidth]{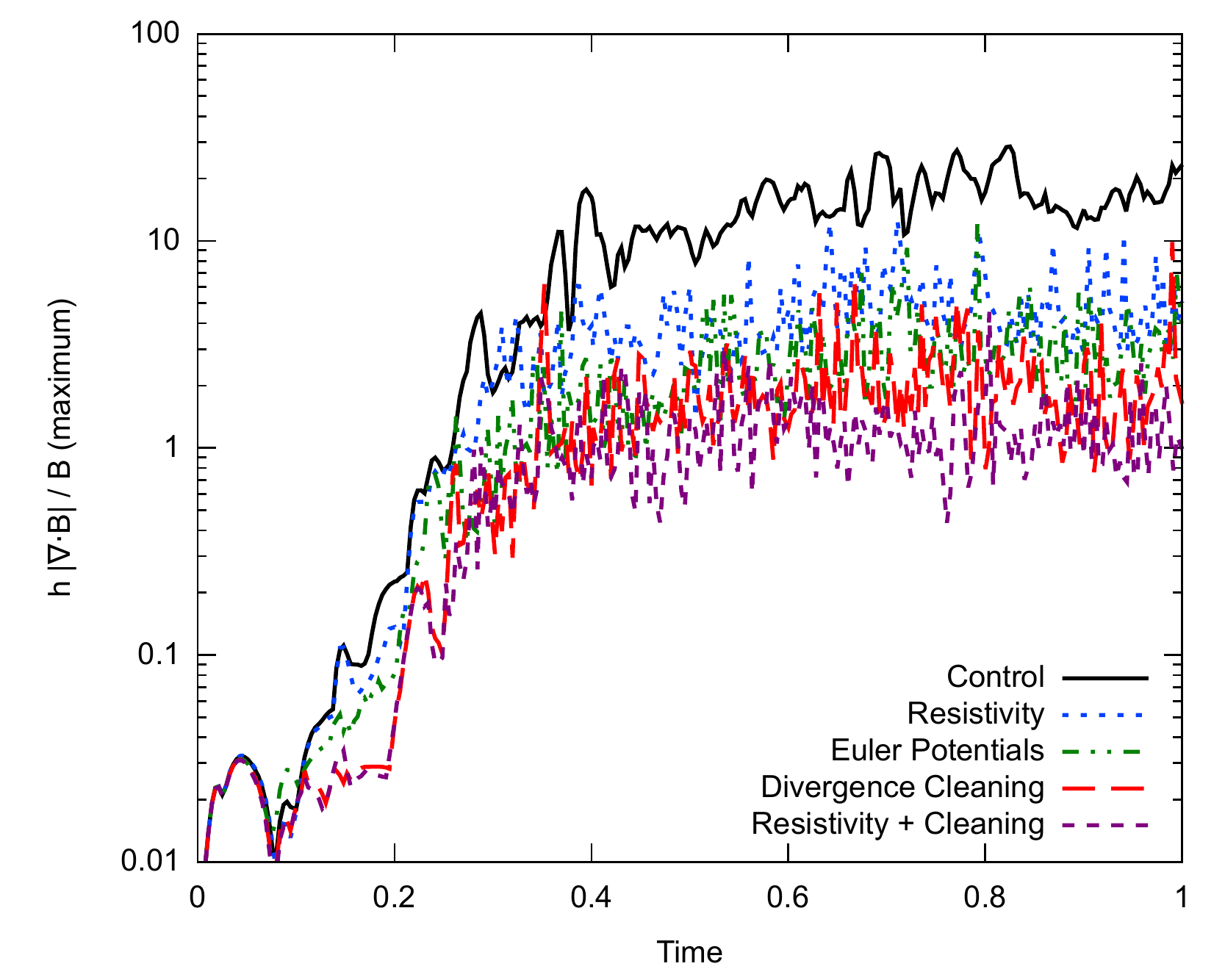}
\caption{Average (left) and maximum (right) $h \vert \nabla \cdot {\bf B}\vert / \vert{\bf B}\vert$ in the Orszag-Tang vortex problem with (top to bottom in left panel) no divergence control; using Euler Potentials, adding an artificial resistivity, using divergence cleaning, and cleaning while including resistivity.  Divergence cleaning has lower divergence error than when using Euler Potentials or artificial resistivity, and continues to reduce divergence error even when used in combination with artificial resistivity.}
\label{fig:orszag-divb}
\end{figure}

Fig.~\ref{fig:orszag-compilation} shows the density (top), magnetic pressure (middle row), and $\nabla \cdot {\bf B}$ (bottom row) at $t=1.0$ for four cases: i) control, ii) using artificial resistivity, iii) employing Euler Potentials, and iv) applying divergence cleaning.  This time is chosen because the divergence errors in the control case are large enough to produce small scale disturbances in the density and magnetic pressure fields.  By adding resistivity or using Euler Potentials, the average $h |\nabla \cdot {\bf B}| / |{\bf B}|$ is decreased by an order of magnitude (c.f. second and third panels in bottom row of Fig.~\ref{fig:orszag-compilation} and the left panel of Fig.~\ref{fig:orszag-divb}).  When divergence cleaning is used, the average divergence error is reduced by almost two orders of magnitude (red/dashed line in left panel of Fig.~\ref{fig:orszag-divb}).  In addition to the average and maximum divergence error for the above four cases, Fig.~\ref{fig:orszag-divb} also presents the results from a case where artificial resistivity has been applied in tandem with divergence cleaning.  In this case, the average $h |\nabla \cdot {\bf B}| / |{\bf B}|$ is reduced by nearly an order of magnitude compared to resistivity alone, and when compared to the control case, this results in two orders of magnitude reduction in the average together with an order of magnitude reduction in the maximum.

\subsubsection{Cleaning using symmetric $\nabla \cdot {\bf B}$}

\begin{figure}
\centering
\begin{minipage}[t]{0.45\textwidth}
\includegraphics[width=\textwidth]{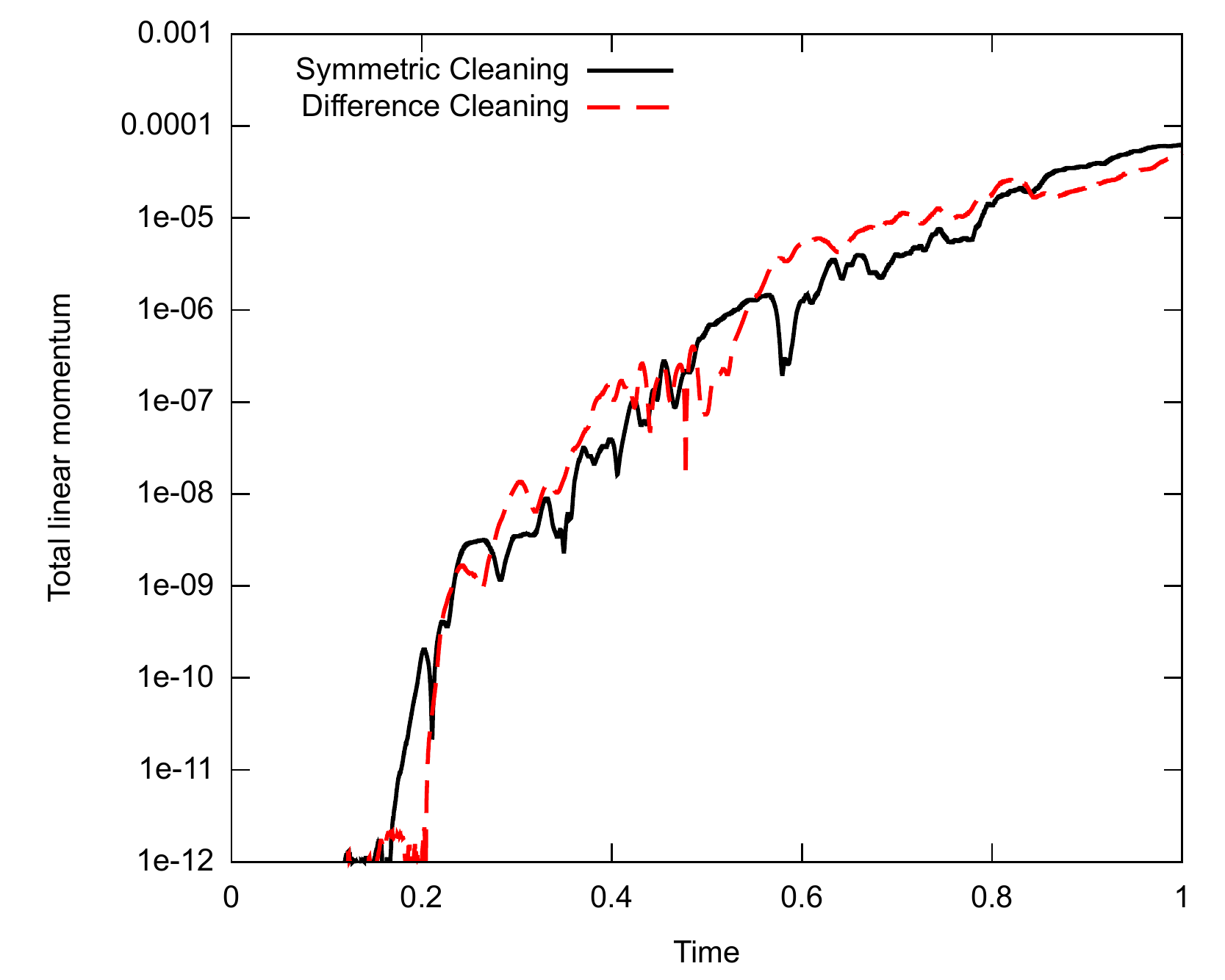}
\caption{Total linear momentum for the Orszag-Tang vortex for divergence cleaning using the difference and symmetric operators of $\nabla \cdot {\bf B}$.  There is no significant distinction between the two.}
\label{fig:orszag-mom}
\end{minipage}
\hspace{0.05\textwidth}
\begin{minipage}[t]{0.45\textwidth}
\includegraphics[width=\textwidth]{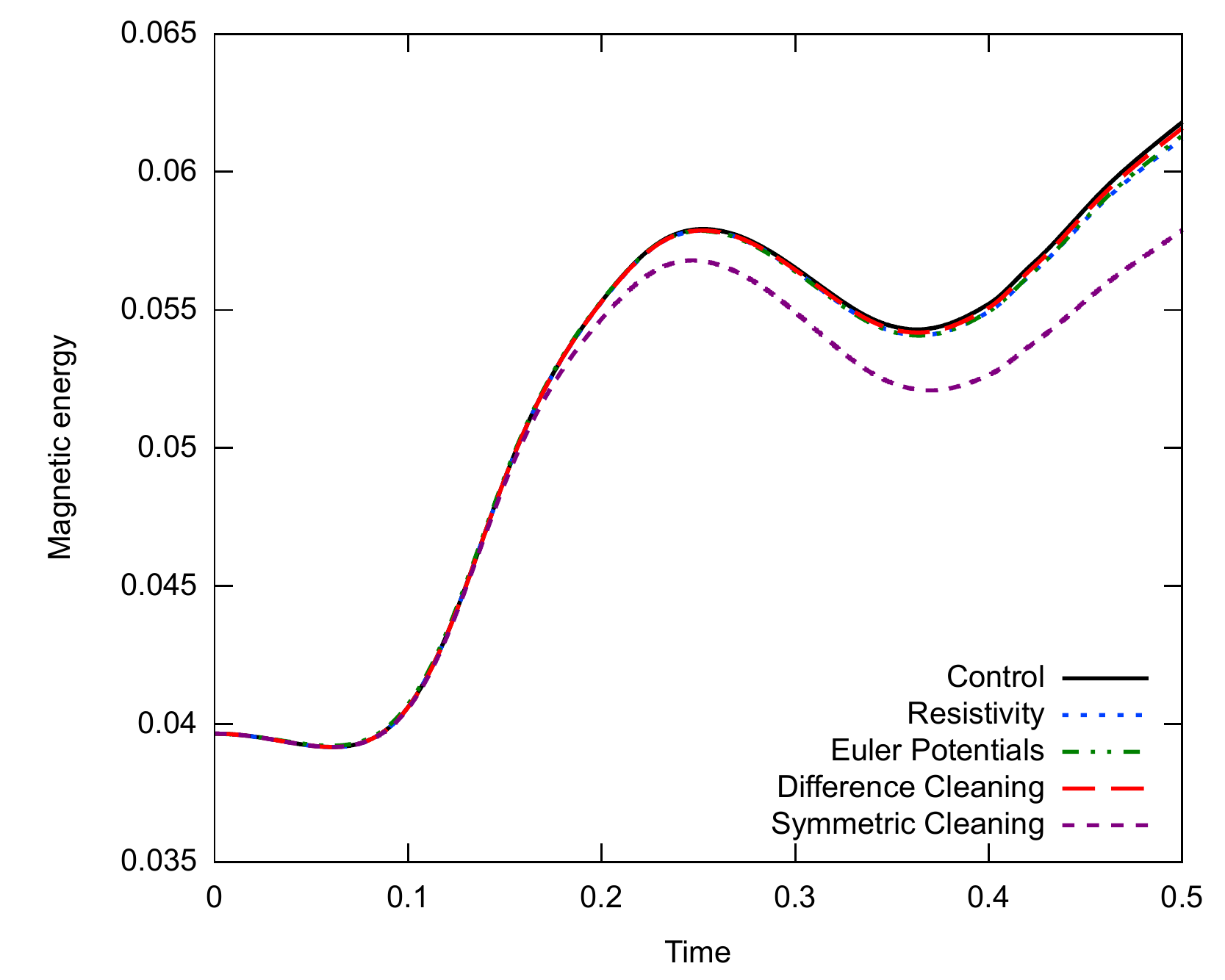}
\caption{Magnetic energy as a function of time in the Orszag-Tang vortex test.  Using the symmetric form of $\nabla \cdot {\bf B}$ for divergence cleaning leads to a $10\%$ reduction in magnetic energy by $t=0.5$ compared to the other schemes.}
\label{fig:orszag-be}
\end{minipage}
\end{figure}

Since the symmetric operator for $\nabla \cdot {\bf B}$ is used in the momentum equation and tensile instability correction term, it was hoped that its use for cleaning would confer some advantage over the difference measure by way of improved momentum conservation.  However, as shown in Fig.~\ref{fig:orszag-mom}, no significant difference in the momentum is found between cleaning with the symmetric operator compared to the difference operator.   Fig.~\ref{fig:orszag-be} shows the magnetic energy profile of the system for $t \le 0.5$, where all test cases (control, resistivity, Euler Potentials, difference cleaning) yield the same profile, except for symmetric cleaning which shows a $\sim 10\%$ reduction in magnetic energy compared to the other solutions. This occurs due to the symmetric operator removing magnetic energy to compensate for irregularities in particle position (which begin to occur at $t \sim 0.15$). Furthermore, although we have already shown in \S\ref{sec:blast-symmdivb} that use of ${\hat{\beta}} = \tfrac{1}{2}$ in the tensile instability correction could result in numerical artefacts in the blast wave test, we also found large errors in the density and magnetic field profiles when ${\hat{\beta}} = \tfrac{1}{2}$ is used in combination with symmetric cleaning on the Orszag-Tang problem.  For these reasons, we recommend using $\hat{\beta} =1$ and applying cleaning only with the difference $\nabla \cdot {\bf B}$ operator.

\subsubsection{Optimal damping values}

\begin{figure}
 \centering
\includegraphics[width=0.45\textwidth]{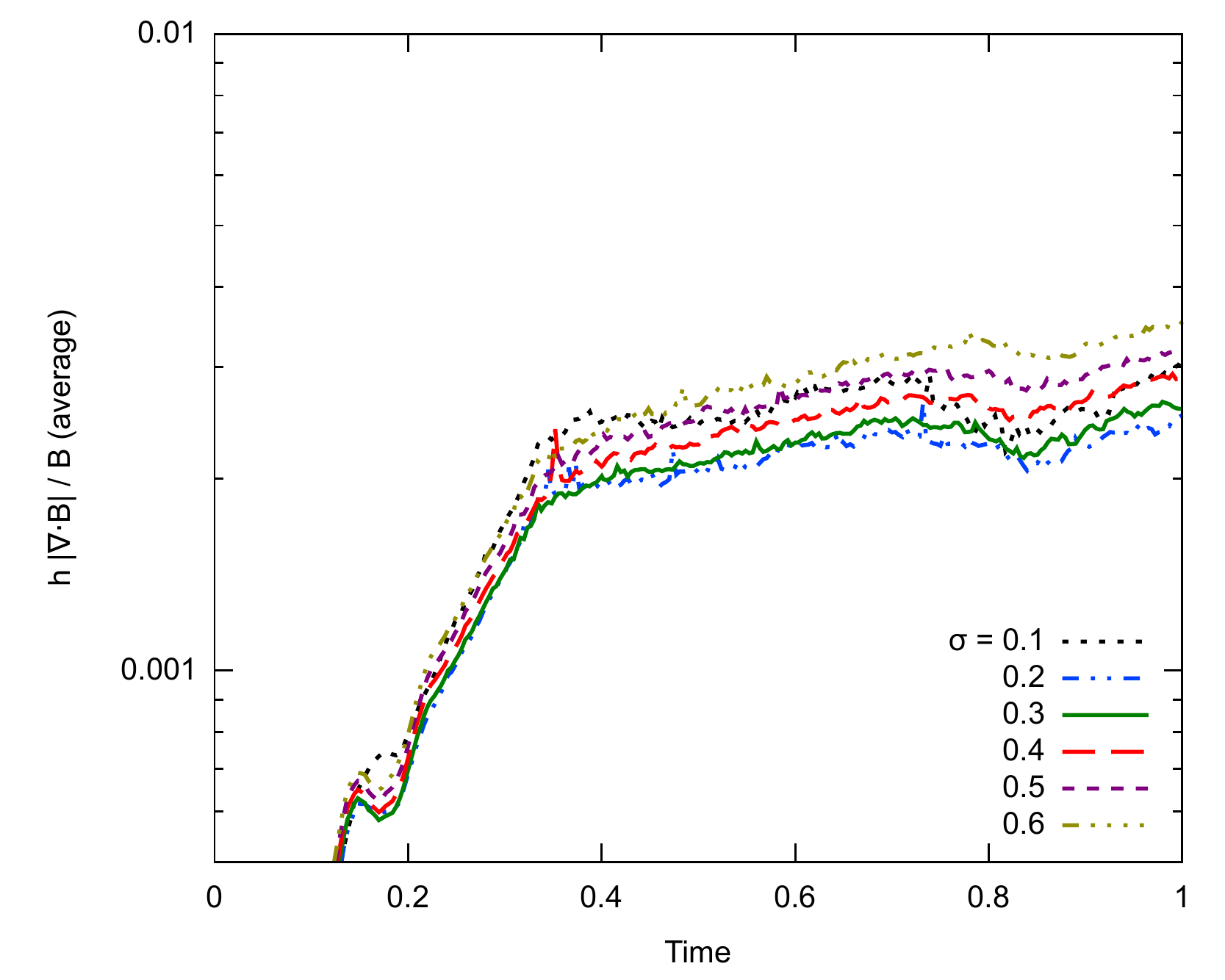}
\includegraphics[width=0.45\textwidth]{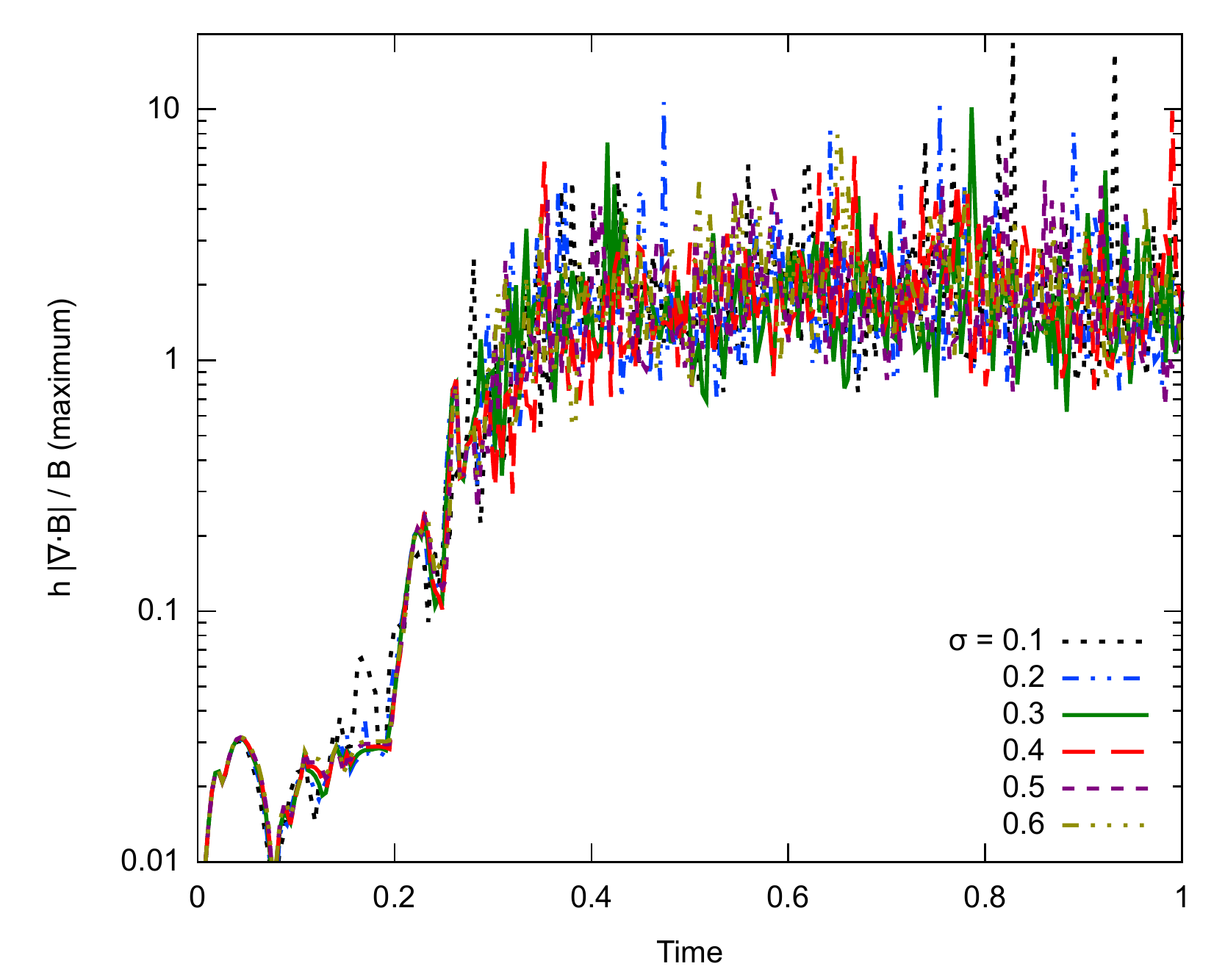}
\caption{Average (left) and maximum (right) divergence error in the Orszag-Tang vortex problem, varying the damping parameter $\sigma$.  The best results are obtained with values $\sim 0.2 - 0.3$.}
\label{fig:orszag-sigma}
\end{figure}

As with the previous tests, the damping parameter $\sigma$ was varied to find the best results (Fig.~\ref{fig:orszag-sigma}), which, as previously, were obtained for $0.2 < \sigma < 0.3$ for this 2D test.

\subsubsection{Resolution study}

\begin{figure}
 \centering
\includegraphics[width=0.18\textwidth]{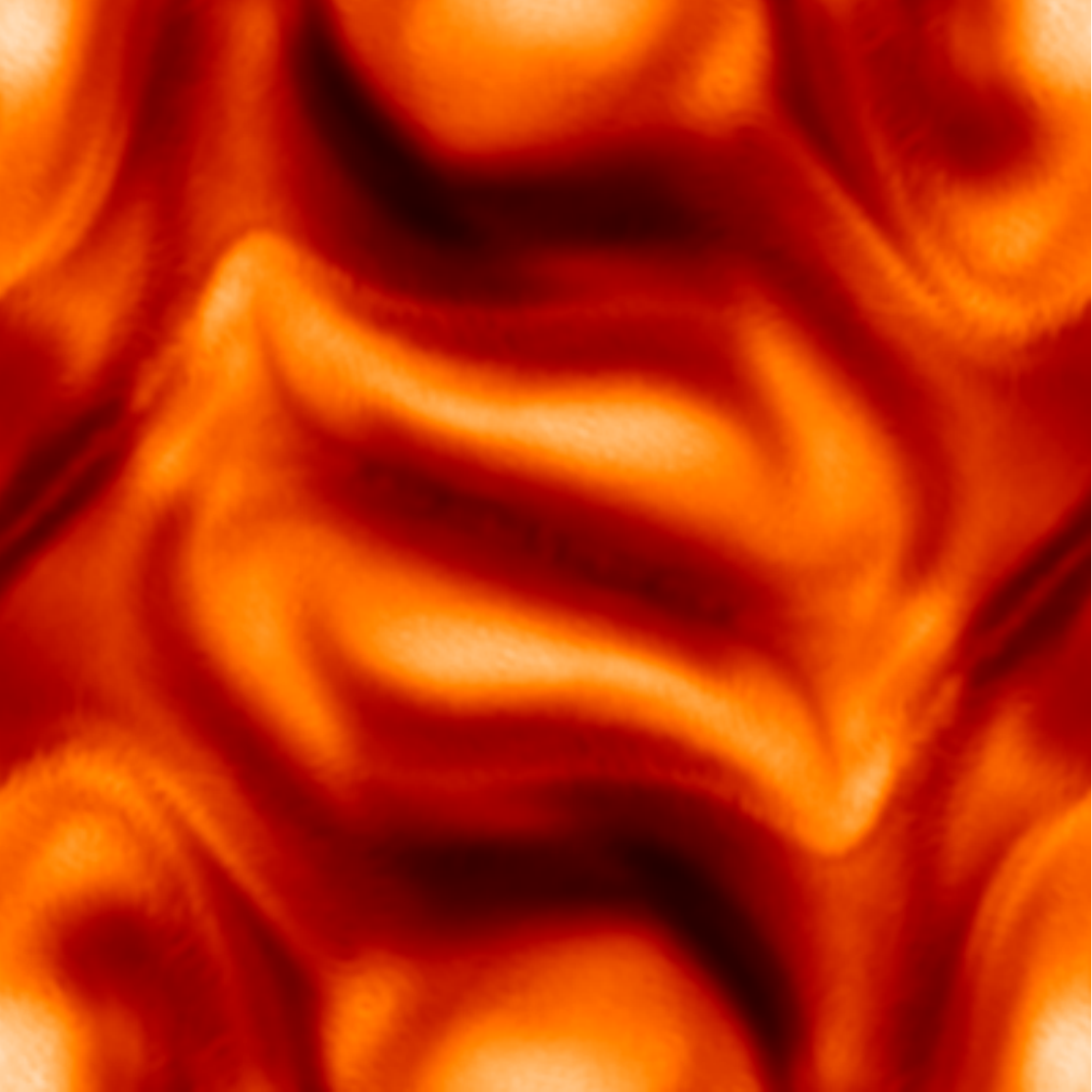}
\includegraphics[width=0.18\textwidth]{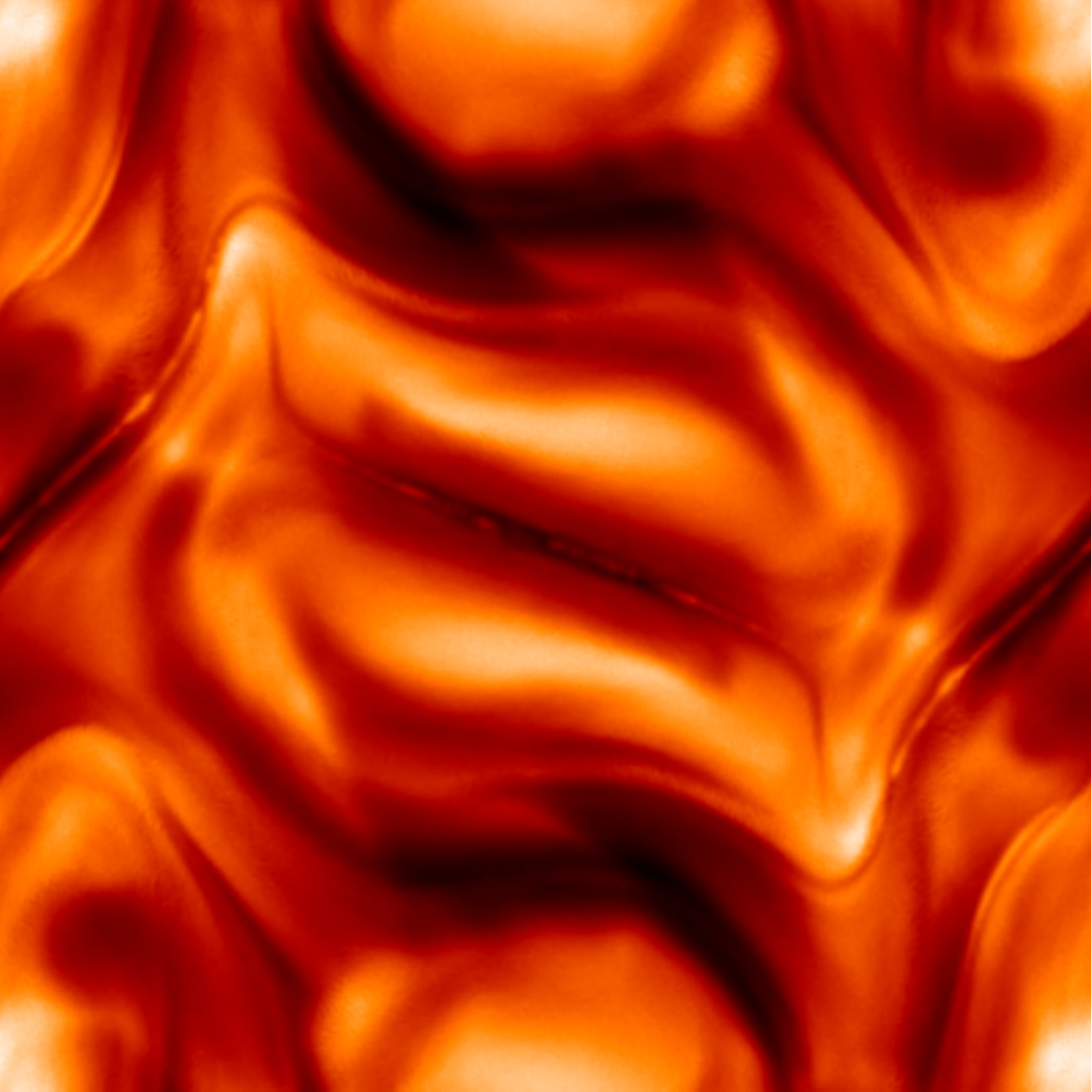}
\includegraphics[width=0.18\textwidth]{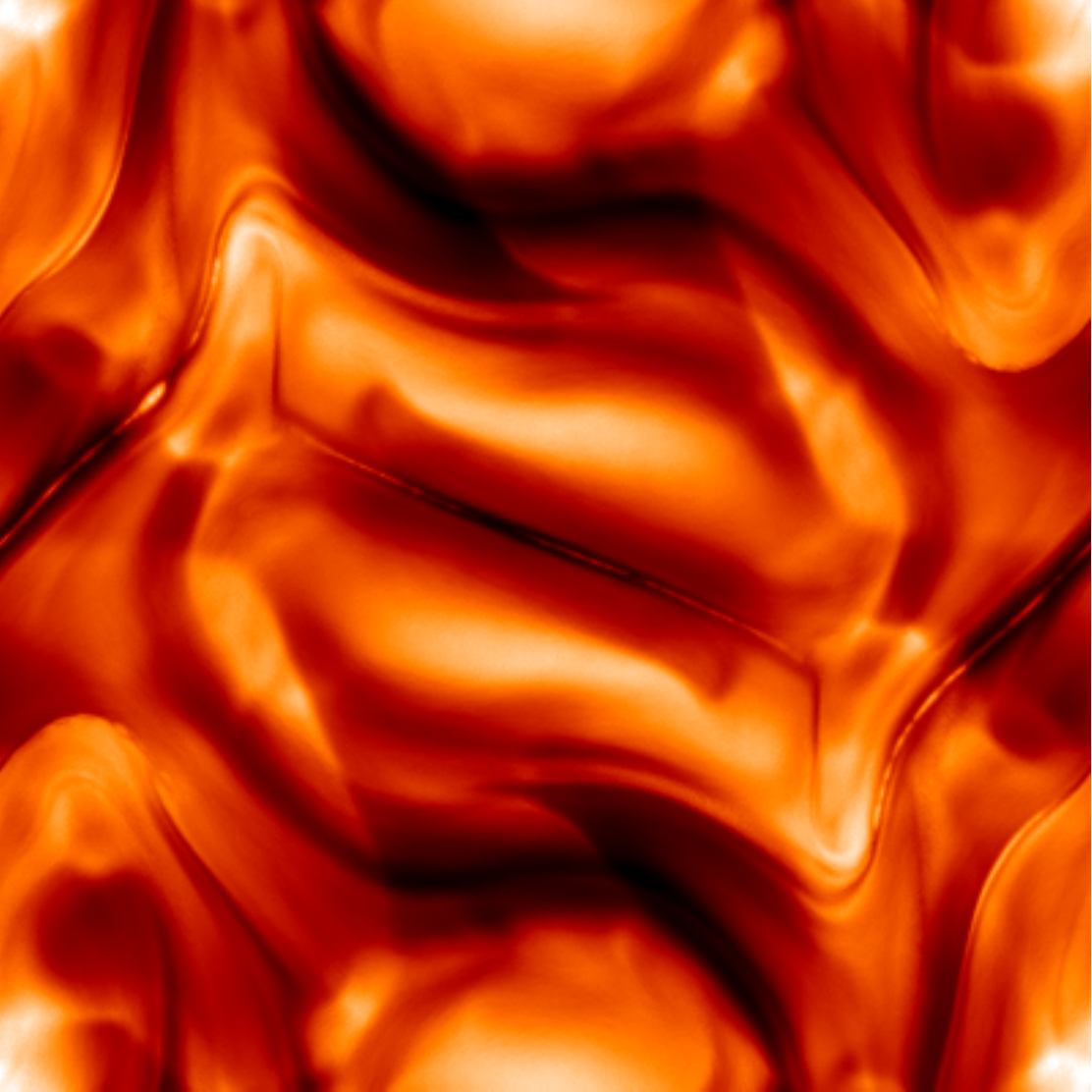}
\includegraphics[width=0.18\textwidth]{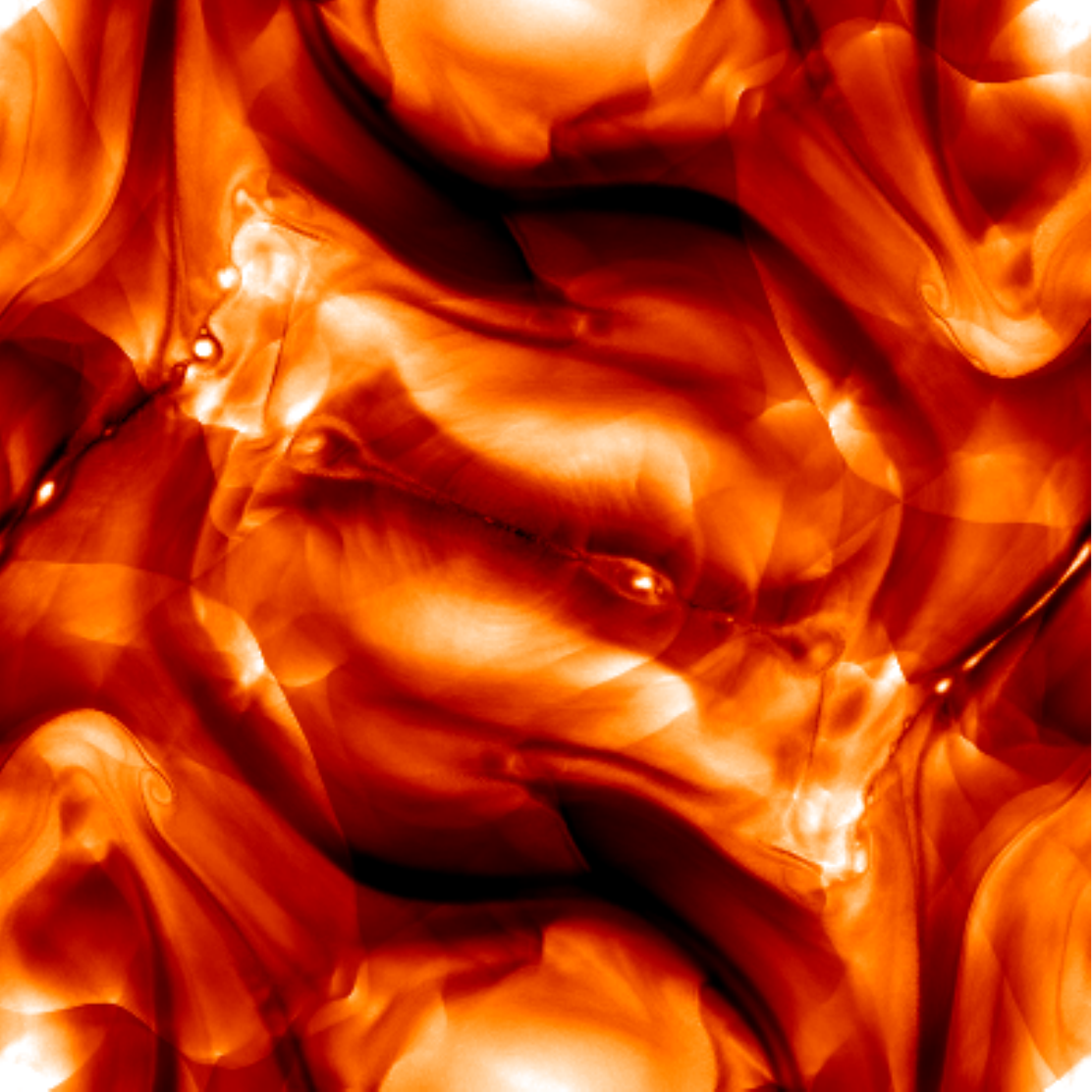}
\includegraphics[width=0.18\textwidth]{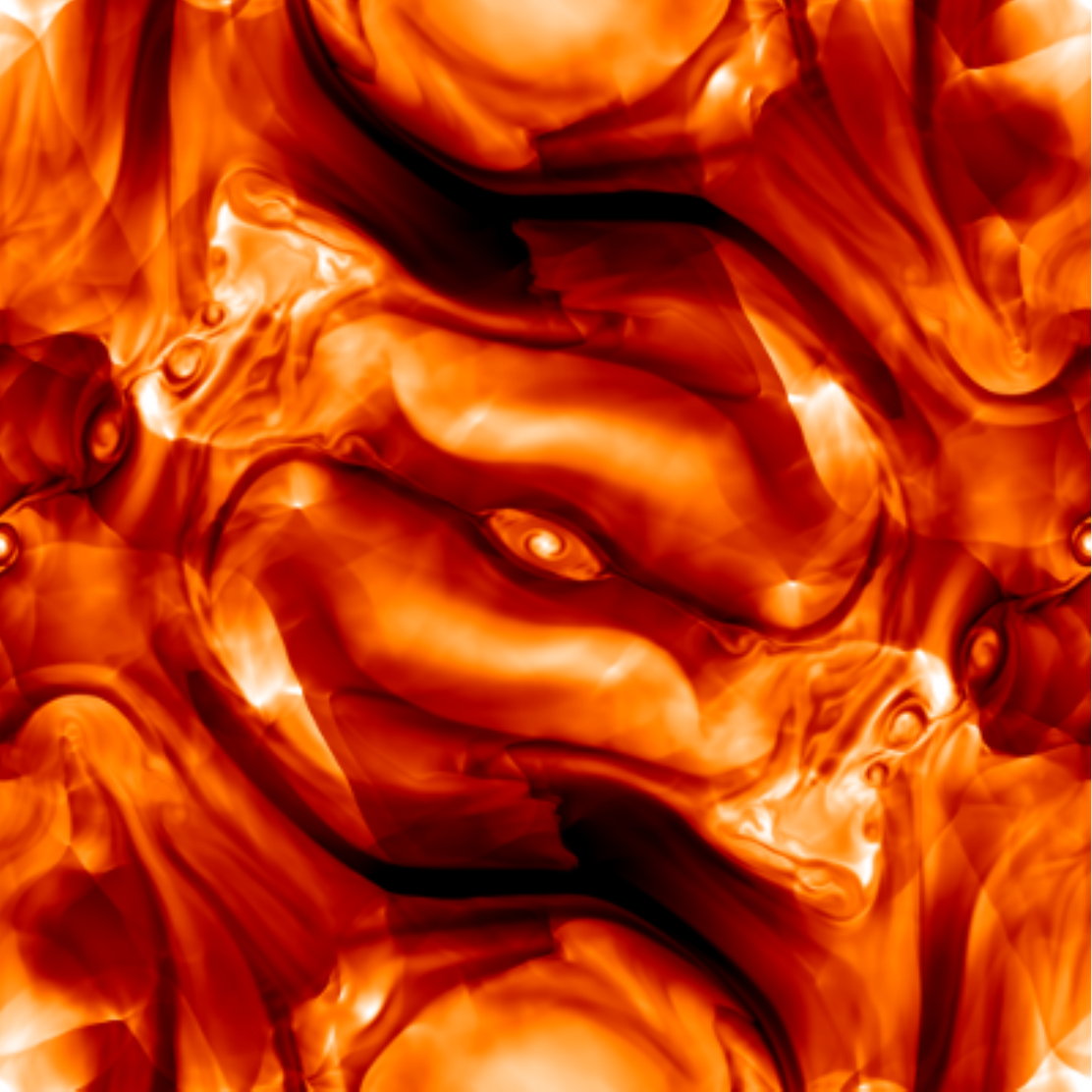}
\includegraphics[height=0.18\textwidth]{colorbar-orszag-density.pdf} \\
\caption{Density of the Orszag-Tang vortex at resolutions of $128\times148$, $256\times296$, $512\times590$, and $1024\times1182$ particles (left to right), with comparison to results obtained using the Athena code for $1024^2$ grid cells (far right).  As the resolution is increased, high density islands begin to form which is also observed in results from the Athena code.}
\label{fig:orszag-resolution-density}
\end{figure}

\begin{figure}
 \centering
\includegraphics[width=0.45\textwidth]{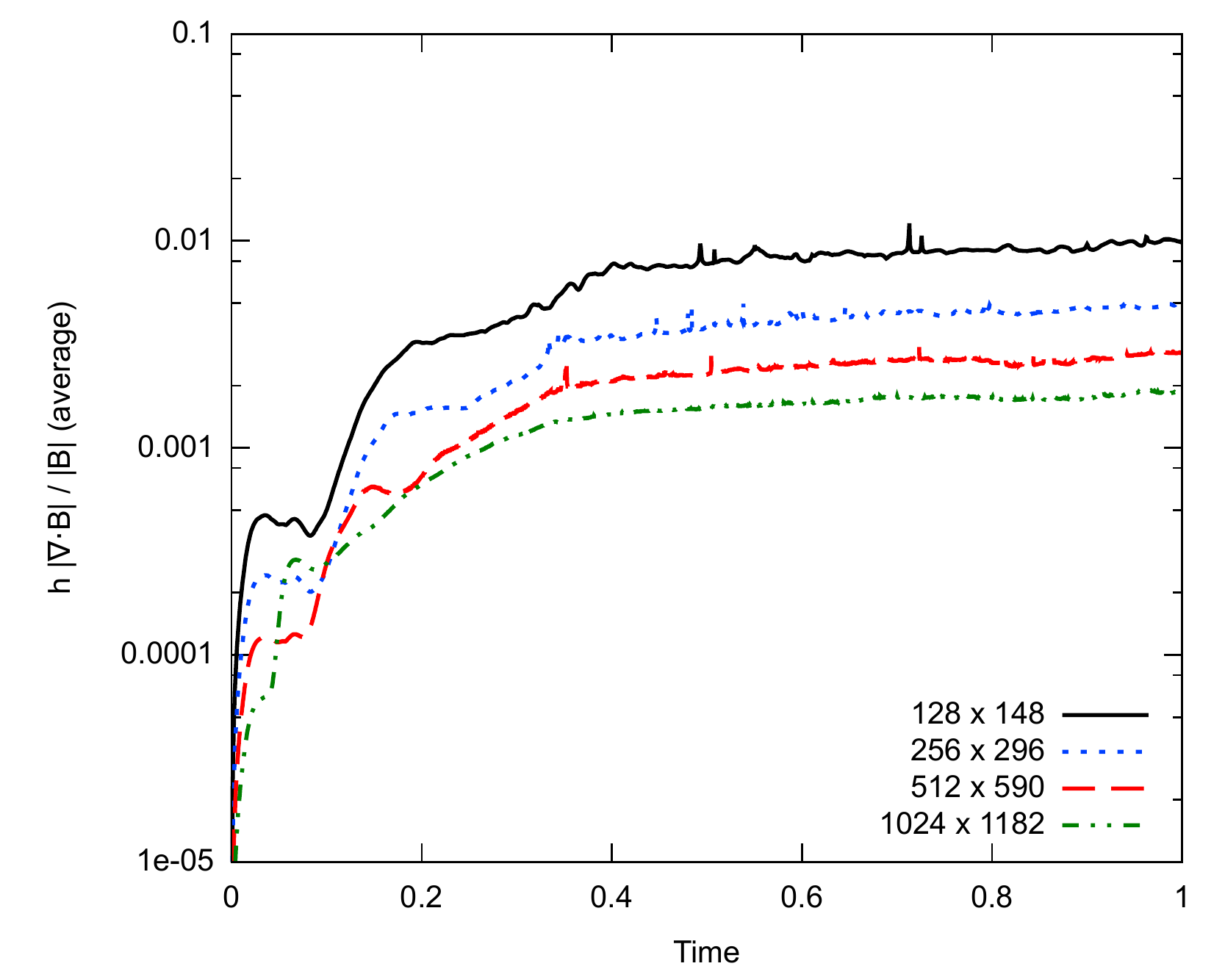}
\includegraphics[width=0.45\textwidth]{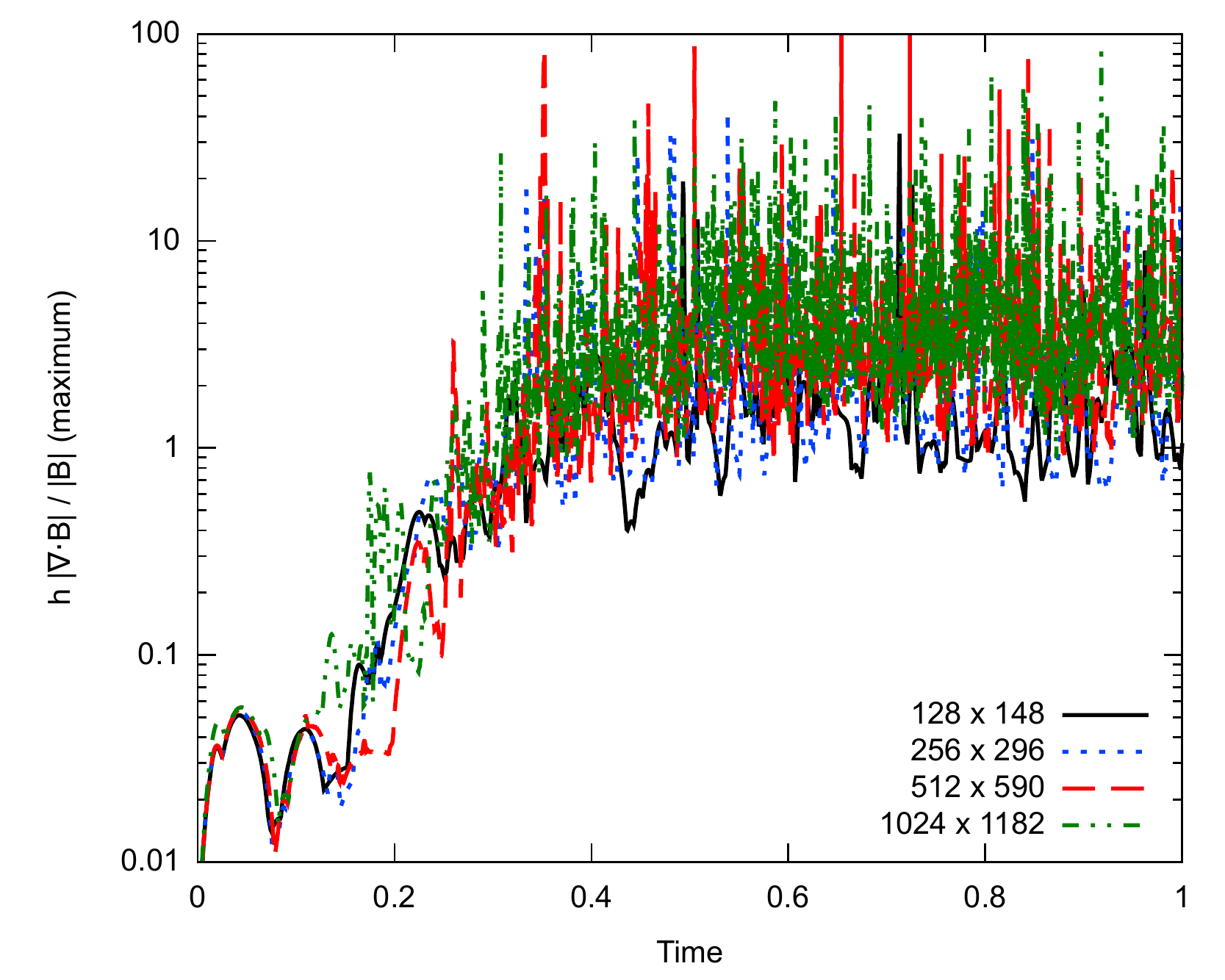}
\caption{Average (left) and maximum (right) divergence error in the Orszag-Tang vortex at resolutions of $128\times148$, $256\times296$, $512\times590$, and $1024\times1182$ particles.  The maximum divergence error remains similar for the different resolutions, but the average divergence error decreases with increasing resolution.}
\label{fig:orszag-resolution}                                            
\end{figure}

Finally, the Orszag-Tang vortex test was performed for a series of increasing resolution: $128\times148$, $256\times296$, $512\times590$, and $1024\times1182$ particles.  Divergence cleaning, without resistivity, was used for all cases.  The densities of these runs at $t=1.0$ are shown in Fig.~\ref{fig:orszag-resolution-density}, along with results obtained using the Athena code \cite{2008ApJS..178..137S} with $1024^2$ grid cells.  In the largest resolution case, high density islands begin to form in the solution.  These features are also exhibited in the results from the Athena code, and can be seen at lower resolutions in SPMHD when the Euler Potentials are used (see Fig.~\ref{fig:orszag-compilation} for an example).  Fig.~\ref{fig:orszag-resolution} shows the average and maximum divergence error (left and right panels, respectively).  Though the maximum error remains similar for all cases, the average is seen to decrease with increasing resolution.

\subsection{Three dimensional divergence advection}
\label{sec:adv3d}
 We now turn to 3D tests, beginning with a three dimensional generalisation of the divergence advection problem.  In particular, we wish to determine the optimal values for $\sigma$ when the divergence waves propagate in three dimensions rather than two.

\subsubsection{Setup}

The principle of the test remains similar to 2D versions, except a cubic volume of fluid is used in the region $x,y,z \in [-0.5, 1.5]$.  The initial velocity field is extended to ${\bf v} = [1,1,1]$ to add drift in the $z$-direction.  The magnetic field remains as previously, $B_z = 1 / \sqrt{4 \pi}$, with a spherical perturbation introduced to the $x$-component of the field as given by Eq.~\ref{eq:adv-divergence-perturbation}, except now using $r = \sqrt{x^2 + y^2 + z^2}$.  The radial extent $r_0 = h$ is chosen to mimic a divergence error at the resolution scale.  The density and pressure remain unchanged, with $\rho = 1$, $P = 6$, and $\gamma = 5/3$.  The problem is set up on a cubic lattice with $50^3$ particles.

\subsubsection{Optimal values of the damping parameter}
\label{sec:adv3d-sigma}

\begin{figure}
 \centering
\includegraphics[width=0.45\textwidth]{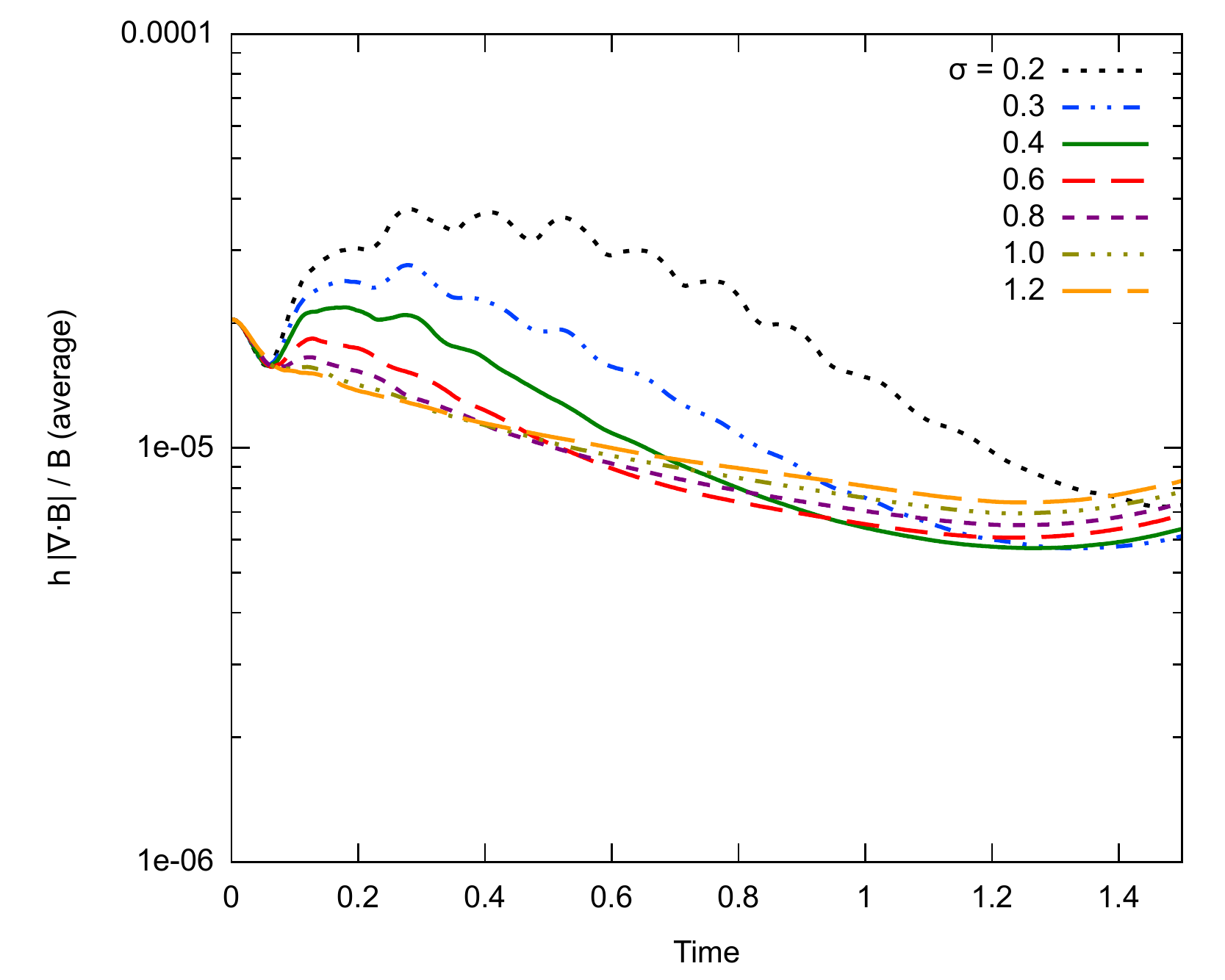}
\includegraphics[width=0.45\textwidth]{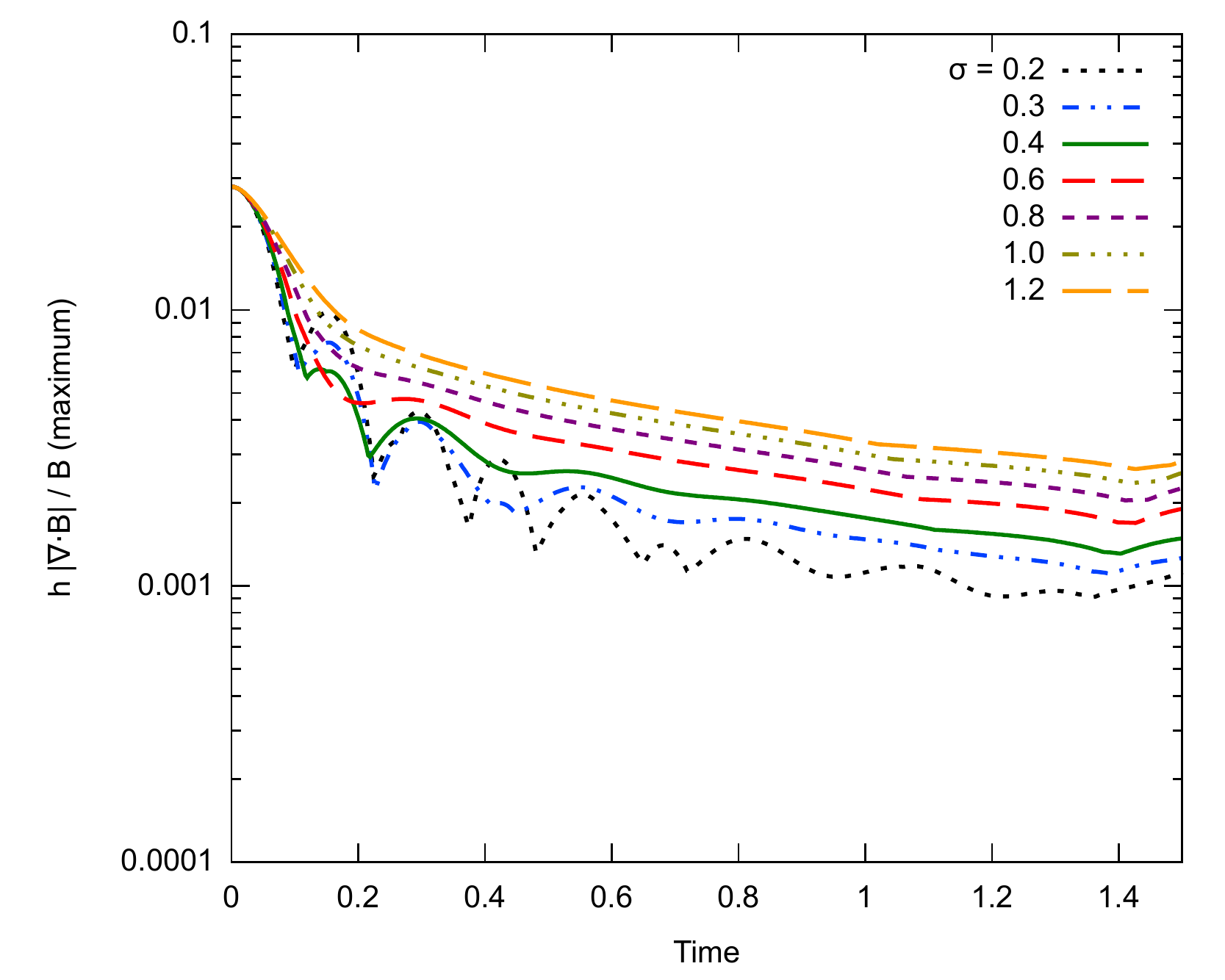}
\caption{Average and maximum divergence error in the 3D advection test for varying strengths of the damping parameter, $\sigma$.  The best results are obtained for $\sigma \sim$ 0.8--1.2.}
\label{fig:adv3d-sigma}                                            
\end{figure}

This test was performed for $\sigma \in [0.2, 1.2]$ with results of the average and maximum divergence given by Fig.~\ref{fig:adv3d-sigma}.  The optimal cleaning is obtained for $\sigma \sim$ 0.8--1.2, which differs from the optimal values obtained for the 2D tests of $\sigma \sim$ 0.2--0.3.  This is attributed to the hyperbolic wave spreading spreading over a volume instead of an area, thus being more effective than in our 2D tests, and therefore requiring a higher value of $\sigma$ to achieve critical damping.

\subsection{Gravitational collapse of a magnetised molecular cloud core}
\label{sec:jet}

Our final test is drawn from our intended application: simulations of star formation that involve magnetic fields \citep{jetletter}.  These simulations follow \cite{2007MNRAS.377...77P}, where an initial one solar mass sphere of gas with uniform magnetic field in the $z$-direction and in solid body rotation contracts under self-gravity to form a protostar with surrounding disc.  However, at times near peak density, the magnetic field in the dense central region becomes strong and can produce high divergence errors.  This has limited the range of initial magnetic field strengths which could be simulated, as if the divergence grows too large, the tensile instability correction term injects enough momentum into the system to erroneously eject the protostar out of its disc \citep{pf10b}.  Thus, this simulation proves an excellent demonstration of the capabilities of the constrained hyperbolic divergence cleaning method to reduce divergence errors in realistic, 3D simulations.

\subsubsection{Setup}

The sphere of gas has radius $R = 4\times10^{16} \text{cm}$ with uniform density $\rho = 7.43\times10^{-18}$ g $\text{cm}^{-3}$ and is set in solid body rotation with $\Omega= 1.77\times10^{-13}$ rad $\text{s}^{-1}$.  A barotropic equation of state is used, as described in \cite{2007MNRAS.377...77P}.  The magnetic field strength is set to give a mass-to-magnetic flux ratio of 5 times the critical value for magnetic fields to provide support against gravitational collapse.  To avoid edge effects with the magnetic field, the sphere is embedded in a periodic box of length $4R$ containing material surrounding the sphere set in pressure equilibrium with density ratio 1:30.  This test uses only a minimal amount of resistivity, with $\alpha_B \in [0, 0.1]$. Self-gravity is simulated using a hierarchical binary tree where each node contains mutual nearest neighbours \cite{1990ApJ...348..647B}, with gravitational force softening using the SPH kernel as described by \cite{pm07}.  The free fall time is $\sim 24000$ years.  A sink particle is inserted once the gas density surpasses $\rho_{\text{sink}} = 10^{-10}$ g $\text{cm}^{-3}$, and accretes particles within a radius of $6.7$ AU.  

\subsubsection{Results}
\label{sec:star-results}
\begin{figure}
\centering
\includegraphics[width=0.9\textwidth]{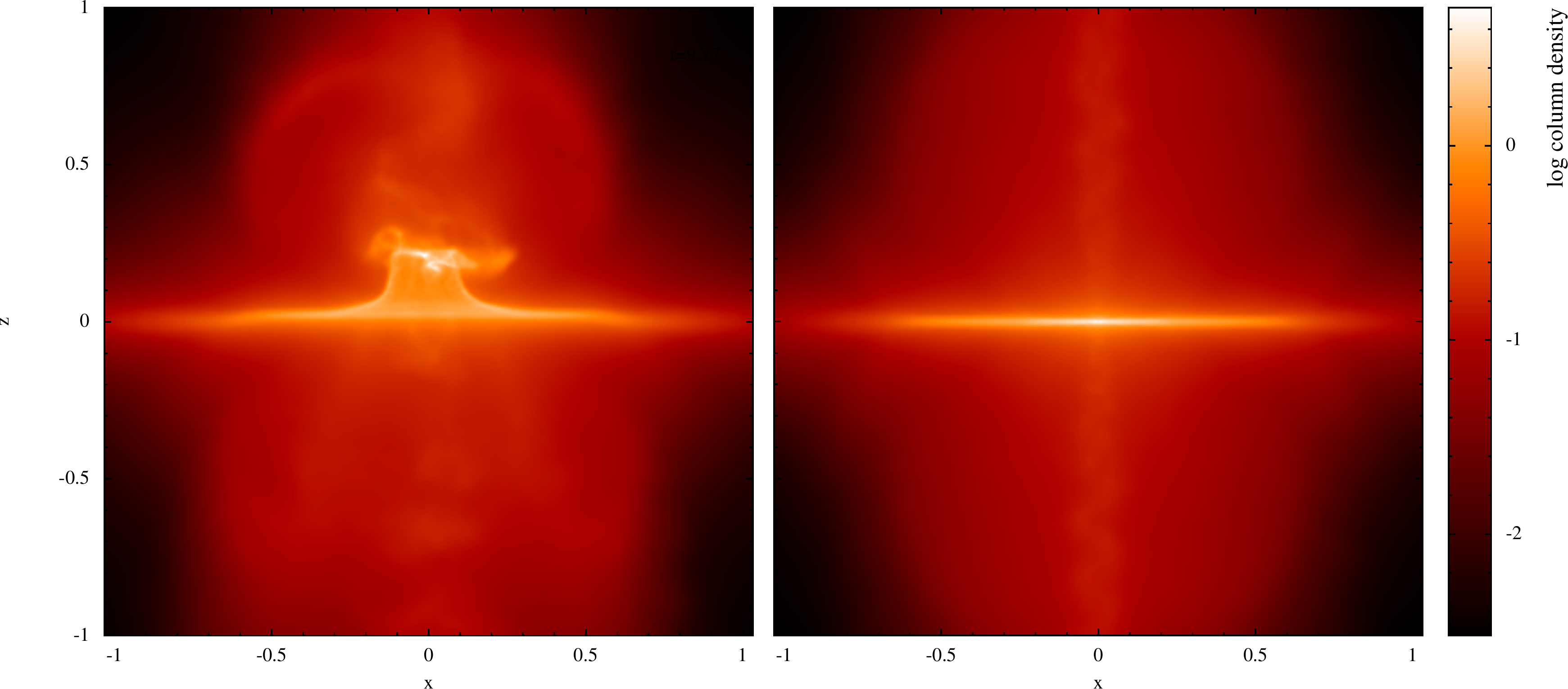}
\caption{Renderings of the column density of the star formation simulation at $t=1.1$ free fall times. The simulation without cleaning (left) suffers a dramatic loss of momentum conservation (c.f. Fig.~\ref{fig:star-mom}) induced by high divergence errors (c.f. Fig.~\ref{fig:star-divb}). By contrast, the simulation with our new divergence cleaning scheme applied (right) remains stable and launches a steady, collimated outflow \citep{jetletter}.}
\label{fig:star-column-density}
\end{figure}

\begin{figure}
\centering
\begin{minipage}[t]{0.45\textwidth}
\includegraphics[width=\textwidth]{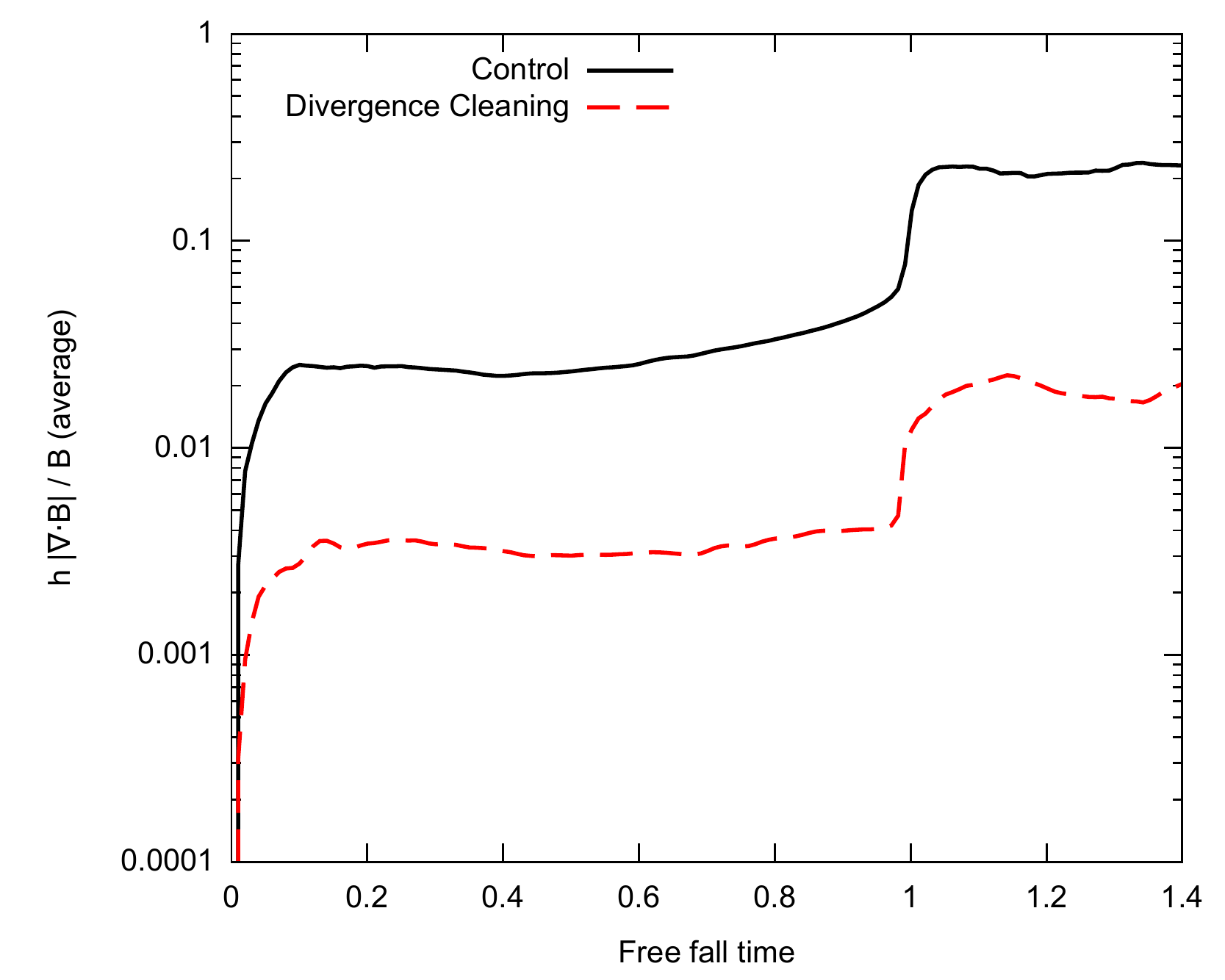}
\caption{Average divergence error as a function of time for the star formation simulation, which shows that adding divergence cleaning reduces the divergence error by an order of magnitude.}
\label{fig:star-divb}
\end{minipage}
\hspace{0.05\textwidth}
\begin{minipage}[t]{0.45\textwidth}
\includegraphics[width=\textwidth]{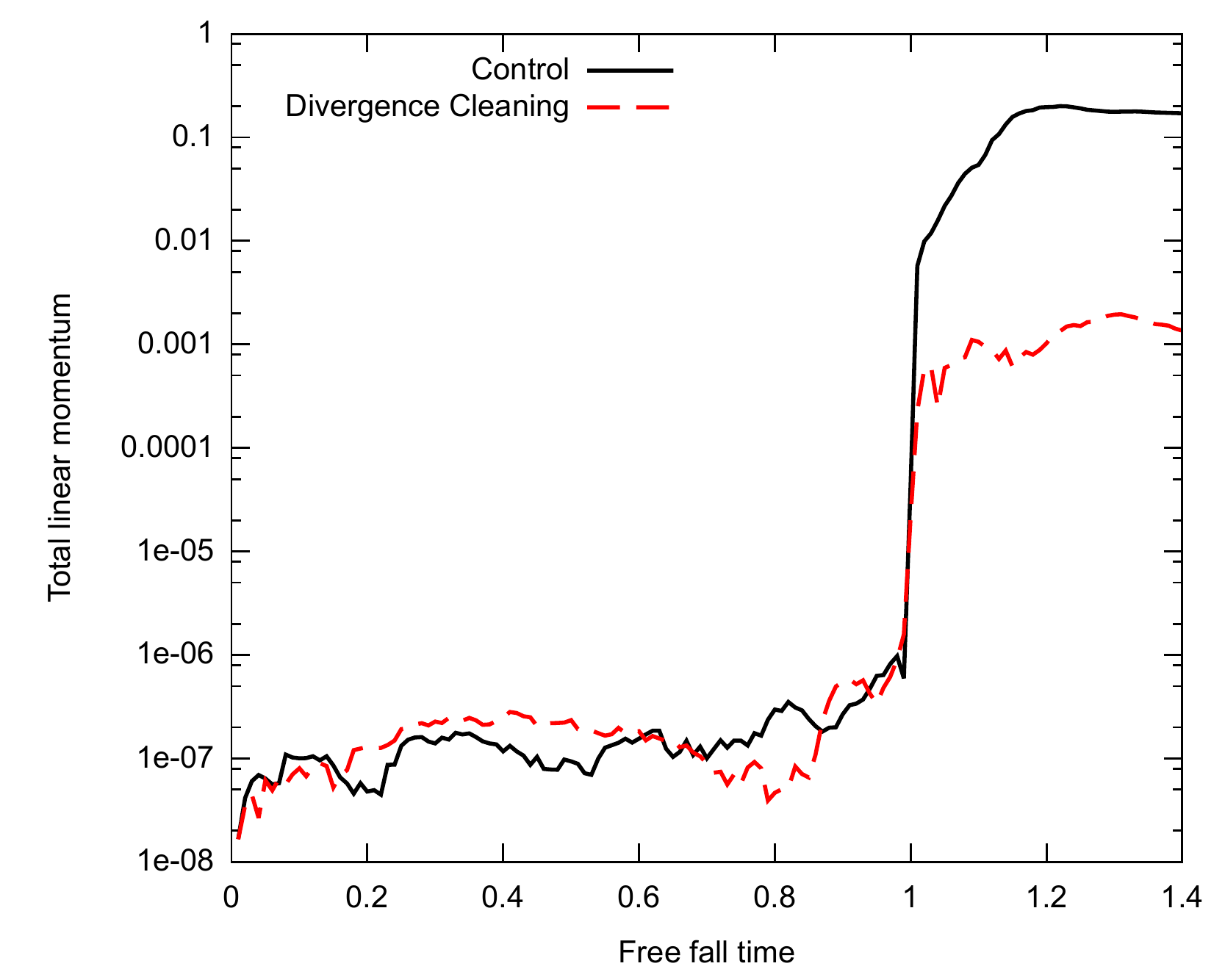}
\caption{Magnitude of the total linear momentum in the star formation simulation. The system initially has zero net momentum, which increases due to the magnetic tensile instability correction and tree-based gravitational forces.  After the protostar forms ($t=1$), the momentum conservation in the divergence cleaning case is improved by two orders of magnitude over the control case.}
\label{fig:star-mom}
\end{minipage}
\end{figure}

Fig.~\ref{fig:star-column-density} shows column density comparisons of simulations with (right) and without (left) divergence cleaning at $t=1.1$ free fall time, showing that drastic improvements to the results are obtained by incorporating divergence cleaning.  Most importantly, the protostar remains stable in its disc.  The average and maximum divergence error are both reduced by roughly an order of magnitude (Fig.~\ref{fig:star-divb}), and this leads to a corresponding improvement in the momentum conservation of around two orders of magnitude (Fig.~\ref{fig:star-mom}).

\subsubsection{Optimal sigma values}

\begin{figure}
 \centering
\includegraphics[width=0.45\textwidth]{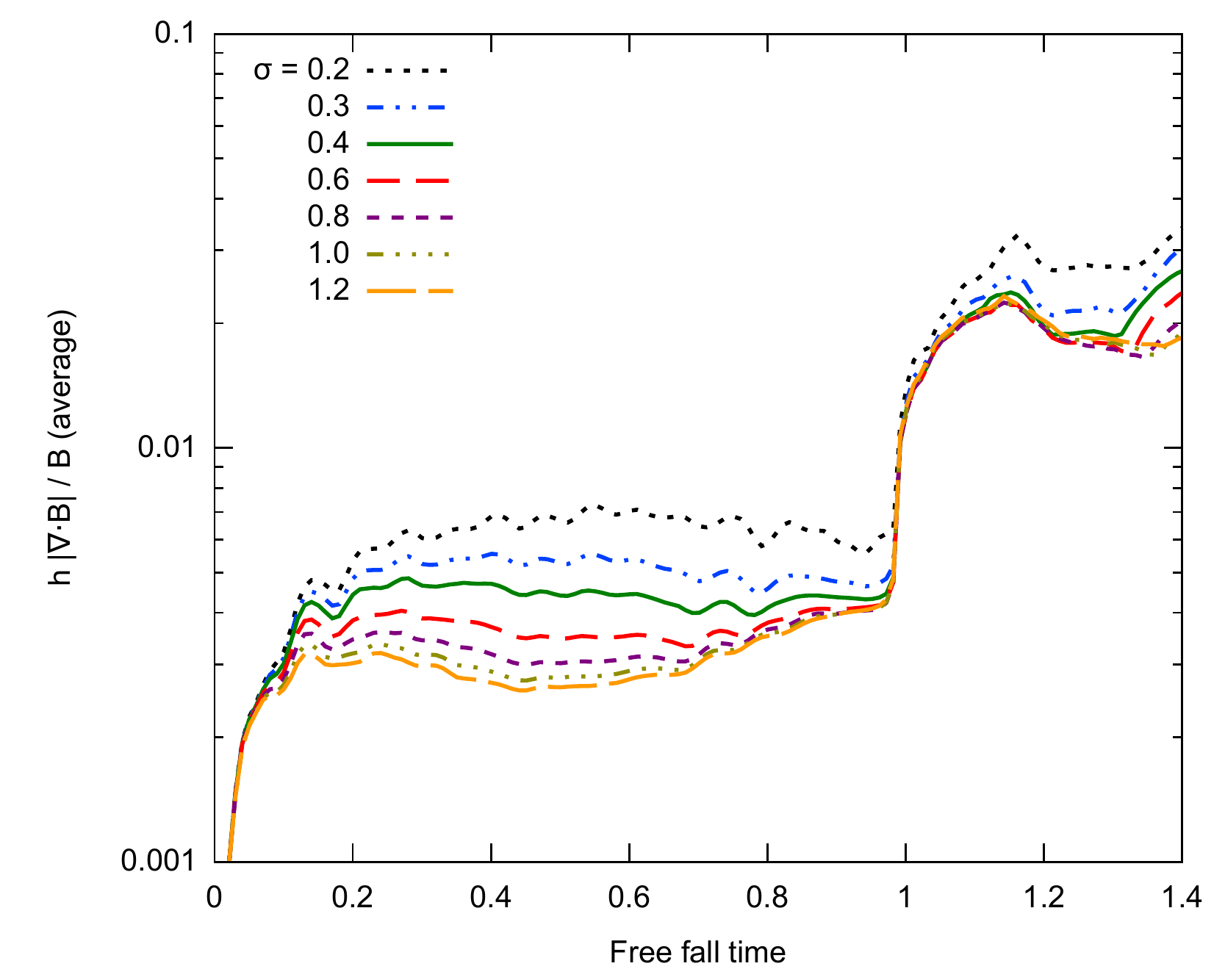}
\includegraphics[width=0.45\textwidth]{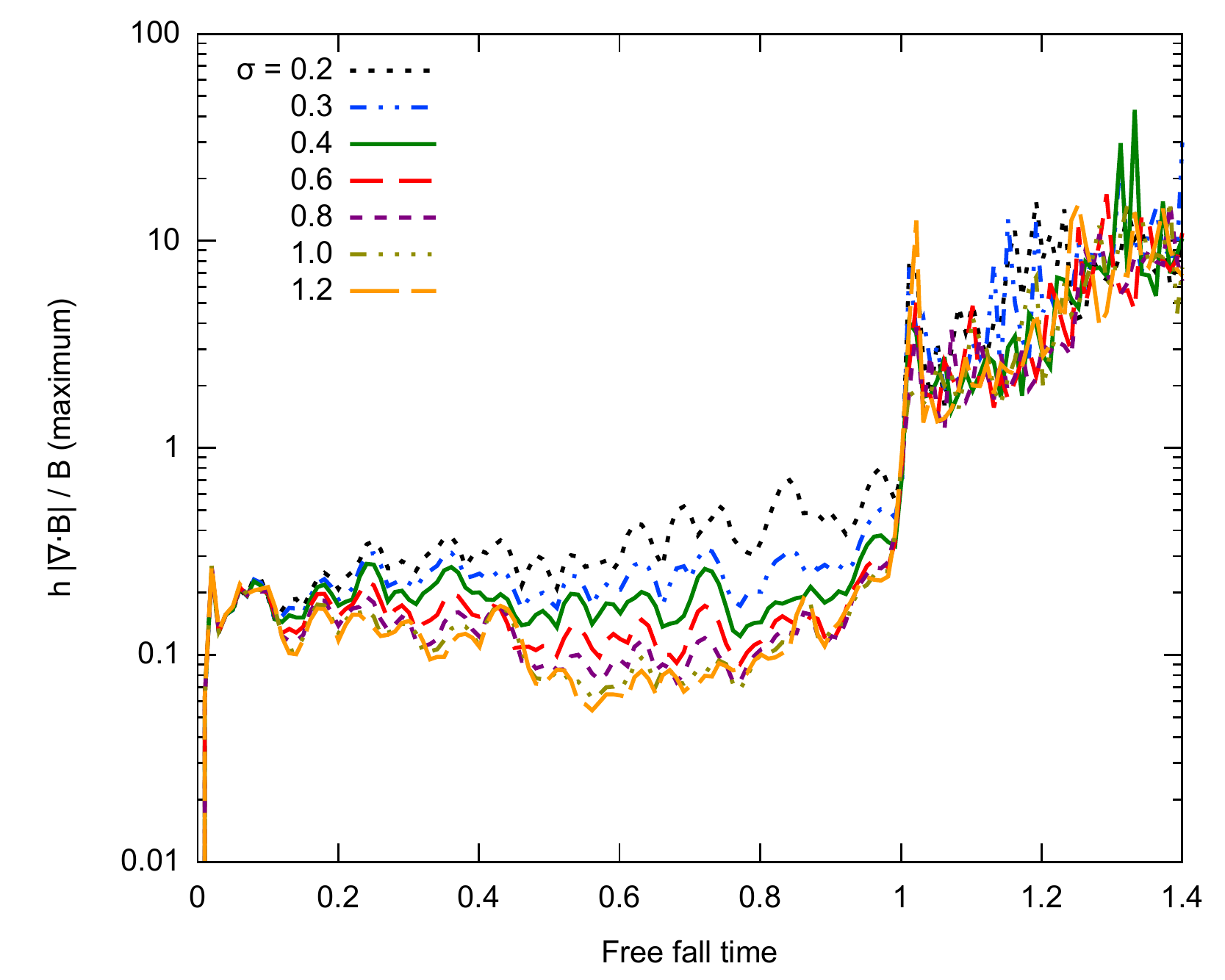}
\caption{Average and maximum divergence error in the star formation simulation, varying the damping parameter in the range $\sigma \in [0.2,1.2]$.  The best results are obtained with $\sigma \sim$ 0.8--1.2.}
\end{figure}

This simulation was repeated for several values of the damping parameter in the range $\sigma \in [0.2-1.2]$.  Optimal results were obtained for $\sigma \sim$ 0.8--1.2, which agrees with values found for the 3D advection test (\S\ref{sec:adv3d-sigma}).

\subsubsection{Inclusion of the $\tfrac{1}{2} \psi (\nabla \cdot {\bf v})$ term}

Adding $\tfrac{1}{2} \psi (\nabla \cdot {\bf v})$ to the evolution equation for $\psi$ was motivated by energy conservation considerations, but the resulting question is what effect this has on divergence cleaning. The star formation simulation represents the ideal test case with which to examine this, with a large $\nabla\cdot{\bf v}$ present due to the gravitational collapse of the gas.  We have performed this simulation both with and without this term, using $\sigma=0.8$, and found no distinguishable difference in the linear momentum, and average and maximum $h \vert \nabla\cdot{\bf B} \vert / \vert {\bf B} \vert$ profiles. Similar results were obtained also found in the other tests. We conclude that, although this term is necessary for strict energy conservation, it has almost zero effect on the effectiveness of the cleaning scheme.

\section{Summary}
\label{sec:summary}
In this paper we have developed an implementation of Dedner et al's hyperbolic divergence cleaning for SPMHD that is constrained to be numerically stable and to always decrease the magnetic energy.  To achieve this, we first defined the energy associated with the scalar $\psi$ field (\S\ref{sec:continuum-energy-conservation}). This term was used to show that when the density varies over time, the evolution equation of $\psi$ should be modified to include a $-\tfrac{1}{2} \psi (\nabla\cdot {\bf v})$ term in order to conserve energy.

 In \S\ref{sec:sph_energy_conserv} we derived an energy conserving formulation of divergence cleaning for SPMHD.  By using the $\psi$ energy term, we showed that if a difference operator is chosen to discretise $\nabla\cdot{\bf B}$ in the ${\rm d}\psi/{\rm d}t$ equation, then the conjugate, symmetric operator for $\nabla\psi$ should be used (\S\ref{sec:spmhd-clean-diff}).  Similarly, with symmetric $\nabla\cdot{\bf B}$, difference $\nabla\psi$ should be used in the induction equation (\S\ref{sec:spmhd-clean-symm}).  Use of conjugate operators was found to be the key to a numerically stable formulation.  In \S\ref{sec:spmhdenergy}, we presented the correct SPMHD form of the $-\tfrac{1}{2}\psi(\nabla\cdot{\bf v})$ term, and in \S\ref{sec:negdef}, demonstrated that parabolic damping will always lead to negative definite changes of energy.

Tests of our constrained hyperbolic divergence cleaning were presented in \S\ref{sec:tests}.  The selection of tests were for both 2 and 3D, and were designed to evaluate our method in isolation using simple, idealised systems and also in more realistic applications. Our idealised 2D tests consisted of a divergence advection test (\S\ref{sec:divBadvection}), and variants involving a density jump (\S\ref{sec:test-density-jump}) and free boundaries (\S\ref{sec:test-free-boundaries}).  The more complex 2D tests were an MHD blast wave (\S\ref{sec:blast}) and the Orszag-Tang vortex (\S\ref{sec:ot}).  A version of the divergence advection test extended to 3D was used in \S\ref{sec:adv3d}.  Results from the gravitational collapse of a molecular cloud core, representing our most challenging test case and a gauge of divergence cleaning applied to ``real'' applications, were presented in \S\ref{sec:jet}.  From the results of these tests, we draw the following conclusions:
\begin{enumerate}
\item[i)] Constrained hyperbolic/parabolic divergence cleaning provides an effective method of maintaining the divergence constraint in SPMHD, typically maintaining the average $h \vert \nabla\cdot{\bf B} \vert / \vert {\bf B} \vert$ to between 0.1--1\%.
\item[ii)] The constrained formulation using conjugate operators for $\nabla\cdot{\bf B}$ and $\nabla\psi$ is stable at density jumps and free boundaries, in contrast to previous implementations.
\item[iii)] We strongly recommend cleaning using the difference operator for $\nabla\cdot{\bf B}$. Cleaning using the symmetric operator was not found to provide any advantage over the difference operator in terms of momentum conservation and was found to dissipate physical components of the magnetic field as well as the divergence error.
\item[iv)] Constrained divergence cleaning is more effective than artificial resistivity at reducing the divergence error, and still reduces the divergence error further when used in combination with resistivity.
\item[v)] Divergence cleaning can provide an improvement of up to two orders of magnitude in momentum conservation when applied to realistic, 3D simulations.
\item[vi)] Optimal values for the damping parameter $\sigma$ were found to be $\sigma =$ 0.2--0.3 in 2D and $\sigma =$ 0.8--1.2 in 3D for all of the test problems considered in this paper.
\item[vii)] Addition of the $-\tfrac{1}{2}\psi \nabla\cdot{\bf v}$ term to the ${\rm d}\psi/{\rm d}t$ equation, while necessary for strict energy conservation of the hyperbolic cleaning equations, was found to have no noticeable effect.  even in simulations where gas is strongly compacted.  
\item[viii)] We found numerical artefacts in several problems when subtracting only $- \tfrac{1}{2}{\bf B} (\nabla \cdot{\bf B})$ in the momentum equation to counteract the tensile instability. Instead, we strongly recommend using the full $-{\bf B} (\nabla \cdot{\bf B})$ correction.
\end{enumerate}

In summary, our constrained hyperbolic divergence cleaning scheme is a robust and effective method for enforcing the divergence constraint in SPMHD simulations, providing a pathway to accurate simulation of a wide range of magnetic phenomena in astrophysics and beyond.

\section*{Acknowledgments}

We thank Matthew Bate and Evghenii Gaburov for useful discussions. This work was inspired by a conversation with Klaus Dolag at the 2010 Cosmic Magnetism conference in Kiama, NSW. T. Tricco is supported by Endeavour IPRS and APA postgraduate research scholarships. DJP acknowledges support from the Australian Research Council via Discovery Project grant DP1094585. We acknowledge the use of \textsc{splash}/\textsc{giza} \citep{splashpaper}.

%\section*{References}
\bibliographystyle{elsarticle-harv}
\bibliography{cleaning-bib}

\appendix

\section{Artificial $\psi$-dissipation term}
\label{sec:dissipation_term}
\begin{figure}
 \centering
\includegraphics[width=0.45\textwidth]{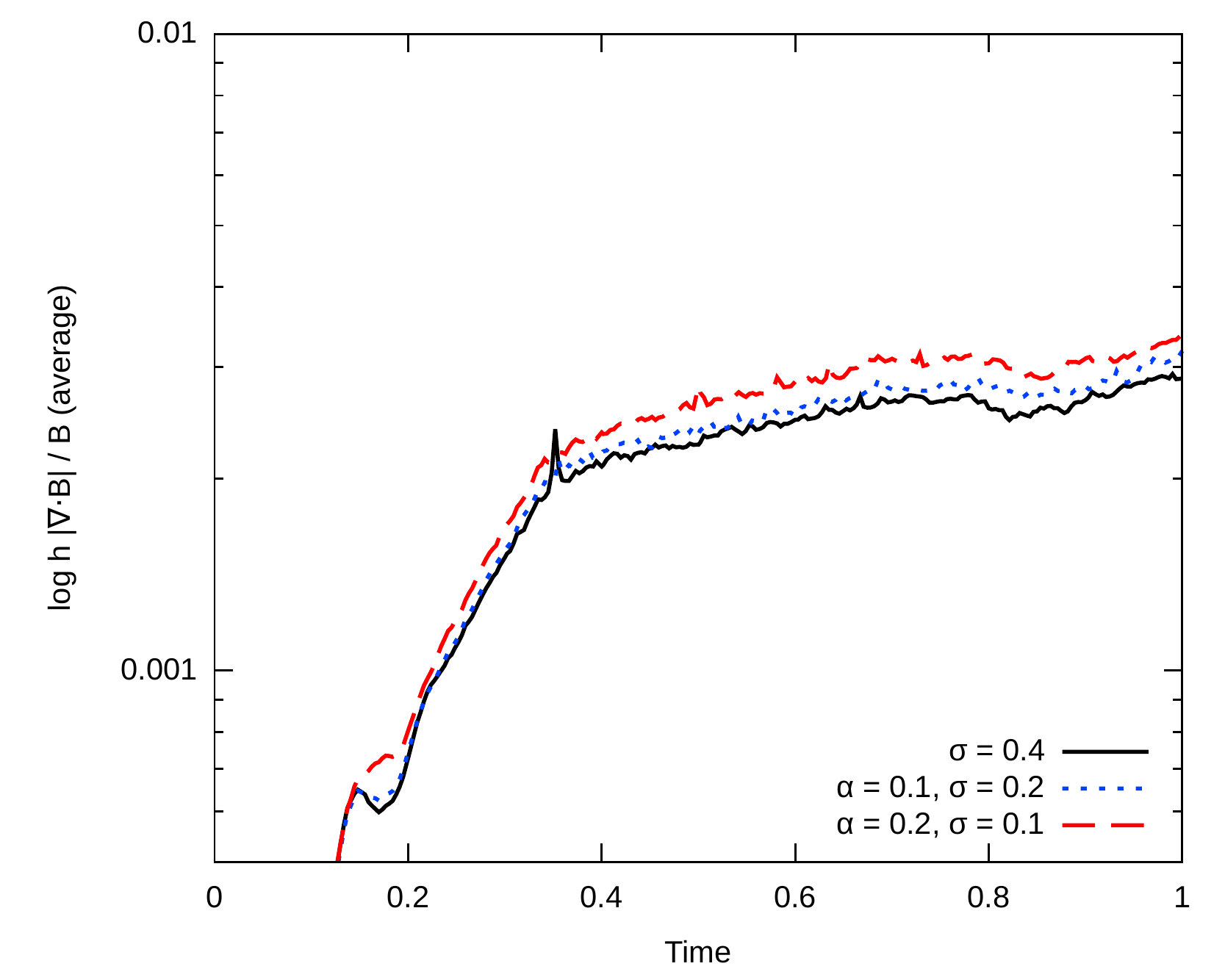}
\includegraphics[width=0.45\textwidth]{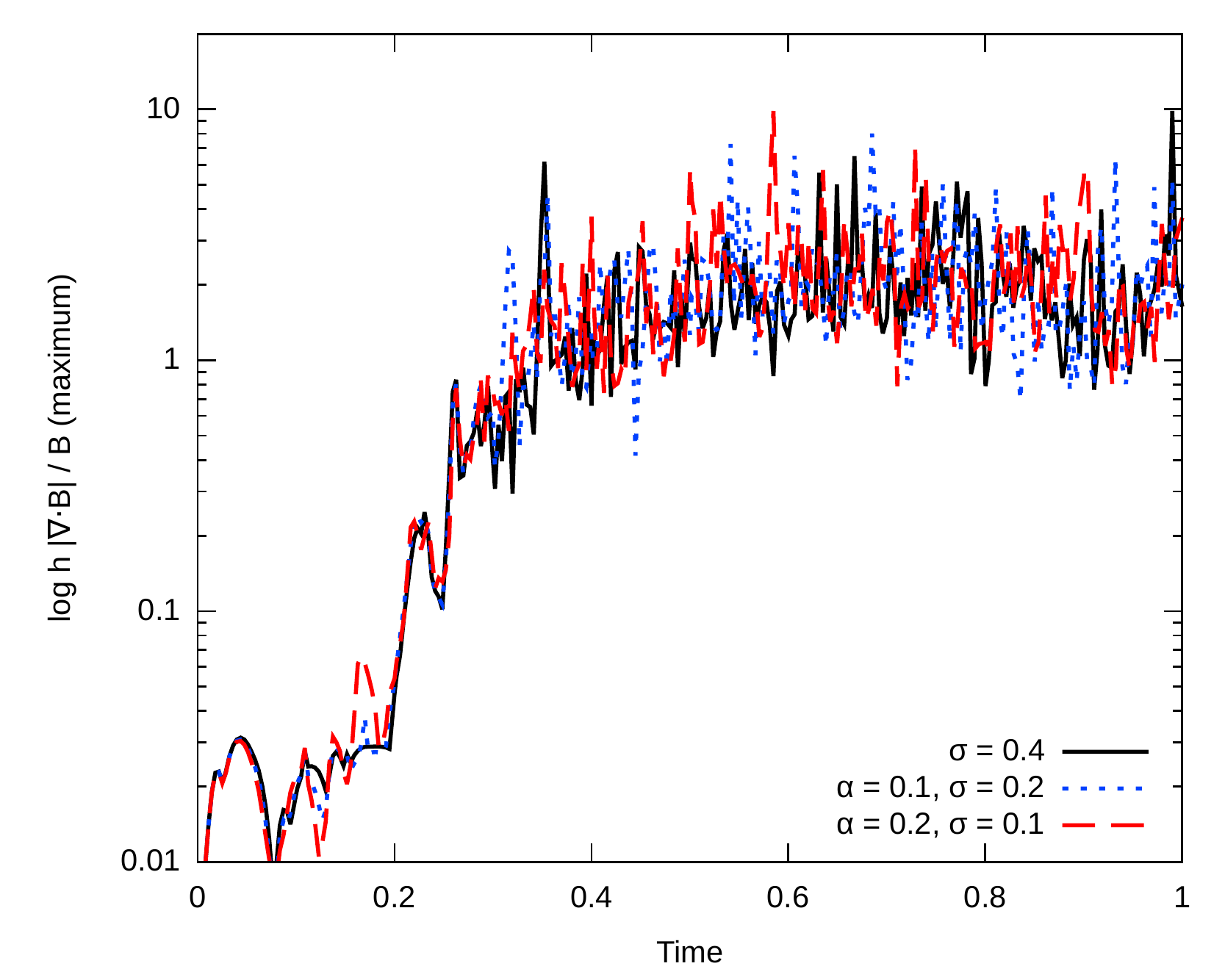}
\caption{Average and maximum divergence error when including the new, artificial $\psi$ dissipation term in the Orszag-Tang vortex test.  Values of $\alpha_\psi$ and $\sigma$ are chosen so that the combination is close to critical damping, however no benefit is noted over use of the regular damping term.}
\label{fig:dissipation}
\end{figure}

 Although the hyperbolic divergence cleaning method already includes a damping term to reduce $\psi$, we have investigated the addition of a new dissipation term, analogous to artificial resistivity or viscosity, of the form
\begin{equation}
 \left( \frac{{\rm d}\psi_a}{{\rm d}t} \right)_\text{diss} = \rho_a \sum_b m_b \frac{c_h \alpha_\psi}{\overline{\rho}_{ab}^2} \left( \psi_a - \psi_b \right) F_{ab},
 \label{eq:artificial-psi-dissipation}
\end{equation}
where $\nabla W_{ab} = \hat{\bf r}_{ab} F_{ab}$. This dissipation term is mainly designed to capture discontinuities in the $\psi$ field, motivated by our neglect of the surface integral term in Eq.~\ref{eq:psi_energy_derivation}. The term is essentially an SPH expression for a diffusion term of the form $\eta_{\psi} \nabla^{2} \psi$, where $\eta_{\psi} \propto \alpha_{\psi} c_{h} h$, which in comparison to the damping term, acts more strongly to smooth relative differences in $\psi$. This artificial $\psi$-dissipation can be used in conjunction with the damping term, however since both the damping and diffusion terms dissipate $\psi$, it is important that values of $\alpha_\psi$ and $\sigma$ be chosen carefully to avoid overdamping the system.  For example, we found that in our two dimensional tests that propagation of divergence waves were damped too severely with $\alpha_\psi=1$, and that using $\alpha_\psi, \sigma = [0.1, 0.2]$ or $[0.2, 0.1]$ yielded near critical damping (see Fig.~\ref{fig:dissipation}).

  For this dissipation term, the energy loss is given by
\begin{equation}
 \sum_a m_a \frac{\psi_a}{\mu_0 \rho_a c_h^2} \left( \frac{{\rm d}\psi_a}{{\rm d}t} \right)_{\text{diss}} = \sum_a m_a \frac{\psi_a}{\mu_0 c_h} \sum_b m_b \frac{\alpha_\psi}{\overline{\rho}_{ab}^2} \left(\psi_a - \psi_b\right) F_{ab}.
\end{equation}
This can be shown to be negative definite by splitting the RHS into two halves, performing a change of summation indices on the second half, then rejoining to obtain
\begin{equation}
 -\frac{1}{2} \sum_a m_a \frac{\alpha_\psi}{\mu_0 c_h} \sum_b m_b \frac{\left(\psi_a - \psi_b\right)^2}{\overline{\rho}_{ab}^2} F_{ab},
\end{equation}
which, since $F_{ab}$ is negative for positive kernels, gives a negative definite contribution to the total energy (and conversely would give a positive definite heat contribution).

 Inclusion of the dissipation term was tried with all test cases presented in this paper.  Similar reductions in the divergence error were obtained, however no results were improved beyond that of using the damping alone (Fig.~\ref{fig:dissipation}).

 \end{document}